%% file: EPJC.tex
\documentclass[epj]{svjour}

\usepackage{graphicx}
\usepackage{fancyhdr}
\def\makeheadbox{{
 }}
\usepackage{amssymb}
\usepackage{setspace}

\input{def.tex}

\def\mytitle{My title}
\def\myauthors{My name}
\def\mytype{My type of session}
\def\mysession{My session}

\def\mytitle{Review of Searches for Higgs Bosons and BSM Physics at the Tevatron - Page \thepage} 
\def\myauthors{Arnaud Duperrin}    
\def\mytype{Review - Page \thepage}
\def\mysession{\myauthors}

\begin{document}

%
\title{Review of Searches for Higgs Bosons and Beyond the Standard Model Physics at the Tevatron}
\author{Arnaud Duperrin\thanks{\emph{e-mail:} duperrin@cppm.in2p3.fr}
}                     
%
\institute{CPPM, IN2P3-CNRS, Universit\'{e} de la
M\'{e}diterran\'{e}e, F-13288 Marseille, France}
%
\date{May 22, 2008}
%
\input{abstract.tex}
\maketitle
%
\tableofcontents

\input{Higgs.tex}

\input{BSM.tex}

\section{Conclusion}
The Tevatron Run II collider program is scheduled to run until
October 2009 and possibly extend into 2010 to add an extra 25\%
of data, leading to an expected delivered integrated luminosity
of about 8-9~\invfb. The search for the Higgs boson and physics
beyond the standard model will greatly benefit from this
additional integrated luminosity. The accelerator performance
is excellent and provides a great opportunity for the CDF and
D\O\ experiments to meet or exceed their stated physics goals.
Both CDF and D\O\ experiments have now surpassed the 4~\invfb\
in delivered luminosity.

While the ATLAS and CMS experiments at the LHC should be in
good position to discover the Higgs boson on a time scale
similar to the one considered at the Tevatron, observation in
the \Hbb\ decay channel will be extremely difficult at the LHC.
Given the importance of observing the Higgs boson in its main
decay mode, searches at the Tevatron have thus to be viewed as
complementary to, rather than in competition with, the Higgs
boson search at the LHC. Failure to observe the Higgs boson in
the mass range considered would also be a very important
result, as it would indicate a breakdown of the standard model
and give directions for alternative theories.

Current electroweak data point to the existence of a light
Higgs, which means that the so-far elusive Higgs particle is
within reach of Tevatron. Soon the LHC will also exploit its
potential for discovery of new particles. However, a light
Higgs boson corresponds to the less favorable scenario at LHC
for an early discovery. In addition, it may take some time to
operate the ATLAS and CMS detectors and to understand the new
data. At the Tevatron, a Higgs boson in this mass range can
only be convincingly observed if the integrated luminosity
delivered is sufficiently large (8-9~\invfb) and the current
analyses continue to improve their sensitivity. Furthermore, if
its mass is heavier than $\approx$140~\GeV, the {\tt MSSM} will
be ruled out, a conclusion which applies to the majority of
supersymmetric models.

This review has summarized the searches for Higgs bosons and
beyond-the-standard-model physics at the Tevatron that have
been conducted until May 2008. However, despite all efforts, no
significant deviations from the standard model predictions have
been found to date based on data samples corresponding to
integrated luminosities of up to 2.5~\invfb. Of course, this
should not be taken to mean absence of new physics in these
data, for there are still a number of \invfb\ of frontier
physics ahead of us at the Tevatron.

\section{Acknowledgments}
I would like to thank my colleagues both at Fermilab and at
other collaborating institutions, especially those who operate
the Tevatron accelerator and construct, maintain, and calibrate
the CDF and D\O\ detectors, essential for any physics analysis
reported here.

I wish to thank all the colleagues for providing valuable input
and/or proofreading this document, in particular I would like
to address special thanks to Chris Hays, Gavin Davies, Gregorio
Bernardi, and Matthew Herndon. I am also grateful to Todd
Adams, Volker B\"uscher, Dmitri Denisov, Monica D'onofrio, Wade
Fisher, Aurelio Juste, Stefan S\"oldner-Rembold, Patrice
Verdier, and Darien Wood for their suggestions and comments
during the preparation of this review.

%
%


\end{document}

%% file: def.tex
\def\lsim{\mathrel{\rlap{\lower4pt\hbox{$\sim$}}
    \raise1pt\hbox{$<$}}}                

\newcommand{\raa}       {\mbox{$\rightarrow$}}

\newcommand{\keV}       {\mbox{keV}}
\newcommand{\TeV}       {\mbox{TeV}}

\newcommand{\GeV}       {\mbox{GeV}}

\newcommand{\invfb}     {\mbox{fb$^{-1}$}}
\newcommand{\fb}        {\mbox{fb}}
\newcommand{\invpb}     {\mbox{pb$^{-1}$}}
\newcommand{\pb}        {\mbox{pb}}
\newcommand{\mb}        {\mbox{mb}}

\newcommand{\pp}        {\mbox{$p\bar{p}$}}

\newcommand{\MSB}     {\mbox{$m_{\tilde{b}}$}}

\newcommand{\met}    {\mbox{${\hbox{$E$\kern-0.6em\lower-.1ex\hbox{/}}}_T$}} 
\newcommand{\mht}    {\mbox{${\hbox{$H$\kern-0.75em\lower-.05ex\hbox{/}}}_T$}} 
\newcommand{\HT}{\mbox{$H_T$}}
\newcommand{\pt}    {\mbox{$p_T$}}
\newcommand{\ET}    {\mbox{$E_T$}}

\newcommand{\Hbb}     {\mbox{$H \raa b \bar{b} $}}

\newcommand{\ppWH}     {\mbox{$p\bar{p} \raa WH $}}

\newcommand{\ppWHlnbb}     {\mbox{$p\bar{p} \raa WH \raa \ell \nu b\bar{b}$}}
\newcommand{\ppWHnbb}     {\mbox{$p\bar{p} \raa WH \raa \ell \kern-0.45em\lower-.05ex\hbox{/}\nu b\bar{b}$}}
\newcommand{\ppZHllbb}     {\mbox{$p\bar{p} \raa ZH \raa \ell\ell b\bar{b}$}}
\newcommand{\ppZHnnbb}     {\mbox{$p\bar{p} \raa ZH \raa \nu\bar{\nu} b\bar{b}$}}
\newcommand{\ppWHWWW}     {\mbox{$p\bar{p} \raa WH \raa WWW^{(*)} \raa \ell^{\pm} \ell^{\pm}$}}
\newcommand{\ppHWW}     {\mbox{$p\bar{p} \raa H \raa WW^{(*)}$}}

\newcommand{\ppHWWlnulnu}     {\mbox{$p\bar{p} \raa H \raa WW^{(*)} \raa \ell\nu \ell^{'}\nu$}}
\newcommand{\ppHZZ}     {\mbox{$p\bar{p} \raa H \raa ZZ^{(*)}$}}
\newcommand{\ppHgamgam} {\mbox{$p\bar{p} \raa H \raa \gamma \gamma$}}

\newcommand{\WHnbb}     {\mbox{$WH \raa \ell \kern-0.45em\lower-.05ex\hbox{/}\nu b\bar{b}$}}
\newcommand{\ZHllbb}     {\mbox{$ZH \raa \ell\ell b\bar{b}$}}
\newcommand{\ZHnnbb}     {\mbox{$ZH \raa \nu\bar{\nu} b\bar{b}$}}

\newcommand{\Hgamgam} {\mbox{$H \raa \gamma \gamma$}}
\newcommand{\Htautau} {\mbox{$H \raa \tau \tau$}}

\newcommand{\ppbbphi} {\mbox{$p\bar{p} \raa b\bar{b}\phi(h/H/A)\raa b\bar{b}b\bar{b}$}}
\newcommand{\pphbb} {\mbox{$p\bar{p} \raa b\phi(h/H/A)\raa bb\bar{b} + X$}}
\newcommand{\pphtautau} {\mbox{$p\bar{p} \raa \phi(h/H/A)\raa \tau^+\tau^-$}}

\def\pp{$p\bar{p}$}

\def\WH{$WH\rightarrow \ell\nu b\bar{b}$}

\def\whl{$WH\rightarrow \ell\nu b\bar{b}$}

\def\lmet{$WH\rightarrow \ell\kern-0.45em\raise0.19ex\hbox{/} \nu b\bar{b}$}
\def\ZH{$ZH\rightarrow \ell\ell/\nu\nu b\bar{b}$}

\def\www{$WH \rightarrow WW^{+} W^{-}$}

\def\hww{$H\rightarrow W^+ W^-$}

\newcommand{\ns}  {\mbox{${\rm ns}$}}
\newcommand{\cms}  {\mbox{${\rm cm^{-2}~s^{-1}}$}}

\newcommand{\lumi}[2]{${#1\times10^{#2}~\cms}$}
\newcommand{\mAh}  {\mbox{${\rm mA/hours}$}}

\newcommand{\DO}   {D\O\ Collaboration}
\newcommand{\CDF}  {CDF Collaboration}

\newcommand{\LEP}  {LEP Collaborations}
\newcommand{\ADLO}  {ALEPH, DELPHI, L3, and OPAL Collaborations}

\newcommand{\DOnote}[3]{D\O\ Note #1, {\it #2}, (#3)}
\newcommand{\CDFnote}[3]{CDF Note #1, {\it #2}, (#3)}
\newcommand{\CDFDOnote}[4]{CDF and D\O\ Collaborations, {\it #3}, CDF Note #1, D\O\ Note #2 (#4)}

\newcommand{\TEVNPHWG}[5]{Tevatron New Phenomena \& Higgs Working Group (TEVNPHWG), {\it #4}, CDF Note #1, D\O\ Note #2, FERMILAB-PUB-#3-E (#5)}

\newcommand{\prl}[3]{{\sl Phys.\ Rev.\ Lett.} {\bf #1}, #2 (#3)}
\newcommand{\nim}[3]{{\sl Nucl.\ Instrum.\ Methods} {\bf A#1}, #2 (#3)}
\newcommand{\plB}[3]{{\sl Phys.\ Lett.} {\bf B#1}, #2 (#3)}
\newcommand{\pl}[3]{{\sl Phys.\ Lett.} {\bf #1}, #2 (#3)}
\newcommand{\jhep}[3]{{\sl J.\ High Energy Phys.} {\bf #1}, #2 (#3)}
\newcommand{\cpc}[3]{{\sl Comput.\ Phys.\ Commun.} {\bf #1}, #2 (#3)}
\newcommand{\prD}[3]{{\sl Phys.\ Rev.} {\bf D#1}, #2 (#3)}
\newcommand{\prDR}[3]{{\sl Phys.\ Rev.} {\bf D#1}, #2(R) (#3)}
\newcommand{\npB}[3]{{\sl Nucl.\ Phys.} {\bf B#1}, #2 (#3)}

\newcommand{\epjC}[3]{{\sl Eur.\ Phys.\ J.} {\bf C#1}, #2 (#3)}
\newcommand{\pr}[3]{{\sl Phys.\ Rep.} {\bf #1}, #2 (#3)}

\newcommand{\mpl}[3]{{\sl Mod.\ Phys.\ Lett.} {\bf A#1}, #2 (#3)}
\newcommand{\ppnp}[3]{{\sl Prog.\ Part.\ Nucl.\ Phys.} {\bf #1}, #2 (#3)}
\newcommand{\ijmp}[3]{{\sl Int.\ J.\ Mod.\ Phys.} {\bf A#1}, #2 (#3)}
\newcommand{\nc}[3]{{\sl Nuovo.\ Cimento.} {\bf A#1}, #2 (#3)}
\newcommand{\jpG}[3]{{\sl J.\ Phys.} {\bf G#1}, #2 (#3)}

\newcommand{\hepph}[1]{arXiv:hep-ph/#1}
\newcommand{\hepex}[1]{arXiv:hep-ex/#1}


%% file: abstract.tex
\abstract{The energy frontier is currently at the Fermilab
Tevatron accelerator, which collides protons and antiprotons at
a center-of-mass energy of 1.96~\TeV. The luminosity delivered
to the CDF and D\O\ experiments has now surpassed the 4~\invfb.
This paper reviews the most recent direct searches for Higgs
bosons and beyond-the-standard-model ({\tt BSM}) physics at the
Tevatron. The results reported correspond to an integrated
luminosity of up to 2.5~\invfb\ of Run II data collected by the
two Collaborations. Searches covered include: the standard
model ({\tt SM}) Higgs boson (including sensitivity
projections), the neutral Higgs bosons in the minimal
supersymmetric extension of the standard model ({\tt MSSM}),
charged Higgs bosons and extended Higgs models, supersymmetric
decays that conserve or violate {\it R}-parity, gauge-mediated
supersymmetric breaking models, long-lived particles,
leptoquarks, compositeness, extra gauge bosons, extra
dimensions, and finally signature-based searches. Given the
excellent performance of the collider and the continued
productivity of the experiments, the Tevatron physics potential
looks promising for discovery with the coming larger data sets.
In particular, evidence for the {\tt SM} Higgs boson could be
obtained if its mass is light or near 160~\GeV. The observed
(expected) upper limits are currently a factor of 3.7~(3.3)
higher than the expected {\tt SM} Higgs boson cross section at
$m_H=115~\GeV$ and 1.1~(1.6) at $m_H=160~\GeV$ at 95\%~C.L.
%
} 

%% file: Higgs.tex
\begin{figure*}
 \begin{minipage}{0.7\linewidth}
 \hspace{-2cm}
 \includegraphics[width=1.6\textwidth,height=0.8\textwidth,angle=0]{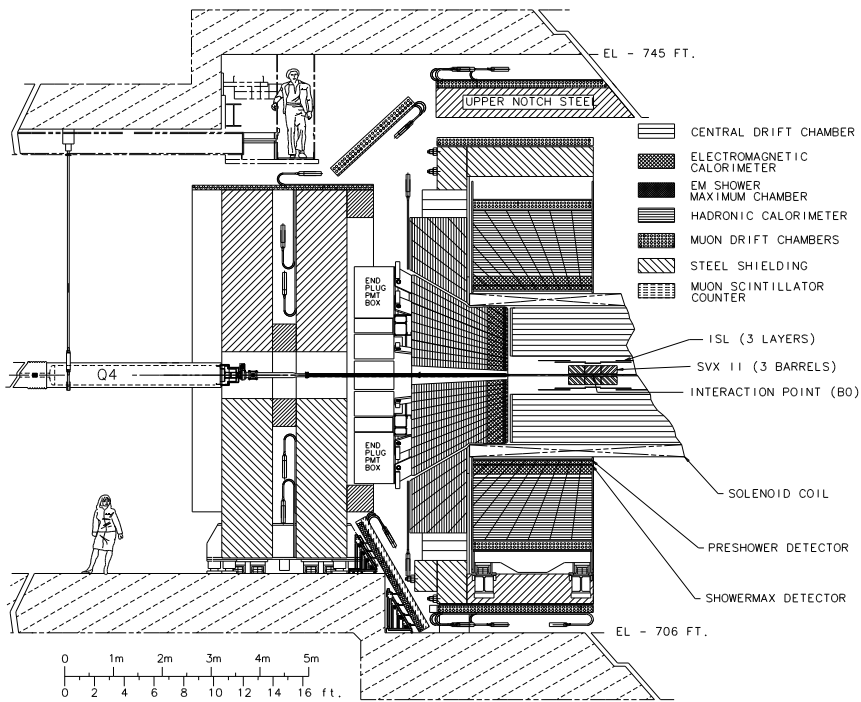}
\end{minipage} \hfill
\begin{minipage}{0.27\linewidth}
\begin{flushleft}
\vspace{8.5cm} \caption{\label{fig:CDF} Elevation view of the
CDF detector in Run~II~\cite{CDF}.}
\end{flushleft}
\end{minipage}
\end{figure*}

\section{Introduction}
The standard model ({\tt SM}) predicts experimental observables
at the weak scale with high precision. Despite the great
success of this model, the electroweak symmetry breaking
mechanism by which weak vector bosons acquire non-zero masses
remains unknown. The simplest mechanism involves the
introduction of a complex doublet of scalar fields that
generate particle masses via their mutual interactions leading
to the so-called {\tt SM} Higgs boson with an unpredicted
mass~\cite{ref:Higgs}. Furthermore, the {\tt SM} fails to
explain, for instance, cosmological phenomena like the nature
of dark matter in the universe~\cite{ref:Munoz}. These
outstanding issues are strong evidence for the presence of new
physics beyond the standard model. Among the possible
extensions of the standard model, supersymmetric ({\tt SUSY})
models~\cite{theo:SUSY_ref1,theo:SUSY_ref2} provide mechanisms
allowing for the unification of the forces and a solution to
the hierarchy problem. Particularly attractive are models that
conserve {\it R}-parity, in which {\tt SUSY} particles are
produced in pairs and the lightest supersymmetric particle
({\tt LSP}) is stable. \ In \ supergravity-inspired models
({\tt SUGRA})~\cite{theo:sugra}, the lightest neutralino
$\tilde{\chi}^0_1$ arises as the natural {\tt LSP}, which being
neutral and weakly interacting could be responsible for the
dark matter in the universe.

This paper reports recent experimental results of direct
searches for the Higgs boson and beyond-the-standard-model
({\tt BSM}) physics based on data collected by the CDF and D\O\
Collaborations at the Fermilab Tevatron collider. The dataset
analyzed corresponds to an integrated luminosity of up to
2.5~\invfb. More details on the analyses can be found in
Ref.~\cite{CDF-PHYS,D0-PHYS}.

\section{The Tevatron accelerator}
The Tevatron is performing extremely well. For Run~II, which
started in March 2001, a series of improvements were made to
the accelerator to operate at a center-of-mass energy of 1.96
\TeV\ with a bunch spacing of 396~\ns. Before the 2007
shutdown, monthly integrated and peak luminosities of up to 167
\invpb\ and $2.6\times10^{32}$ \cms, respectively, have been
achieved. Since the most recent shutdown, beams with peak
luminosity of \lumi{3.2}{32} have been delivered, with weekly
integrated luminosity and antiproton hourly production rate
reaching 58 \invpb\ and 26 \mAh, respectively. The consequence,
in terms of the number of interactions per crossing, is that
the Tevatron is running in a mode similar to that expected at
the Large Hadron Collider (LHC).

The D\O\ integrated luminosity delivered and recorded since the
beginning of Run~II is given in Table~\ref{tab:D0_IntLumi}
(with similar values for CDF).

\begin{table}[b]
\begin{center}
\caption{Run II luminosity delivered by the Tevatron
accelerator, and luminosity recorded by the D\O\
experiment.}\label{tab:D0_IntLumi}
\begin{tabular}{lll}
\hline
        & Delivered  & Recorded   \\ \hline
Run~IIa & 1.6~\invfb & 1.3~\invfb \\
Run~IIb & 2.5~\invfb & 2.2~\invfb \\ \hline
Total   & 4.1~\invfb & 3.5~\invfb \\ \hline
\end{tabular}
\end{center}
\end{table}

\section{The CDF and D\O\ detectors}
\begin{figure*}
 \begin{minipage}{0.7\linewidth}
 \includegraphics[width=.99\textwidth,height=0.54\textwidth,angle=0]{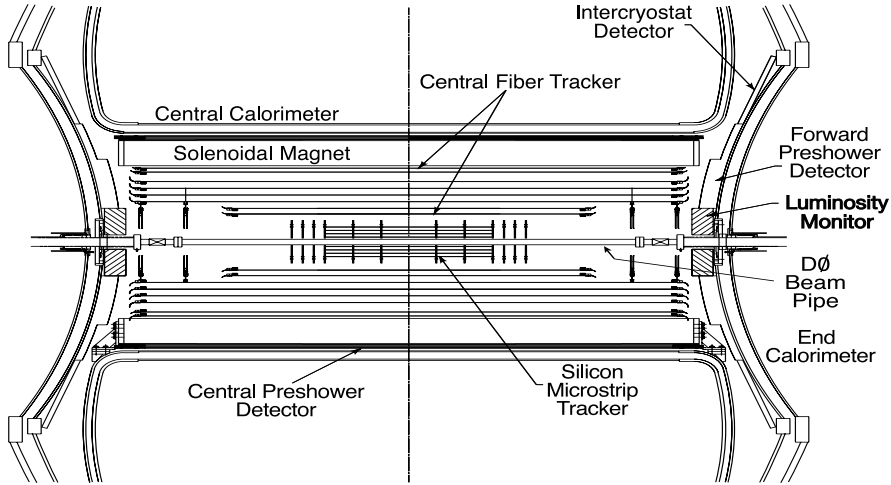}
\end{minipage} \hfill
\begin{minipage}{0.27\linewidth}
\begin{flushleft}
\vspace{3cm} \caption{\label{fig:D0} Cross section of the
central tracking system in the $x - z$ plane of the D\O\
detector in Run~II. Also shown are the locations of the
solenoid, the preshower detectors, luminosity monitor, and the
calorimeters~\cite{D0}.}
\end{flushleft}
\end{minipage}
\end{figure*}

A full description of the CDF (Fig.~\ref{fig:CDF}) and D\O\
(Fig.~\ref{fig:D0}) Run II detectors in operation since 2001 is
available in Ref.~\cite{CDF,D0}. Both experiments are
multipurpose detectors and are in a steady state of running.
Detectors take data with an average efficiency of 85\%. An
upgrade of the detectors to improve their performance for
Run~IIb was successfully concluded in 2006. The D\O\ upgrade
included the challenging insertion of an additional layer close
to the beam pipe of radiation-hard silicon (L0) to improve the
tracking performance. CDF and D\O\ completed calorimeter and
track trigger upgrades to significantly reduce the jet, missing
energy, muon and electron trigger rates at high luminosity,
while maintaining good efficiency for physics.

In the following, the CDF and D\O\ detector components used in
the analyses are briefly described. Both experiments use a
cylindrical coordinate system around the proton beam axis in
which $\theta$ and $\phi$ are the polar and azimuthal angles,
respectively, and the pseudorapidity $\eta$ is defined as
$\eta=-\ln \left[ \tan \left( \theta/2 \right) \right]$. The
transverse momenta and energy of a particle are defined as $\pt
= p \sin \theta$ and $\ET = E \sin \theta$, respectively. In
the following, imbalance in transverse momentum is referred to
as missing transverse energy or \met. The trigger and data
acquisition systems are designed to accommodate the high rates
and large data volume of Run II. It comprise three levels of
increasing complexity with a rate of accepted events written to
permanent storage of about 50-150~Hz. The beam luminosity is
determined by using counters located in the forward
pseudorapidity region that measure the average number of
inelastic \pp\ collisions per bunch crossing.

\subsection{CDF}
The tracking system consists of a cylindrical open-cell drift
chamber and silicon microstrip detectors in a 1.4~T magnetic
field parallel to the beam axis. The silicon
detectors~\cite{CDF_SMT} provide tracking information for
$|\eta|<2$ and are used to detect collision and decay points.
The drift chamber~\cite{CDF_CFT} surrounds the silicon
detectors and covers the central rapidity region $|\eta|<1$.
The energies of electrons and jets are measured in calorimeters
covering the region $|\eta| < 3.6$ and segmented into towers
pointing toward the center of the detector. Jets are
reconstructed from energy depositions in the calorimeter towers
using a jet clustering cone algorithm~\cite{CDF_jetalgo} with a
cone size of radius $\Delta R = \sqrt{(\Delta \phi)^2+(\Delta
\eta)^2}=0.4$. Corrections are applied to account for effects
that can cause mismeasurements in the jet energy such as
non-linear calorimeter response, multiple beam interactions, or
displacement of the event vertex from the nominal $z = 0$
position. Both the magnitude and direction of the \met\ are
recomputed after the jet energies have been corrected. Outside
the calorimeters, layers of steel absorb the remaining hadrons
leaving only muons, which are detected by drift chambers and
scintillation counters up to $|\eta|<1.5$.

\subsection{D\O}
The central tracking system consists of a silicon microstrip
tracker and a central fiber tracker, both located within a
1.9~T superconducting solenoid. A liquid-argon and uranium
calorimeter covers pseudorapidities up to $|\eta|$ $\approx
4.2$. The calorimeter has three sections, housed in separate
cryostats: the central one covers $|\eta|$ $\lsim 1.1$, and the
two end sections extend the coverage to larger
$\vert\eta\vert$. The calorimeter is segmented in depth, with
four electromagnetic layers followed by up to five hadronic
layers. It is also segmented into projective towers of
$0.1\times 0.1$ size in $\eta - \phi$ space. An outer muon
system, covering $|\eta|<2$, consists of a layer of tracking
detectors and scintillation trigger counters positioned in
front of 1.8~T toroids, followed by two similar layers after
the toroids. Jet reconstruction is based on the Run~II cone
algorithm~\cite{D0_jetalgo} with a cone size of 0.5 that uses
energies deposited in calorimeter towers. Jet energies are
calibrated using transverse momentum balance in photon+jet
events. The missing transverse energy in an event is based on
all calorimeter cells, and is corrected for the jet energy
calibration and for reconstructed muons.

\begin{figure}[b]
\begin{center}
\includegraphics[width=.93\linewidth,height=0.75\linewidth]{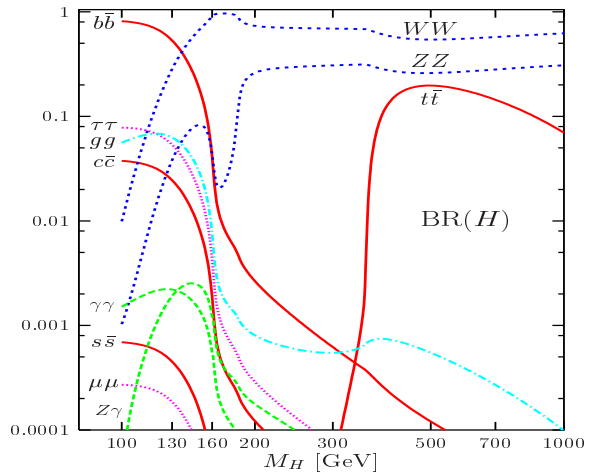}
\caption{\label{fig:br_H}
Standard model Higgs boson branching fractions as a function of its mass~\cite{ref:Djouadi}.}
\end{center}\end{figure}

\section{Standard model Higgs boson}
The discovery of the Higgs boson is commonly considered to be
the highest priority of particle physics today. This
fundamental ingredient of the theory has not yet been observed
and could be reachable at the Tevatron if the {\tt SM} Higgs
boson mass $m_H$ is light or near 160~\GeV. The goal at the
Tevatron is to find evidence for the Higgs boson using the full
dataset, expected to correspond to about 7~\invfb\ by 2010. A
considerable effort has been devoted in recent years to improve
the theoretical predictions and they are now known to good
precision. The Higgs boson couples preferentially to the
heaviest particles. As shown in Fig.~\ref{fig:br_H}, the decay
mode \Hbb\ is the dominant one in the mass range
$m_H<135~\GeV$, with a branching fraction ($Br$) of 73\% at
$m_H=115~\GeV$. For Higgs masses above $\approx135~\GeV$ the
main decay mode is into $WW$ pairs, where one of the vector
bosons is off-shell below the corresponding kinematic
threshold.

The discovery of the Higgs boson is also among the main reasons
for the construction of the Large Hadron Collider (LHC) at
CERN, which is expected to begin operation in summer 2008. The
LHC was designed such that the discovery of the {\tt SM} Higgs
boson, if it exists, would be guaranteed up to
${\cal{O}}$(1~\TeV) \cite{ref:Kolda}, the highest energy
consistent with general theoretical principles. Precision
measurements, most notably of the top mass $m_t = 172.6 \pm 1.4
~\GeV$ \cite{CDF_DO_mtop} at the Tevatron and of the W mass
$m_W = 80.398 \pm 0.025~\GeV$ \cite{LEP_mw,CDF_mw} at LEP and
at the Tevatron, however provide strong indications that the
{\tt SM} Higgs boson should be much lighter than that upper
bound, having a mass smaller than 160~\GeV\ at 95\% confidence
level (C.L.) \cite{ref:LEPEWWG}. Direct searches for the Higgs
boson at LEP in the $e^{+} e^{-} \raa$ $Z^{*}$ $\raa ZH$
reaction provide a lower limit of 114.4
\GeV~\cite{ref:LEP_Higgs}, but also revealed several
interesting candidate events with masses just above that lower
bound. If the direct lower limit from LEP is taken into account
to extract an upper bound from precision measurements, the {\tt
SM} Higgs boson mass upper limit becomes 190~\GeV\ at 95\% C.L.
Such a mass range is favorable to the Tevatron's reach.

\begin{figure}\begin{center}
        \begin{minipage}{.49\linewidth}
          \begin{center}
             \includegraphics[width=.9\linewidth,height=.7\linewidth]
             {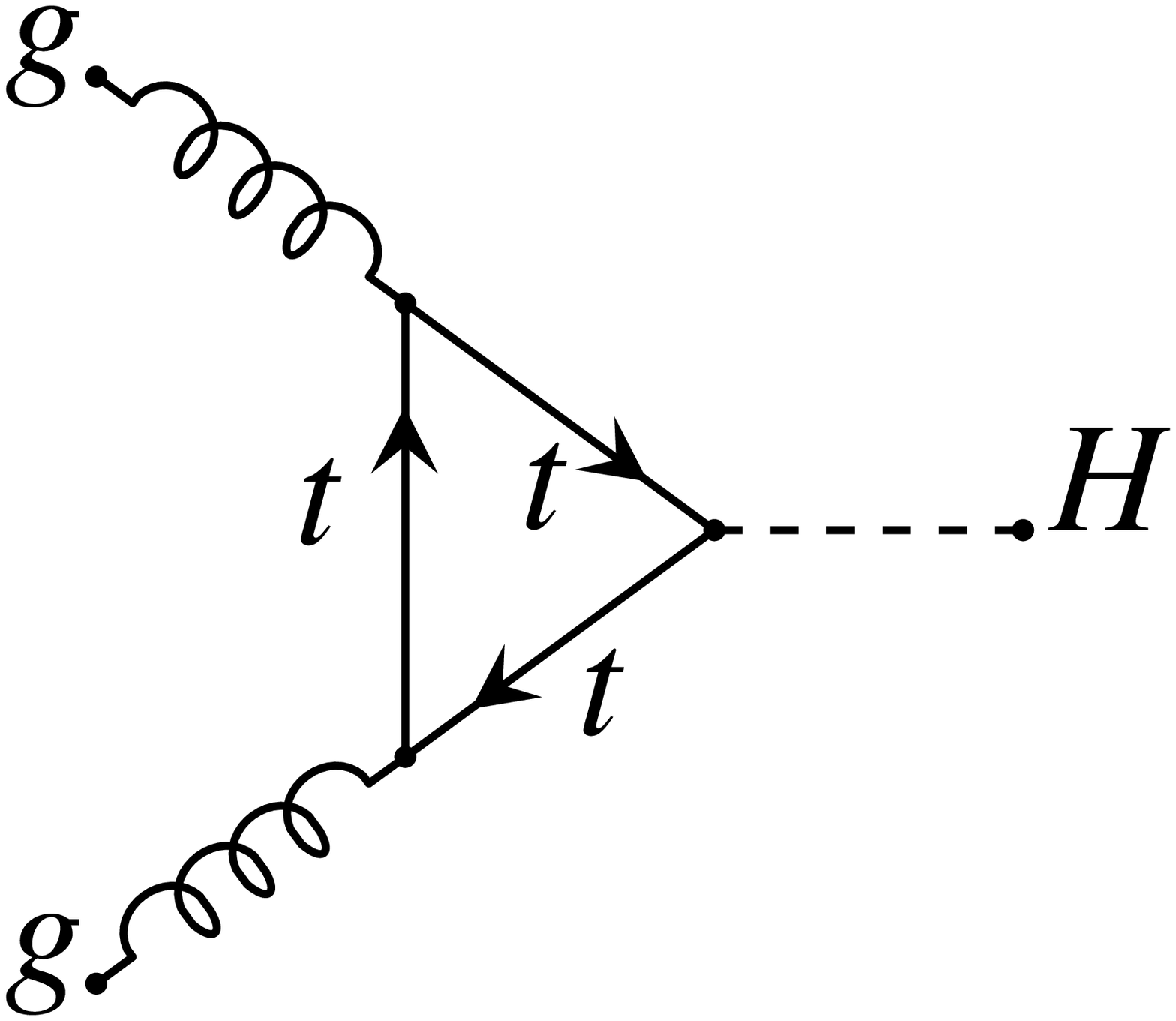}
          \end{center}
        \end{minipage} \hfill
        \begin{minipage}{.49\linewidth}
          \begin{center}
             \includegraphics[width=.99\linewidth,height=.7\linewidth]
             {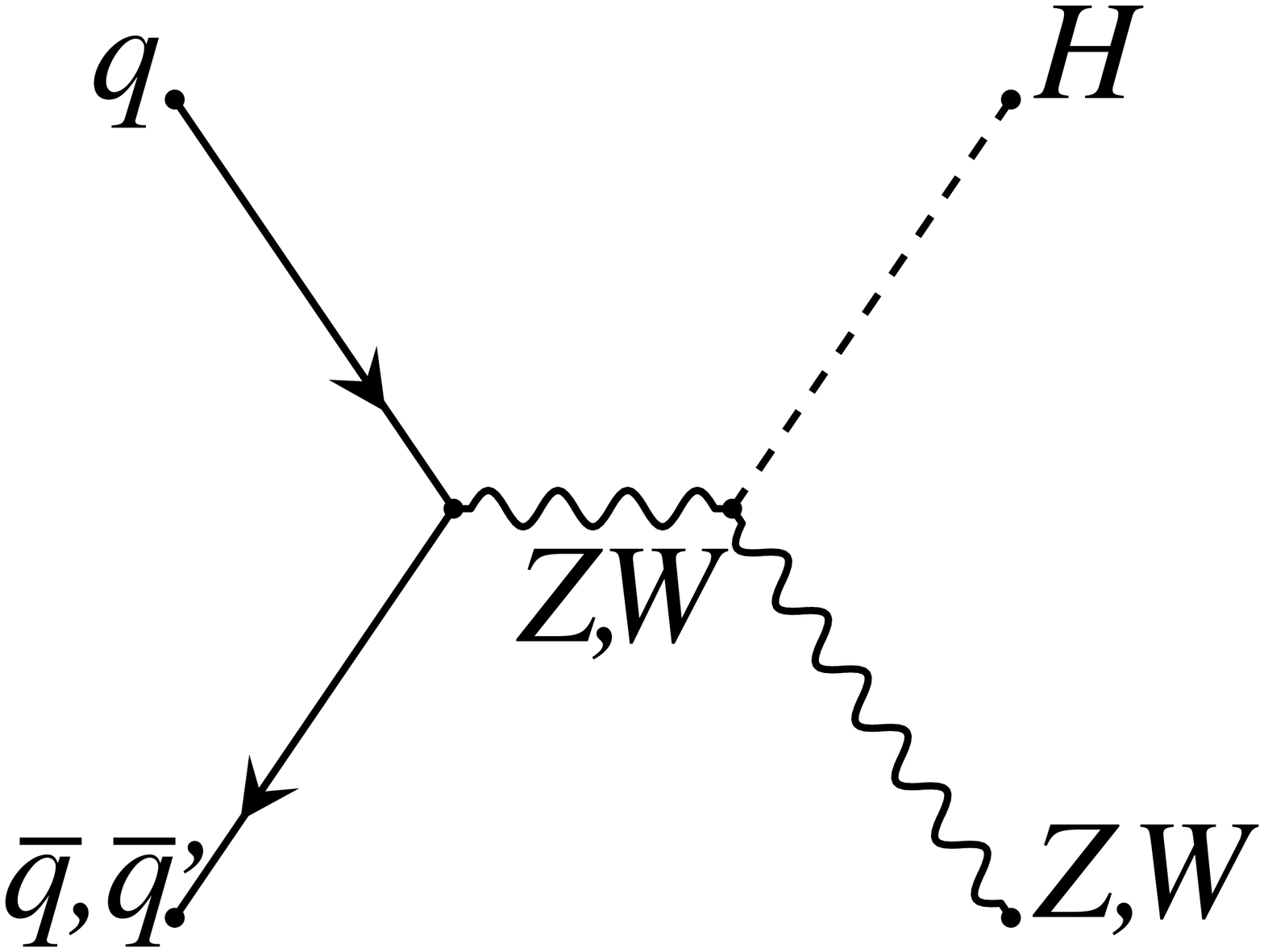}
          \end{center}
        \end{minipage}
        \begin{minipage}{.49\linewidth}
          \begin{center}
             \includegraphics[width=.9\linewidth,height=.7\linewidth]
             {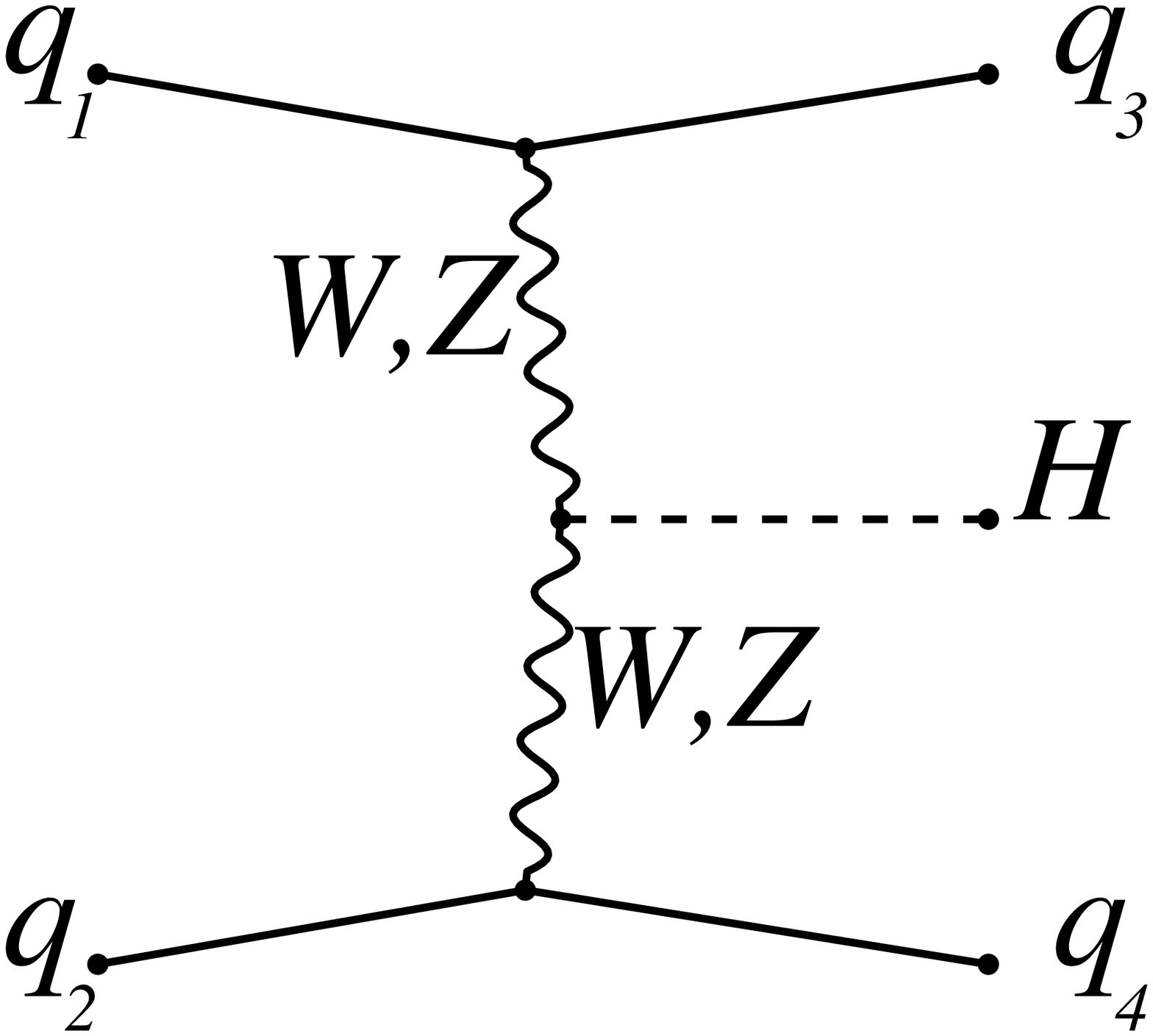}
          \end{center}
        \end{minipage} \hfill
        \begin{minipage}{.49\linewidth}
          \begin{center}
             \includegraphics[width=.9\linewidth,height=.7\linewidth]
             {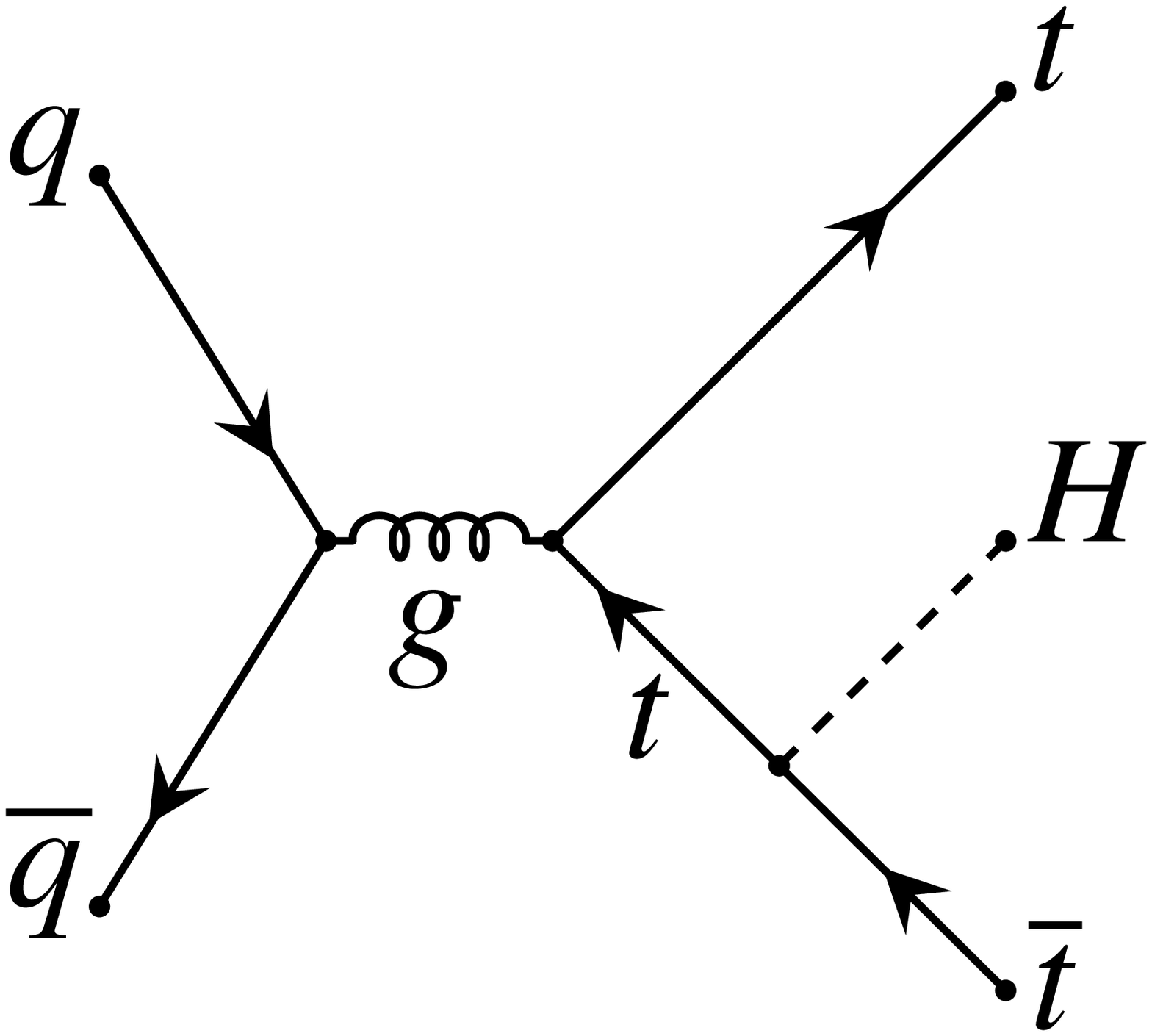}
          \end{center}
        \end{minipage}
\caption{\label{fig:feynman_mH} Main diagrams for Higgs boson production at
leading order: (top left) gluon fusion, (top right) Higgs associated production,
(bottom left) vector boson fusion, (bottom right) associated production with heavy quark (top).}
\end{center}\end{figure}

\subsection{Higgs boson production}
\begin{figure}\begin{center}
\includegraphics[width=.40\textwidth,height=0.49\textwidth,angle=-90]{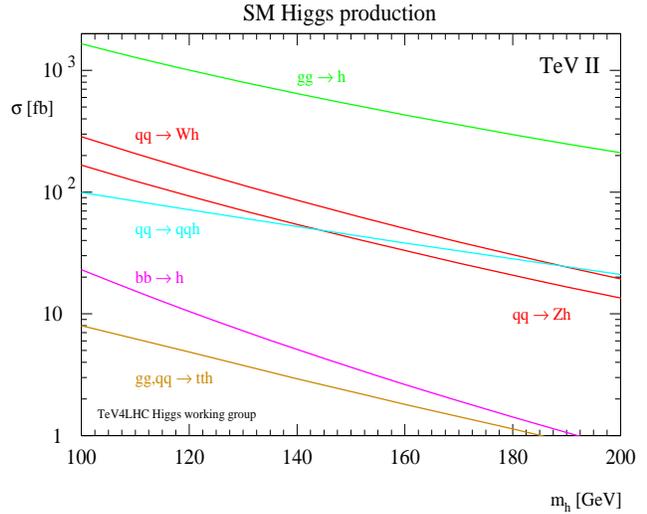}
\caption{\label{fig:sig_H_tev}
Standard model Higgs boson production cross sections (fb) at the Tevatron
  ($\sqrt{s}=1.96$~\TeV) for the most relevant production mechanisms as
  a function of the Higgs boson mass~\cite{ref:TeV4LHC}.}
\end{center}\end{figure}

The main diagrams for Higgs production are displayed at leading
order in Fig.~\ref{fig:feynman_mH}. The production cross
section of the Higgs boson are summarized in
Fig.~\ref{fig:sig_H_tev} for \pp\ collisions at the Tevatron.
They are small, of the order of 0.1~\pb, not including the
decay branching fractions, while typical backgrounds such as
$W+b\bar{b}$ or $Z+b\bar{b}$ have cross sections three orders
of magnitude larger. The $q\bar{q} \raa qqH$ vector boson
fusion ({\tt VBF}) process is known at next-to-leading order
({\tt NLO}) in {\tt QCD} and is marginal at the Tevatron with
cross sections between 0.1-0.02~\pb\ for masses $100~\GeV < m_H
< 200~\GeV$ ~\cite{ref:qqH_VBF}. The associated production
process $t\bar{t}H$ is also known at {\tt NLO} in {\tt QCD} and
can be exploited only at the LHC~\cite{ref:ttH}. The bottom
fusion $bb\raa H$ process is known at next-to-next-to-leading
order ({\tt NNLO}) in {\tt QCD} in the five-flavor
scheme~\cite{ref:bbH,ref:Higgs_WG_2003} and has a cross section
of 25~\fb\ for $m_H \approx 100$~\GeV. The two main production
modes at the Tevatron are therefore gluon fusion $gg \raa H +
X$ and associated production $q\bar{q} \raa WH + X, q\bar{q}
\raa ZH + X$. The associated production processes have cross
sections known at {\tt NNLO} in {\tt QCD} and {\tt NLO} for the
electroweak corrections with a rather small residual
theoretical uncertainty that is less than
5\%~\cite{ref:Higgs_WG_2003,ref:Higgs_asso_prod}. The gluon
fusion process $gg \raa H + X$ is known at {\tt NNLO} in {\tt
QCD} (in the large top-mass limit) and at {\tt NLO} in {\tt
QCD} for arbitrary top mass, with an overall residual
theoretical uncertainty estimated to be around
10\%~\cite{ref:Higgs_gg_fusion}.

All Higgs signals are simulated using {\sc
pythia}~\cite{ref:pythia}, and CTEQ5L or
CTEQ6L~\cite{ref:pdflib} leading-order ({\tt LO}) parton
distribution functions. The signal cross sections are
normalized to next-to-next-to-leading order ({\tt NNLO})
calculations~\cite{ref:Higgs_WG_2003,ref:Higgs_gg_fusion}, and
branching fractions from {\sc hdecay}~\cite{ref:hdecay}.

\subsection{SM backgrounds}
The dominant backgrounds to Higgs analyses at low mass comprise
$W/Z$+jets, $t\bar{t}$, single top, and multijet (instrumental)
events. The latter are sometimes referred to as ``{\tt QCD}
background''. At high mass, di-boson processes represent the
main contribution to the backgrounds.

For both CDF and D\O, the multijet background is estimated from
data, usually in orthogonal samples to those used for the
analyses. For CDF, backgrounds from other {\tt SM} processes
are generated using the {\sc pythia}, {\sc alpgen}
\cite{ref:ALPGEN}, {\sc mc@nlo} \cite{ref:MCNLO} and {\sc
herwig} \cite{ref:herwig} programs. For D\O, these backgrounds
were generated using {\sc pythia}, {\sc alpgen} and {\sc
comphep} \cite{ref:comphep}, with {\sc pythia} providing
parton-showering and hadronization for all the generators.
Background processes are normalized using either experimental
data or next-to-leading order calculations from {\sc MCFM}
\cite{ref:mcfm}.

\begin{figure}
\begin{minipage}{\linewidth}
  \begin{center}
   \includegraphics[width=1.\linewidth,height=0.9\linewidth]{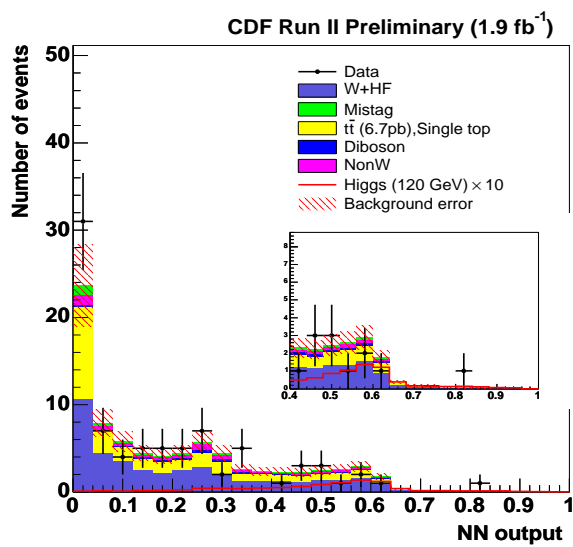}
  \end{center}
    \caption{\label{fig:CDF_NN_WZ}
    Neural network output distribution obtained after final selection in the search for associated Higgs
    boson production (\whl\ channel) using 1.9~\invfb\ of CDF Run II data~\cite{CDF_WH_1.9fb}.
    Two of the jets in the event are required to be identified as containing b-quarks ($b$-tagging).
    The background expectations and the observed data are shown. The expected Higgs signals,
    scaled as indicated, is represented for a neural network
    trained with a Higgs boson mass of $m_H=120~\GeV$. The insert shows a zoom-in the signal region.}
\end{minipage}
\end{figure}

\subsection{Search strategy}
At the Tevatron, the most sensitive channels to search for a
low mass {\tt SM} Higgs boson ($m_H < 135~\GeV$) are those from
associated production, with $W\raa l \nu$ and $H \raa
b\bar{b}$, or $Z \raa ll$ or $\nu\nu$ and $H \raa b\bar{b}$. At
high mass, $m_H \approx 160~\GeV$, the branching fraction is
mainly into $WW$ boson pairs, and the leptonic decays of the
$W$ are exploited through gluon-gluon fusion \ppHWW.

There are secondary modes that provide additional sensitivity.
For intermediate masses around 135~\GeV, all branching
fractions are below 40\%; however, $p\bar{p}$ $\raa WH$ $\raa
WWW^{*}$ $\raa \ell^{\pm} \ell^{\pm}$ and \ppHgamgam\ are used to
strengthen the discovery potential. At the Tevatron, the decay
channel \ppHZZ\ does not significantly contribute due to
combination of lower branching ratios for $H \raa ZZ$ and
$Z\raa \ell \ell$ in addition to the acceptance for four
leptons. Though both Collaborations do intend to add such
analysis in the future, this mode has not yet been studied for
the current combination. At low mass, a 5\% contribution from
the addition of the $H\raa \tau \tau$ decay mode is included.
The CDF experiment contributes to the most recent {\tt SM}
Higgs combination with this new analysis searching for the
Higgs bosons decay to a tau lepton pair, in three production
channels: gluon fusion, associated production, and {\tt VBF}
(details are given later).

\begin{figure}
\begin{minipage}{\linewidth}
  \begin{center}
   \includegraphics[width=1.0\linewidth,height=0.85\linewidth]{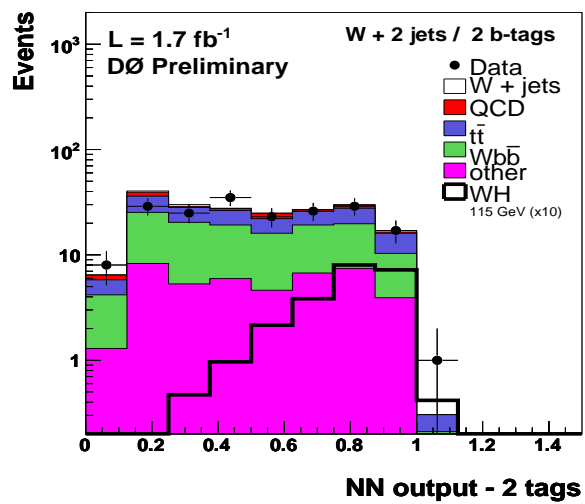}
  \end{center}
    \caption{\label{fig:D0_NN_WZ} D\O\ final analysis variable
    distribution (neural network output) used by the associated Higgs
    boson production search in the \whl\ channel after requiring two
    $b$-tagged jets in the event~\cite{DO_WH_1.7fb}. The result is shown
    for an integrated luminosity of 1.7~\invfb\ of data.
    The background expectations and the observed
    data are shown. The expected Higgs signals,
    scaled as indicated, is represented for a neural network
    trained with a Higgs boson mass of $m_H=115~\GeV$.}
\end{minipage}
\end{figure}

\begin{figure*}
\begin{minipage}{\linewidth}
  \begin{center}
   \includegraphics[width=0.49\linewidth,height=0.40\linewidth]{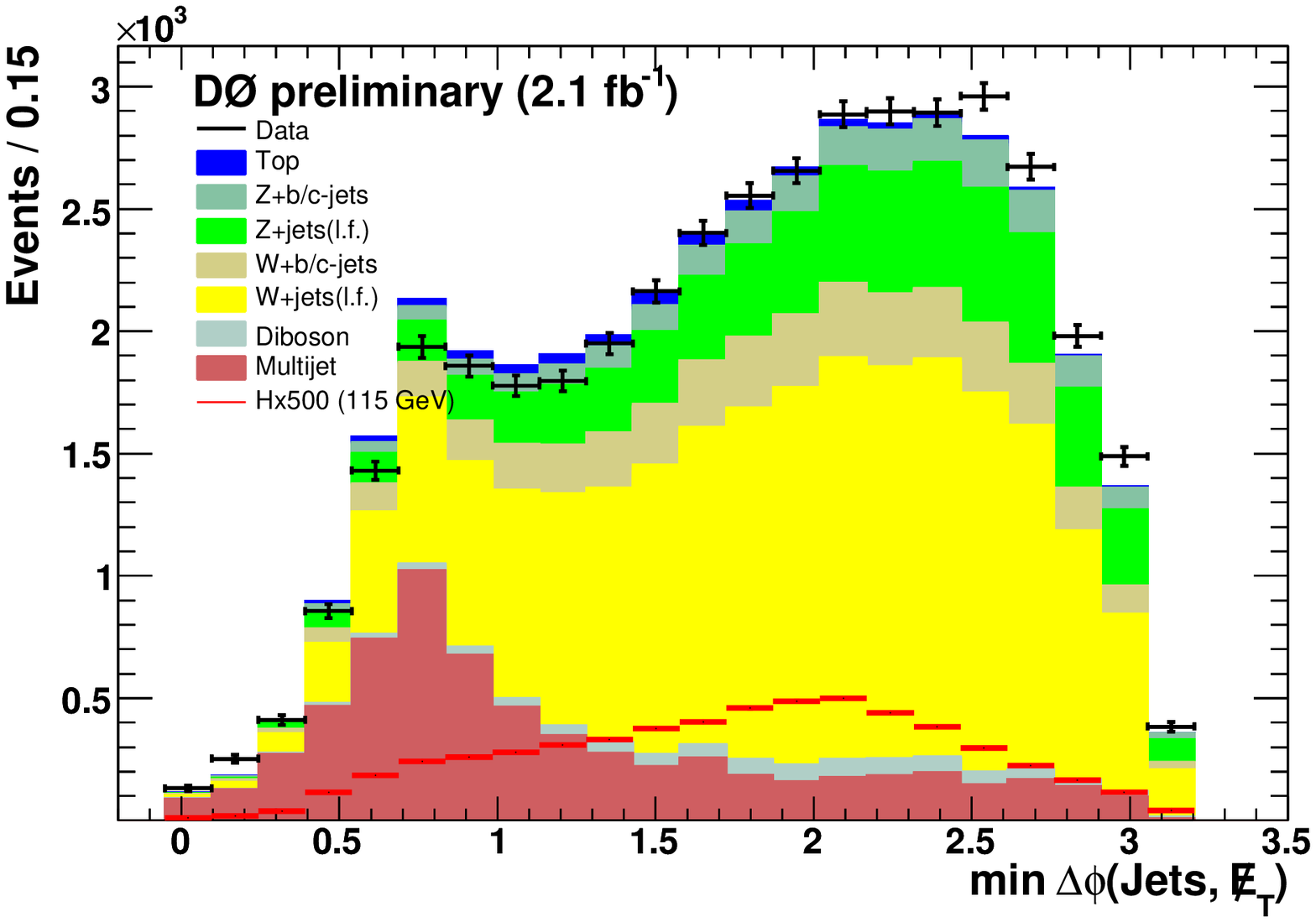}
   \includegraphics[width=0.49\linewidth,height=0.40\linewidth]{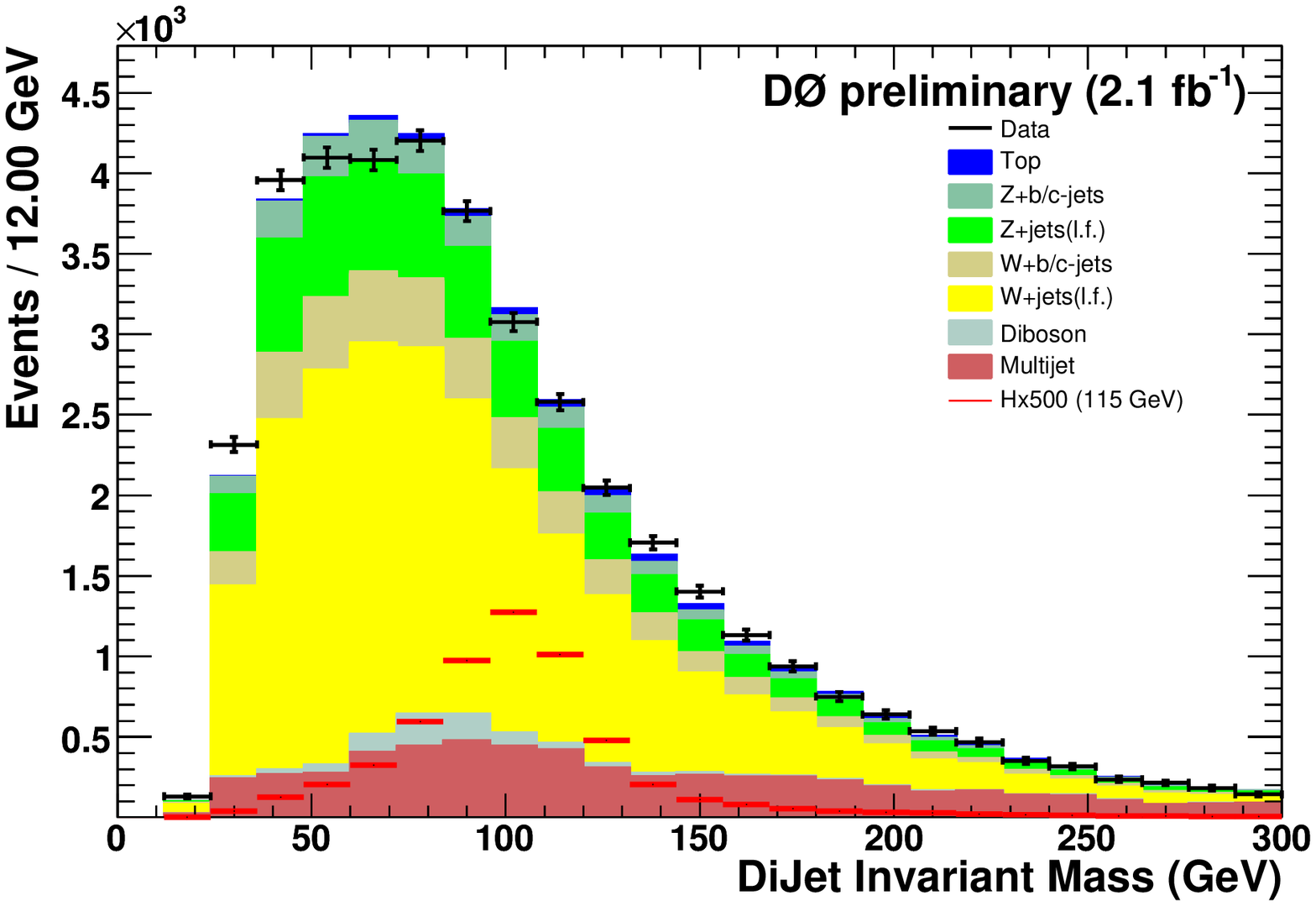}
  \end{center}
    \caption{\label{fig:DO_HZ_nunubb} Distributions of the
    $\Delta\phi_{min}$(\met,jets) (left), the minimum of the
    differences in azimuth between the direction of \met\ and
    the direction of any jet, and of the dijet invariant mass of the
    two leading jets (right), in the analysis sample before
    $b$-tagging for the D\O\ search in the \ppZHnnbb\ channel
    using 2.1~\invfb\ of data \cite{D0_ZH_2.1fb}. The
    various background contributions ({\tt SM} and multijet) are
    shown. Distributions for a signal with a Higgs boson mass
    of 115~\GeV\ are also shown, scaled as indicated.}
\end{minipage}
\end{figure*}

\subsubsection{b-identification:}
At low mass, the searches primarily focus on the dominant
$H\raa b\bar{b}$ decay, which leads to the presence of $b$ jets
for the signal. The sensitivity of the Higgs boson searches is
significantly increased by using a lifetime-based heavy-flavor
tagging algorithm ($b$-tagging) which computes a probability
for a jet to be light-flavored based on the impact parameters
of the tracks in the jet. Another method is based on secondary
vertex reconstruction.

For instance, D\O\ combines in a neural network (NN)
discriminant~\cite{ref:TIM_SCALON} several kinematic variables
sensitive to transversely-displaced jet vertices and jet tracks
with large transverse impact parameters relative to the
hard-scatter vertices. The NN is trained to identify
heavy-flavor quark decays and reject jets arising from
light-flavor quarks or gluons. By adjusting the minimum
requirement on the $b$-tagging NN output, a spectrum of
increasingly stringent $b$-tagging operating points is
achieved, each with a different signal efficiency and purity.
The analyses are usually separated into those where two of the
jets are $b$-tagged with a loose tagging requirement and those
where only one jet is tagged with a tight requirement.
Typically, the $b$-tagging tight (loose) operating point is
selected such that, for jets with \pt\ of $\approx50$~\GeV,
0.5\% (1.5\%) of the light-flavored jets are tagged while the
tagging efficiency for $b$-quark jets is 50\% (60\%).

\subsubsection{Advanced analysis techniques:}

After kinematic selection and $b$-tagging, the remaining
backgrounds in a Higgs boson search would ideally be due solely
to the associated production of a $W$ or $Z$ boson with a pair
of $b$ quarks. The Higgs boson would then appear as a
$b\bar{b}$ resonance over a broad continuum. The significance
of a Higgs boson signal is therefore directly related to the
mass resolution of a system of two $b$ quark jets, and thus to
the jet energy resolution. The dijet mass is however not the
only feature which allows discrimination between signal and
background. Other, more subtle, differences in the kinematic
properties of the signal and the backgrounds can be used.
Because no single variable provides sufficient discriminating
power, advanced analysis techniques have to be used, such as
neural networks or decision trees~\cite{ref:Dtree}. The matrix
element approach, which makes full use of the event properties
at leading order and in which there is already experience in
the CDF and D\O\ Collaborations, can also be introduced to
enhance the discriminating power of, for example, a neural
network. The performance of even the most elaborate
multivariate analysis, however, critically depends on an
accurate understanding of the characteristics of the signal and
background processes.

\subsubsection{Standard model background understanding:}
One of the major goals of the searches is to first carry out a
detailed assessment of the generators used for the simulation
of those processes, by confronting them with measurements,
which are becoming increasingly more precise as the datasets
grow.

Monte Carlo simulations, mostly based on fixed-order matrix
elements, are used to extrapolate these measurements into the
Higgs signal regions. In order to validate the whole analysis
chain, the associated production of vector bosons, $WZ$ and
$ZZ$, can be used. With $WZ\raa \ell \nu b\bar{b}$ and $ZZ\raa
\ell\ell/\nu\nu \, b\bar{b}$, the final states are actually
almost identical to those considered in the search for the
Higgs boson, up to the mass difference between the $Z$ boson
and the Higgs boson. Observation of these reactions will
therefore be among the main goals of the coming year, and their
detailed analysis will pave the way for a definitive
calibration of the Higgs boson searches, with a caveat that
$Z\raa b\bar{b}$ branching fraction is a factor of 3 or less
than the $H\raa b\bar{b}$. The next-to-leading order ({\tt
NLO}) $ZZ$ cross section at the Tevatron is
$1.4~\pm0.1~\pb$~\cite{ref:mcfm}, an order of magnitude above
some of the expected {\tt SM} Higgs production cross sections.
Additionally, this process forms an irreducible background to
Higgs searches in the $ZH$ channel. For instance, the CDF
Collaboration recently submitted for publication the first
measurement at a hadron collider of the cross section for $Z$
boson pair production in the leptonic decay channels with a
significance of 4.4 standard deviations based on 1.9~\invfb\ of
data~\cite{ref:CDF_ZZ}. The measured cross section is
$\sigma(p\bar{p} \raa ZZ) = 1.4^{+0.7}_{-0.6} (stat.+syst.)$
\pb. The D\O\ Collaboration also conducted such a measurement
based on 2.2~\invfb\ of data. Using the final state decay
$ZZ\raa \ell\ell \nu {\bar\nu}$, D\O\ observes a signal with a
$2.4\sigma$ significance and measures a cross section
$\sigma(p\bar{p} \raa ZZ) =$ $2.1 \pm 1.1(stat.) \pm 0.4(sys.)$
\pb\ \cite{ref:D0_ZZ}.

\begin{figure}
\begin{minipage}{\linewidth}
  \begin{center}
   \includegraphics[width=1.\linewidth,height=0.9\linewidth]{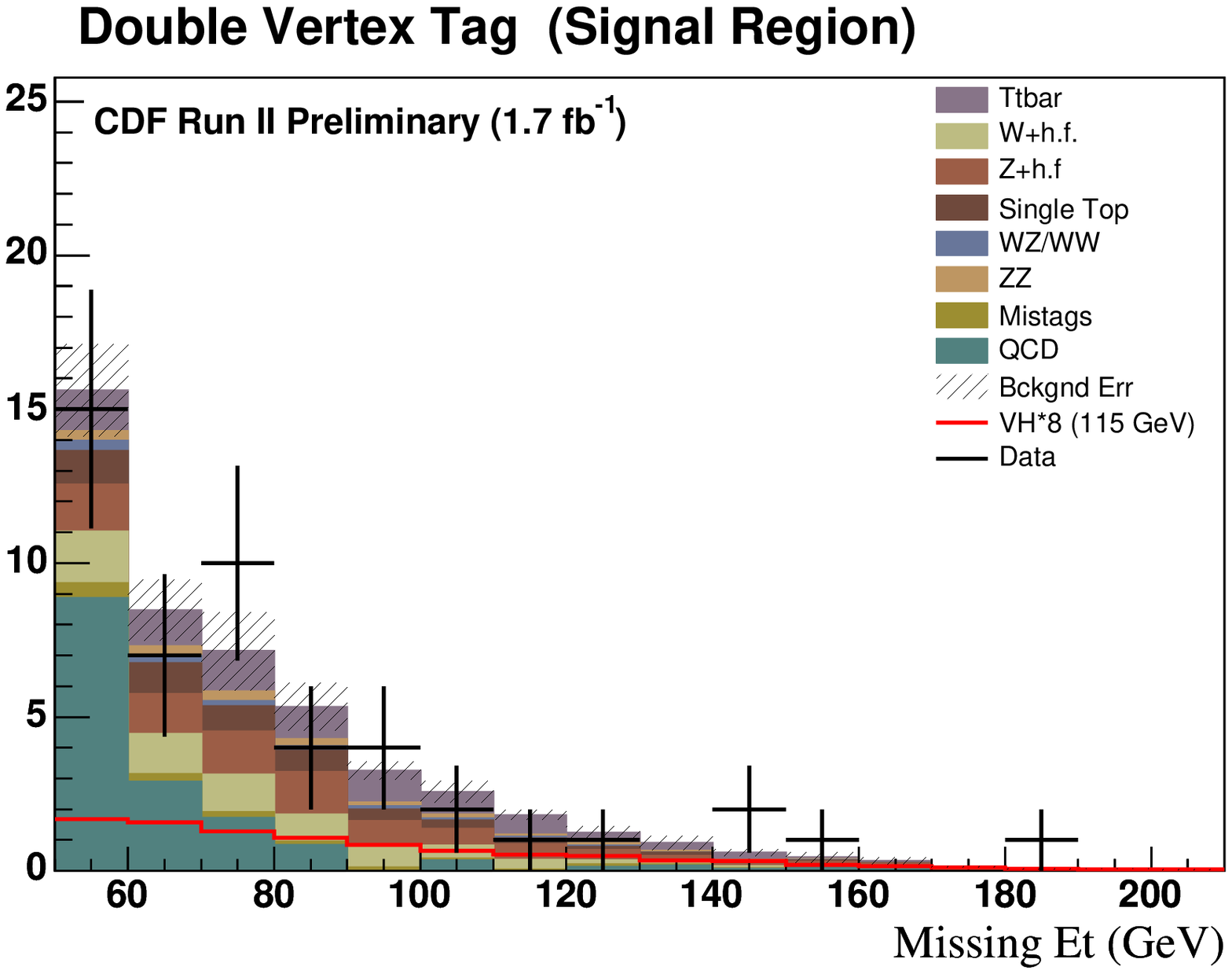}\vfill
   \includegraphics[width=1.\linewidth,height=0.9\linewidth]{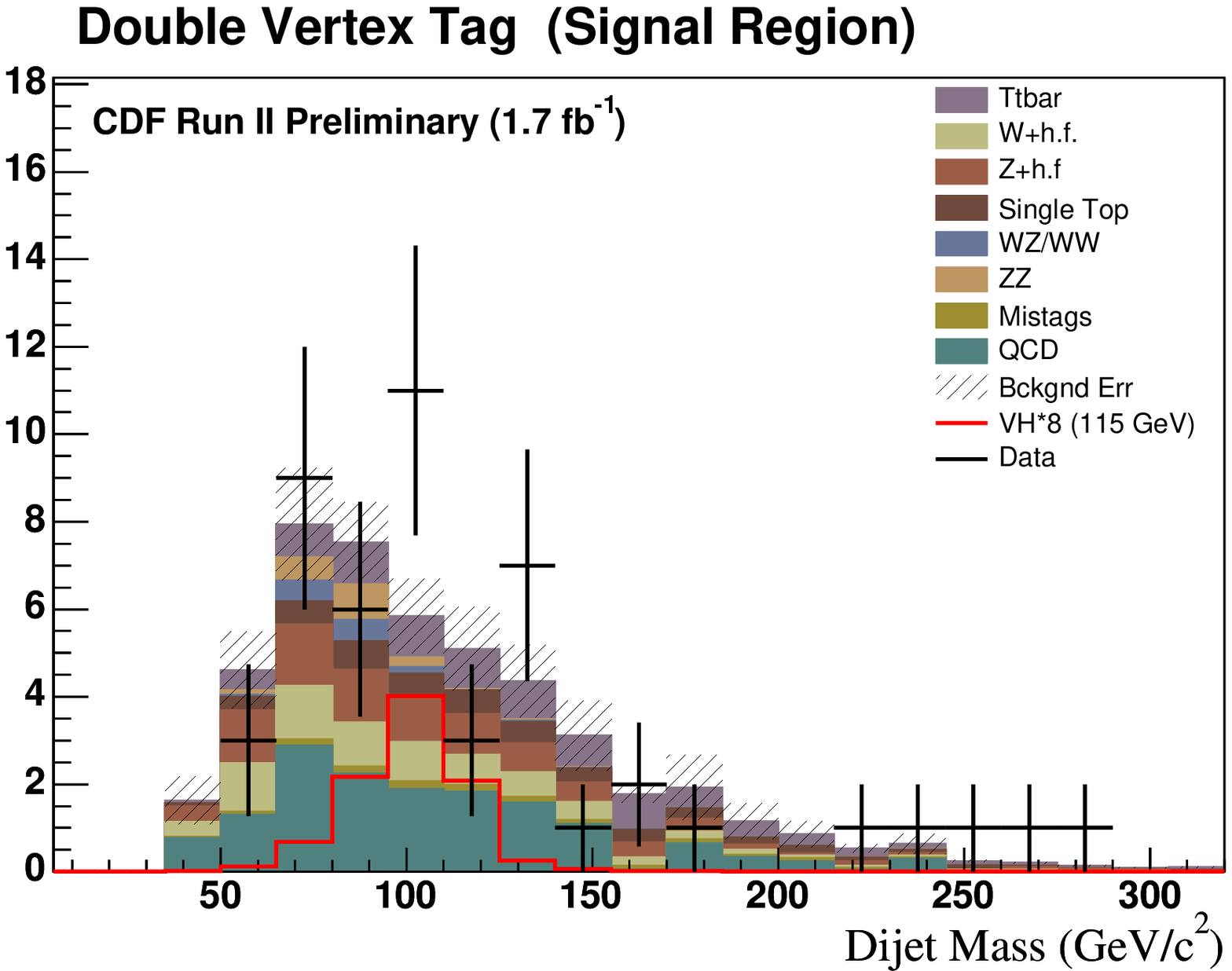}
  \end{center}
    \caption{\label{fig:CDF_HZ_nunubb} Distributions of the
    \met\ (top) and dijet invariant mass (bottom) in the
    signal region after requiring 2 $b$-tagged jets for the CDF
    search in the \ppZHnnbb\ channel using 1.7~\invfb\ of data \cite{CDF_ZH_1.7fb}. The expected Higgs signal at
    $m_H=115~\GeV$ is scaled as indicated.}
\end{minipage}
\end{figure}
\subsection{Associated production }
Associated $VH$ production ($V=W,Z$) can be distinguished from
multijet and electroweak backgrounds by exploiting the leptons
and/or the missing transverse energy in the final state.

\subsubsection{\ppWHlnbb:}
The final state from \ppWH\ production, where the $W$ boson
decays leptonically, provides an ideal trigger signature to
collect as many of the produced Higgs bosons as possible. The
most recent published $WH$ searches have been performed with
integrated luminosities of 955 \invpb\ for
CDF \cite{CDF_WH_1fb} and 440 \invpb\ for
D\O\ \cite{D0_Higgs_Pub0.4fb}. The latest updates with 1.9
\invfb\ from CDF \cite{CDF_WH_1.9fb} (winter 2008) and
1.7 \invfb\ from D\O\ \cite{DO_WH_1.7fb} (summer 2007) use
similar search strategies. The high-\pt\ electron or muon is
required to be isolated with \ET\ (or \pt) greater than
20~\GeV\ and events with more than one isolated lepton are
vetoed. Selected events must also display a significant missing
transverse energy ($\met>20~\GeV$). {\tt QCD} events with false
$W$ signatures are estimated with the data by computing the
ratio of isolated to non-isolated leptons in a control region.
Both CDF and D\O\ use neural-network discriminants to separate
the signal from the {\tt SM} background events. As an
illustration for these searches, Fig.~\ref{fig:CDF_NN_WZ} and
Fig.~\ref{fig:D0_NN_WZ} show some distributions of the final
variables used by the CDF and D\O\ experiments for limit
setting. The D\O\ search requires at least one tight $b$-tagged
jet or exactly two loose $b$-tagged jets in the event, and two
$b$-tagged jets are required in the CDF search. At
$m_H=115~\GeV$, the ratio between the expected (observed) cross
section limit and the {\tt SM} value is 8.2 (7.3) for CDF and
9.1 (11.1) for D\O.
\subsubsection{\ppZHnnbb:}
Triggering on \ppZHnnbb\ is more challenging than on the
charged leptons from $W$ or $Z$ decays. This is because only
hadronic jets are visible in the final state, which makes it
difficult to distinguish it from standard multijet production
via the strong interaction. The distinctive feature is the
missing transverse energy carried by the neutrinos from the $Z$
decay. However, standard multijet events can also exhibit
missing transverse energy due to fluctuations in the jet energy
measurement by the calorimeter, or to semileptonic decay in
heavy flavor jets. Complex topological triggers have therefore
been designed, combining measurements of jet energies and
directions, and missing transverse energy. The D\O\
Collaboration has published a result based on 260~\invpb\ of
data~\cite{D0_ZH_260pb} and the CDF Collaboration recently
submitted for publication a search corresponding to an
integrated luminosity of $1~\invfb$~\cite{CDF_ZH_1fb}. Both
experiments updated these results for the winter 2008
conferences with 2.1~\invfb~\cite{D0_ZH_2.1fb} and
1.7~\invfb~\cite{CDF_ZH_1.7fb} of data for D\O\ and CDF,
respectively.

One of the largest backgrounds in the \met+jets channel
involves heavy flavor multijet production. Although the
probability for multijet events to create artificial missing
energy at a significant level is small, the huge cross section
of multijet production renders this background overwhelming at
the initial stages of the analysis. Currently, severe selection
criteria are used (mainly $\met>50~\GeV$) in order to
practically eliminate that background, thus introducing
substantial inefficiencies. The normalization and shape of
multijet events are obtained from the data to take into account
all relevant instrumental effects and biases. This multijet
control sample is extracted in the data sample where the \met\
is aligned with the second jet. The distributions of the
minimum of the differences in azimuth between the direction of
the \met\ and the direction of any jet, and the dijet invariant
mass of the two leading jets, are shown in
Fig.~\ref{fig:DO_HZ_nunubb} for the D\O\ experiment. It is seen
that the combination of the multijet and {\tt SM} background
provides a good description of the data in the pre-tag sample.

Advantage of the large branching fractions for $H\raa$
$b\bar{b}$ is used by requiring the two leading jets to be
$b$-tagged. Figure~\ref{fig:CDF_HZ_nunubb} shows the
distributions of \met\ and dijet invariant mass for the CDF
analysis after all selection requirements are imposed. In the
case of $WH \raa$ $\ell \nu b\bar{b}$ production where the
primary lepton from the $W$ boson decay falls outside of the
detector acceptance and is not identified, the final state $WH
\raa \ell \kern-0.45em\lower-.05ex\hbox{/}\nu b\bar{b}$ is the
same as the \ZHnnbb. The \WHnbb\ events contribute to
significantly enhance the \met+jets analysis sensitivity.
Finally, a boosted decision tree technique was used for the
D\O\ search and a NN discriminant for the CDF search to
calculate the cross section limit. For a 115~\GeV\ Higgs boson
mass and requiring two $b$-tagged jets in the event, the
observed/expected limits on the cross section of combined
\ZHnnbb\ and \WHnbb\ production are 8/8.3 (7.5/8.4) times
larger than the {\tt SM} value for CDF (D\O).

\begin{figure}\begin{center}
\begin{minipage}{\linewidth}
  \begin{center}
   \includegraphics[width=1.\linewidth,height=0.9\linewidth]{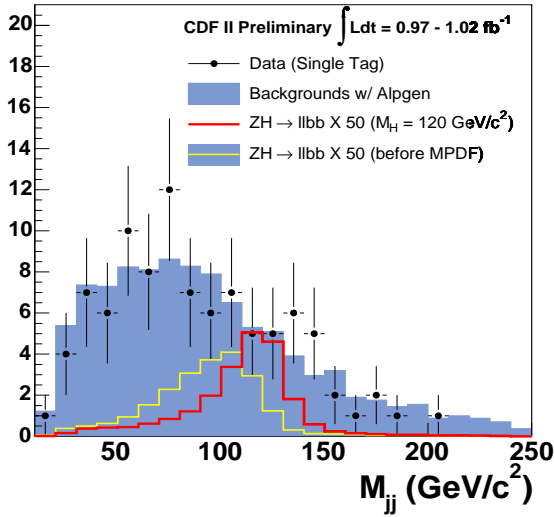}
  \end{center}
    \caption{\label{fig:CDF_ZH_ll} Distributions of the
    dijet invariant mass in the signal
    region after requiring 2 $b$-tagged jets for the CDF search
    in the \ppZHllbb\ channel using $\approx$ 1~\invfb\ of data \cite{CDF_ZH_ll_1fb}. A correction (MPDF)
    is applied to reassign missing energy to the
    jets since the dominant source of \met\ are jet energy mismeasurement.
    The expected Higgs signal at
    $m_H=115~\GeV$ is scaled as indicated.}
\end{minipage}
\end{center}\end{figure}

\subsubsection{\ppZHllbb:}
\begin{figure}\begin{center}
\begin{minipage}{\linewidth}
  \begin{center}
   \includegraphics[width=1.\linewidth,height=.8\linewidth]{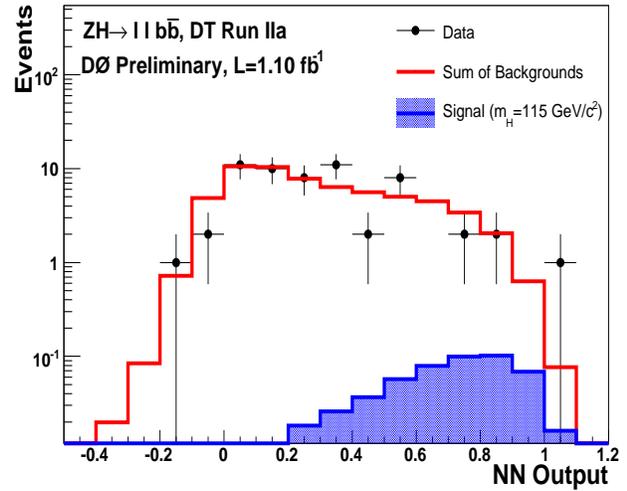}
  \end{center}
    \caption{\label{D0_ZH_ll} Distributions of the neural net output variable
    after requiring 2 $b$-tagged jets for the D\O\ search
    in the \ppZHllbb\ channel using 1.1~\invfb\ of data \cite{DO_ZH_ll_1fb}. The total predicted background, data
    and expected Higgs signal at $m_H=115~\GeV$ are shown.}
\end{minipage}
\end{center}\end{figure}

Searches for the Higgs boson from the process $p\bar{p} \raa
ZH$ $\raa \ell\ell b\bar{b}$ in both $e^+e^-$ and $\mu^+\mu^-$
channels have been carried out by CDF \cite{CDF_ZH_ll_1fb} and
D\O\ \cite{DO_ZH_ll_1fb} in 1 \invfb\ of data. The D\O\
experiment has published a result based on 0.45 \invfb\
\cite{DO_ZH_ll_0.45fb}. These channels have a small background
of mostly $Z$+jets events due to the requirement of two leptons
and a $Z$ mass constraint, but suffer from a smaller $Z$
branching fraction. To maintain signal efficiency and improve
discrimination, the experiments employ neural networks trained
to separate $ZH$ events from the main $Z$+jets background and
the kinematically different $t\bar{t}$ background. To improve
sensitivity, the data are divided into single and double
$b$-tagged channels (double $b$-tagged only for CDF); the
results are shown in Fig.~\ref{fig:CDF_ZH_ll} and
Fig.~\ref{D0_ZH_ll}. The expected (observed) limits from the
data as a ratio compared to the expected {\tt SM} cross section
are 20.4 (17.8) for the D0 analyses, and 16 (16) for the CDF
search, at $m_H=115~\GeV$.

\begin{figure*}
\begin{minipage}{\linewidth}
  \begin{center}
   \includegraphics[width=0.49\linewidth,height=0.40\linewidth]{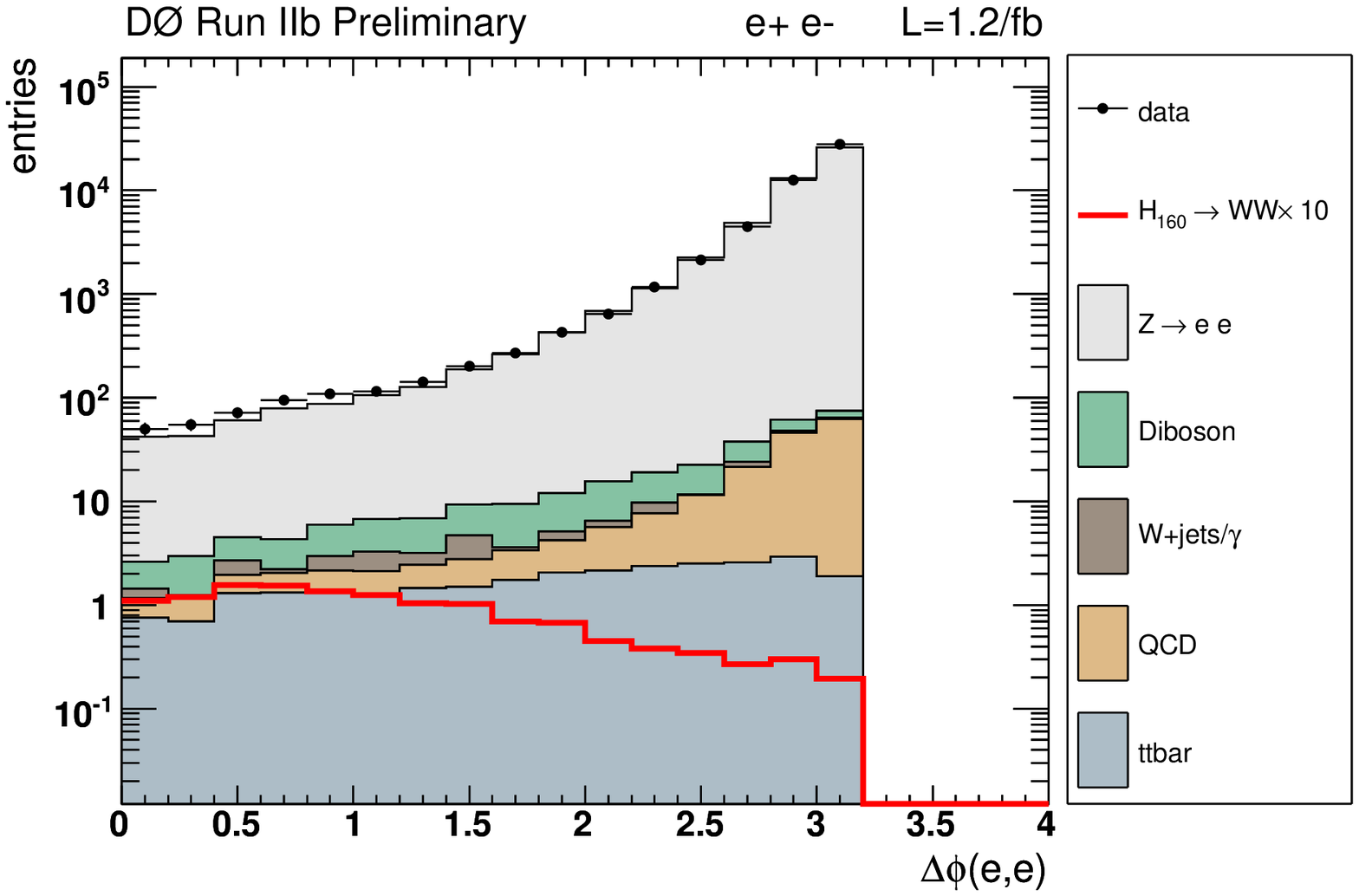}
   \includegraphics[width=0.49\linewidth,height=0.40\linewidth]{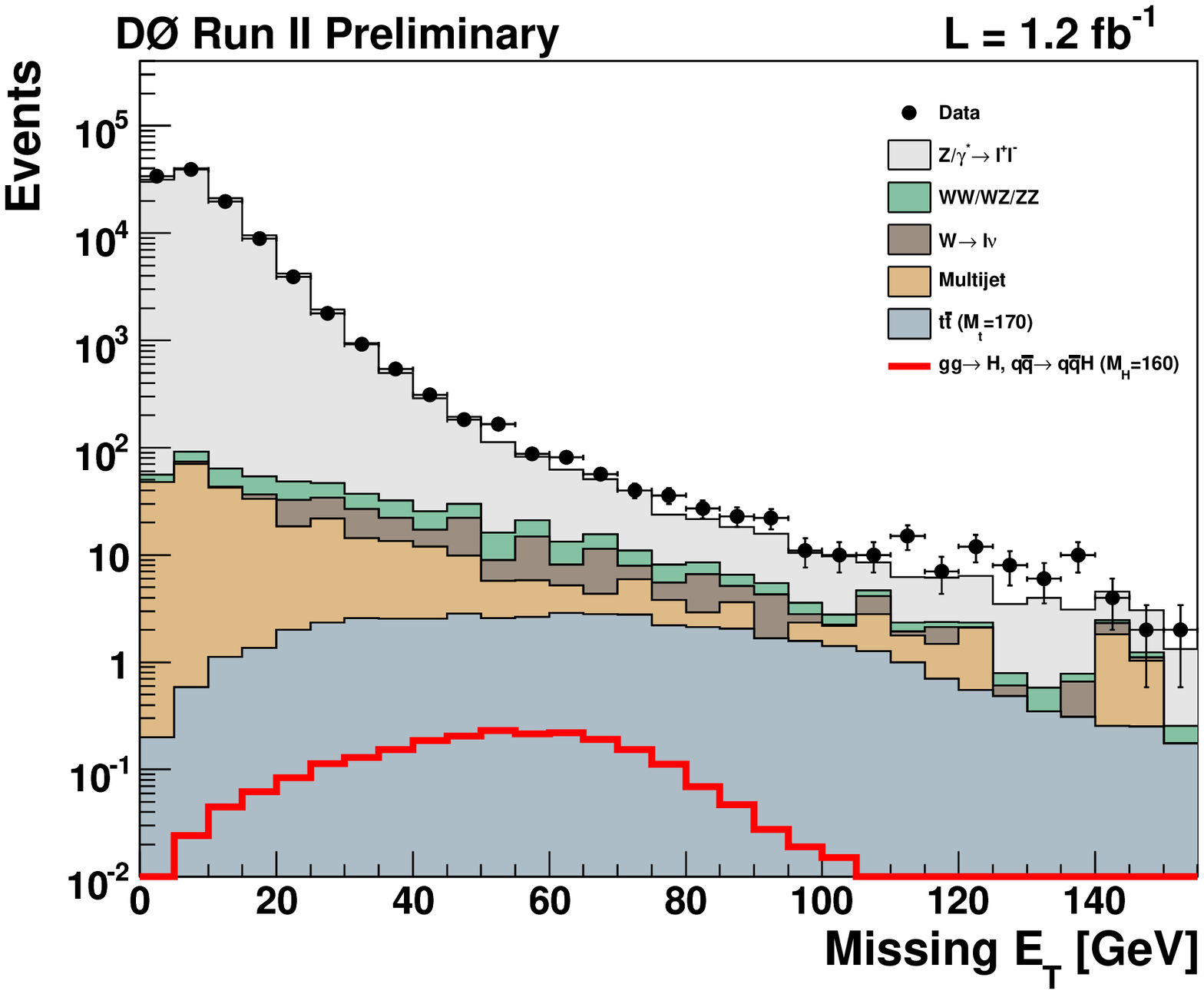}
  \end{center}
    \caption{\label{fig:D0_HWW} Distributions of the opening
    angle $\Delta \phi_{e_1 e_2}$ for the $ee$ system (left)
    and \met\ for the $\mu\mu$ system (right) at
    pre-selection level for the D\O\ search in the \ppHWWlnulnu\
    channel using 1.2 \invfb\ of data \cite{D0_HWW_RunIIb_NN}. The expected Higgs signal
    at $m_H=160~\GeV$, the predicted background, and the data
    events are shown.}
\end{minipage}
\end{figure*}
\begin{table*}
\begin{center}
\caption{\label{tab:CDF_WW} The numbers of signal events
expected for a Higgs boson mass $m_H=160~\GeV$, events expected
from {\tt SM} backgrounds, and data events observed, for the
CDF experiment using 2.4~\invfb\ of data~\cite{CDF_HWW_2.4fb}.
The {\tt SM} Higgs boson production and decay are assumed to be
$gg \rightarrow H \rightarrow WW^{*} \rightarrow l^+l^-\nu\nu
$, where $l^\pm=e,\mu,$~or~$\tau$. The final state $e~trk$
($\mu~trk$) require an electron (a muon) and an additional
track ($trk$).}
\begin{tabular}{lccccccccrr}
\hline\noalign{\smallskip}
Category   & Higgs  & $WW$ & $WZ$ & $ZZ$ & $t\bar{t}$ & DY & $W\gamma$ & $W$+jets & Total & Data \\
           & ($m_H=160~\GeV$)       &    &    &    &     &    &           &          &       &      \\
\noalign{\smallskip} \hline \noalign{\smallskip}
$e~e$      & 1.7 &  55.1 & 6.4 & 7.1 & 3.6 & 33.6 & 33.6 & 29.5 & 169$\pm$14  &  171 \\
$e~\mu$    & 3.8 & 131.7 & 3.9 & 0.4 & 8.7 & 27.4 & 29.4 & 34.1 & 235$\pm$21 & 240 \\
$\mu~\mu$  & 1.6 &  43.4 & 4.9 & 5.8 & 3.3 & 23.1 &  0.0 &  5.2 &  86$\pm$8 &  83 \\
$e~trk$    & 1.6 &  45.5 & 3.1 & 2.9 & 3.3 & 13.1 &  8.0 & 13.3 &  89$\pm$7 & 107 \\
$\mu~trk$  & 1.0 &  24.8 & 2.0 & 2.0 & 1.9 &  7.1 &  1.5 &  7.9 &  47$\pm$4 &  60 \\
\noalign{\smallskip} \hline \noalign{\smallskip}
Total      & 9.5 & 300.3 & 20.6 &18.2&20.8 &104.0 & 72.3 & 90.0 & 626$\pm$54& 661 \\
\noalign{\smallskip}\hline
\end{tabular}
\end{center}
\end{table*}

\subsubsection{\ppWHWWW:}
In the {\tt SM}, the Higgs boson predominantly decays to a
$WW^{(*)}$ pair for Higgs masses above 135~\GeV. Although the
scenario $WH \raa WWW^{(*)}$ $\raa \ell^{\pm} \ell^{\pm} + X$ is not
the most sensitive search for high masses, this process
provides a unique experimental signature with two like-charge
leptons from $W$ decays. Furthermore, in some scenarios with
anomalous couplings, such as fermiophobic Higgs models, the
branching fraction to $WW^{(*)}$ may be close to 100\% for
Higgs masses down to
$\approx$100~\GeV~\cite{ref:Fermiophobic1,ref:Fermiophobic2,ref:Fermiophobic3}.
The D\O\ Collaboration has searched for fermiophobic Higgs
($h_f$) and published~\cite{D0_HW_WWW_fermiophobic_400pb}
results with 380~\invpb\ in the $ee$ channel, 370~\invpb\ in
the $e\mu$ channel, and 360~\invpb\ in the $\mu\mu$ channel,
with the variations related primarily to different trigger
requirements. Upper limits are set on $\sigma(p\bar{p}\raa
Wh_f)$ $\times$ $Br(h_f \raa W^+W^-)$ between 3.2 and 2.8~\pb\
for Higgs boson masses from 115 to 175~\GeV\ at 95\% C.L.
Recently, CDF presented a preliminary result using 1.9~\invfb\
of data~\cite{CDF_HW_WWW_fermiophobic_1.9fb}. This search
expects 0.46 (0.19) event for a fermiophobic Higgs boson mass
of 110 (160)~\GeV, assuming {\tt SM} production cross section.
The expected background is $3.23\pm 0.69$ events, while 3
events are observed in the data. From these results, CDF sets
limits of 2.2~\pb\ for $m_{h_f}=110$~\GeV\ and 1.4~\pb\ for
$m_{h_f}=160$~\GeV\ at 95\% C.L.

For these searches, the main physics background is $WZ$ $\raa
\ell \nu \ell \ell$ production. The irreducible physics
background, which comes from non-resonant $WWW$ triple vector
boson production has a very low cross section (as does
$t\bar{t}$). As the channel involves two neutrinos in the final
state, the reconstruction of the Higgs mass is not feasible and
the potential Higgs signal appears as an excess in the number
of observed events with two like-charge leptons over the
predicted {\tt SM} background. In the absence of such an
excess, upper cross section limits are set by the
CDF~\cite{CDF_HW_WWW_fermiophobic_1.9fb} (D\O~\cite{D0_HW_WWW})
search with 1.9 (1.1) \invfb\ of about 33 (20) times above the
{\tt SM} Higgs boson cross section at $m_H=160$~\GeV.

\subsection{Gluon fusion}
The largest production cross section for the whole Higgs mass
range of interest is the gluon fusion process due to the large
top Yukawa couplings and the gluon densities. At the Tevatron,
however, the gluon fusion process becomes relevant with a clear
experimental signature only at high mass ($m_H \approx
160~\GeV$), where the branching fraction is mainly into $WW$
boson pairs leading to a favorable final state with two leptons
and two neutrinos. At low mass ($m_H \lesssim 135~\GeV$), due
to the large branching fraction of the Higgs boson into
$b\bar{b}$, the gluon fusion Higgs production mode cannot be
disentangled from the multijet background.

\subsubsection{\ppHWW:}
At the Tevatron, the decay mode \ppHWW\ provides the largest
sensitivity for the {\tt SM} Higgs boson search at a Higgs
boson mass of $m_H \approx 160~\GeV$. Upper limits on the cross
section times branching fraction in the leptonic decay modes
$H\raa WW^{*} \raa \ell \ell^{'} (\ell,\ell^{'}=e,\mu,\tau)$
from previous Run~IIa data have already been published by the
CDF (D\O) Collaboration with 360 \invpb\
\cite{CDF_HWW_pub_360pb} (325~\invpb\ \cite{D0_HWW_pub_325pb}).
The most recent data available from Run~IIb have been recently
analyzed and preliminary results have been presented by
CDF~\cite{CDF_HWW_2.4fb} and
D\O~\cite{D0_HWW_RunIIa_NN,D0_HWW_RunIIb_NN} using 2.4~\invfb\
and 2.3~\invfb\ of data, respectively. In all final states, two
isolated leptons of opposite sign originating from the same
primary vertex are required. The background is dominated by
$Z/\gamma^{*}$ and multijet events in which the leptons are
typically back-to-back. It is therefore suppressed by requiring
some missing transverse energy and with a cut on the opening
angle $\Delta \phi_{\ell \ell}$ which is smaller for the signal
than for the background due to the spin-correlation between the
final state leptons in the decay of the spin-0 Higgs boson.
Figure~\ref{fig:D0_HWW} shows the $\Delta \phi_{\ell \ell}$ and
the \met\ distributions at pre-selection level for the D\O\
search.

\begin{figure}
\begin{minipage}{\linewidth}
  \begin{center}
   \includegraphics[width=1.\linewidth,height=0.85\linewidth]{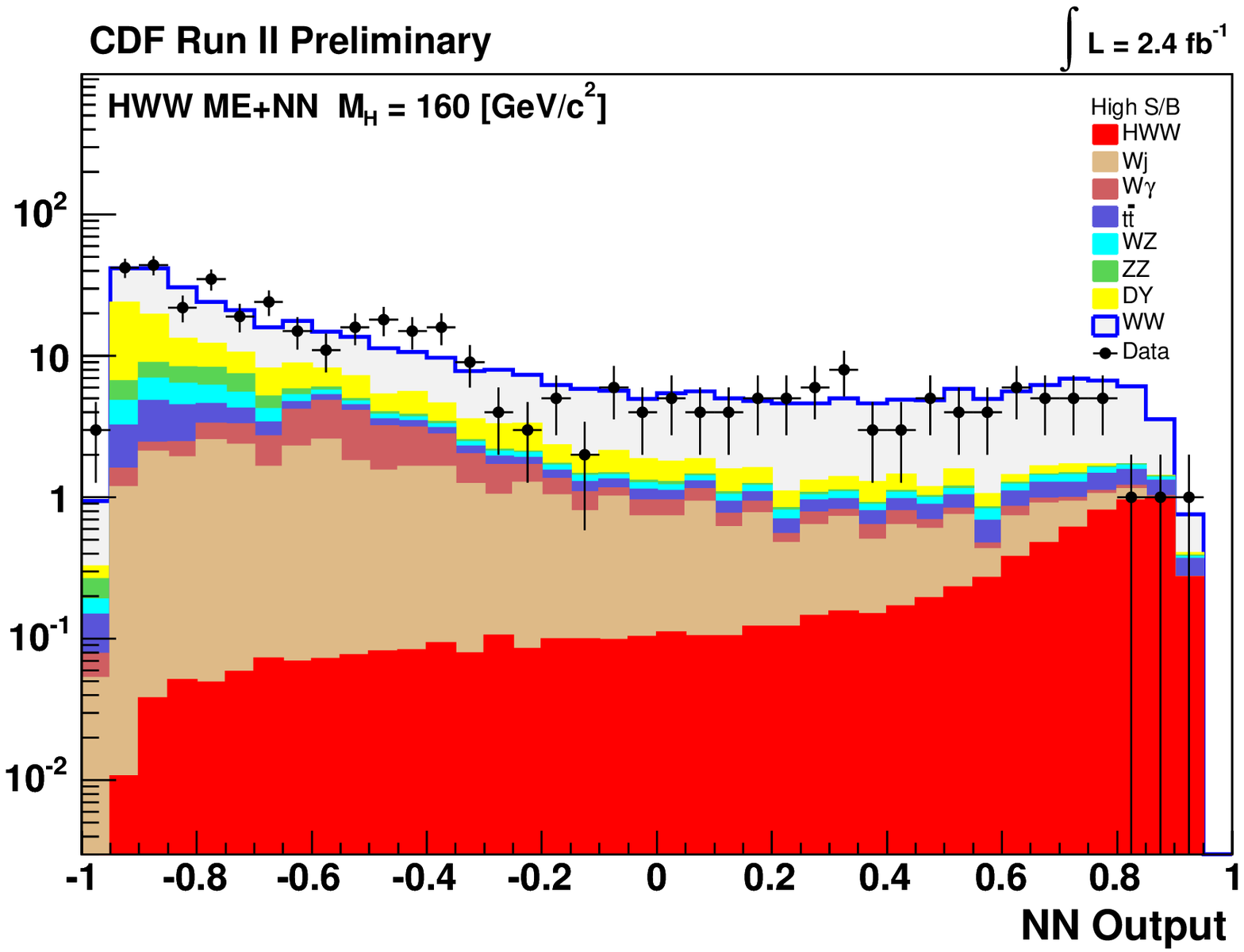}
  \end{center}
    \caption{\label{fig:CDF_HWW} Neural network template used as a final discriminant
   to search for the {\tt SM} Higgs boson in the \ppHWWlnulnu\ decay
    channel using 2.4~\invfb\ of CDF data~\cite{CDF_HWW_2.4fb}. The signal
    is shown for $m_H=160$~\GeV.}
\end{minipage}
\end{figure}

After final selection, the CDF search finds 661 candidates with
an expectation of $626\pm 54$ background events and 9.5 signal
events for a {\tt SM} Higgs at $m_H=160~\GeV$.
Table~\ref{tab:CDF_WW} provides a detailed breakdown of the
signal and background contributions in each final state. The
presence of neutrinos in the final state prevents
reconstruction of the Higgs boson mass. In order to maximize
the signal sensitivity, a combined matrix element method and
neural network approach are utilized to distinguish signal from
background processes. An example of neural network template is
shown in Fig.~\ref{fig:CDF_HWW} for $m_H=160$~\GeV. The median
expected 95\%~C.L. limit at a Higgs mass of 160~\GeV\ is
$2.5^{+1.0}_{-0.7}$ times the {\tt SM} prediction at {\tt
NNLL}, while the observed limit is 1.6 times the {\tt SM}
prediction. To improve the separation of signal from
backgrounds, a neural network is also used for the D\O\ search
in each of the three di-lepton channels. The D\O\ expected and
observed upper limits relative to the {\tt SM} Higgs boson
cross section prediction are 2.4 and 2.1, respectively, for
$m_H=160~\GeV$.

\subsubsection{\ppHgamgam:}
In the {\tt SM}, the diphoton decay of the Higgs boson are
suppressed at tree level and the branching fraction for this
decay is 0.22\% for a 130~\GeV\ Higgs boson mass. However, in
some models beyond the {\tt SM}, this decay can be enhanced
significantly; some examples can be found in
Refs.~\cite{ref:Fermiophobic1,ref:Fermiophobic2,ref:Fermiophobic3}
and section~\ref{sec:fermiophobic} of this review. The D\O\
experiment has recently submitted for publication a search for
a narrow resonance decaying into two photons using 1.1~\invfb\
of data~\cite{D0_Hgamgam}. This result has been updated with
2.3~\invfb~\cite{D0_Hgamgam_2.3fb} and the {\tt SM} Higgs is
used as a possible signal model to set upper limits on the
production cross section times branching fraction ($H\raa
\gamma \gamma$) for different assumed Higgs bosons masses.
There are three major sources of background. The first source
comes from Drell-Yan events where both electrons are
misidentified as photons due to tracking inefficiencies, and is
estimated with Monte Carlo simulations. The second source is
from direct {\tt QED} diphoton events and is also estimated
using simulation. Finally, the background from $\gamma$+jet and
jet+jet events, where the jets are mis-identified as photons,
is obtained from data.

\begin{figure}\begin{center}
\begin{minipage}{\linewidth}
  \begin{center}
   \includegraphics[width=1.0\linewidth,height=0.8\linewidth]{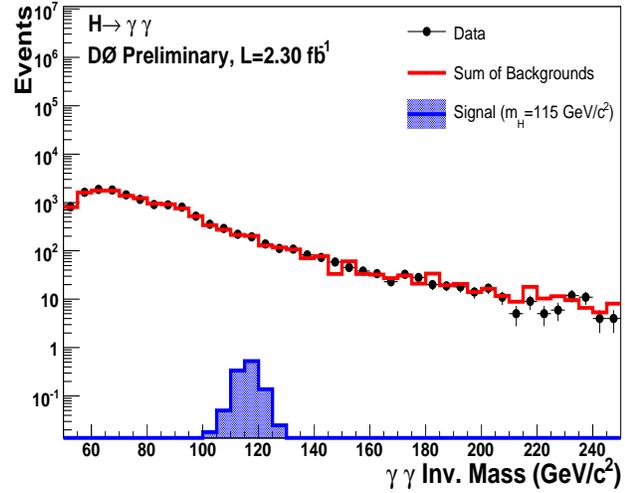}
  \end{center}
    \caption{\label{fig:D0_Hgamgam} Final variable distributions (diphoton invariant mass)
    for the D\O\ Higgs search analysis in \ppHgamgam\ final state using 2.3~\invfb\ of
    Run~II data~\cite{D0_Hgamgam_2.3fb}. The total predicted background, observed number of event
    and expected Higgs signal at $m_H=115~\GeV$ are shown.}
\end{minipage}
\end{center}\end{figure}

The invariant mass of the two photon candidates in the interval
$50~\GeV < m_{\gamma\gamma} < 250~\GeV$, shown in
Fig.~\ref{fig:D0_Hgamgam}, is used as input to the limit
setting program. This search contributes to improve the global
sensitivity in the difficult region around
$m_H\approx130~\GeV$, with a $\sigma \times Br$ ratio to {\tt
SM} of about 45.

\begin{figure*}
\begin{center}
\includegraphics[width=1.\textwidth,height=0.8\textwidth,angle=0]{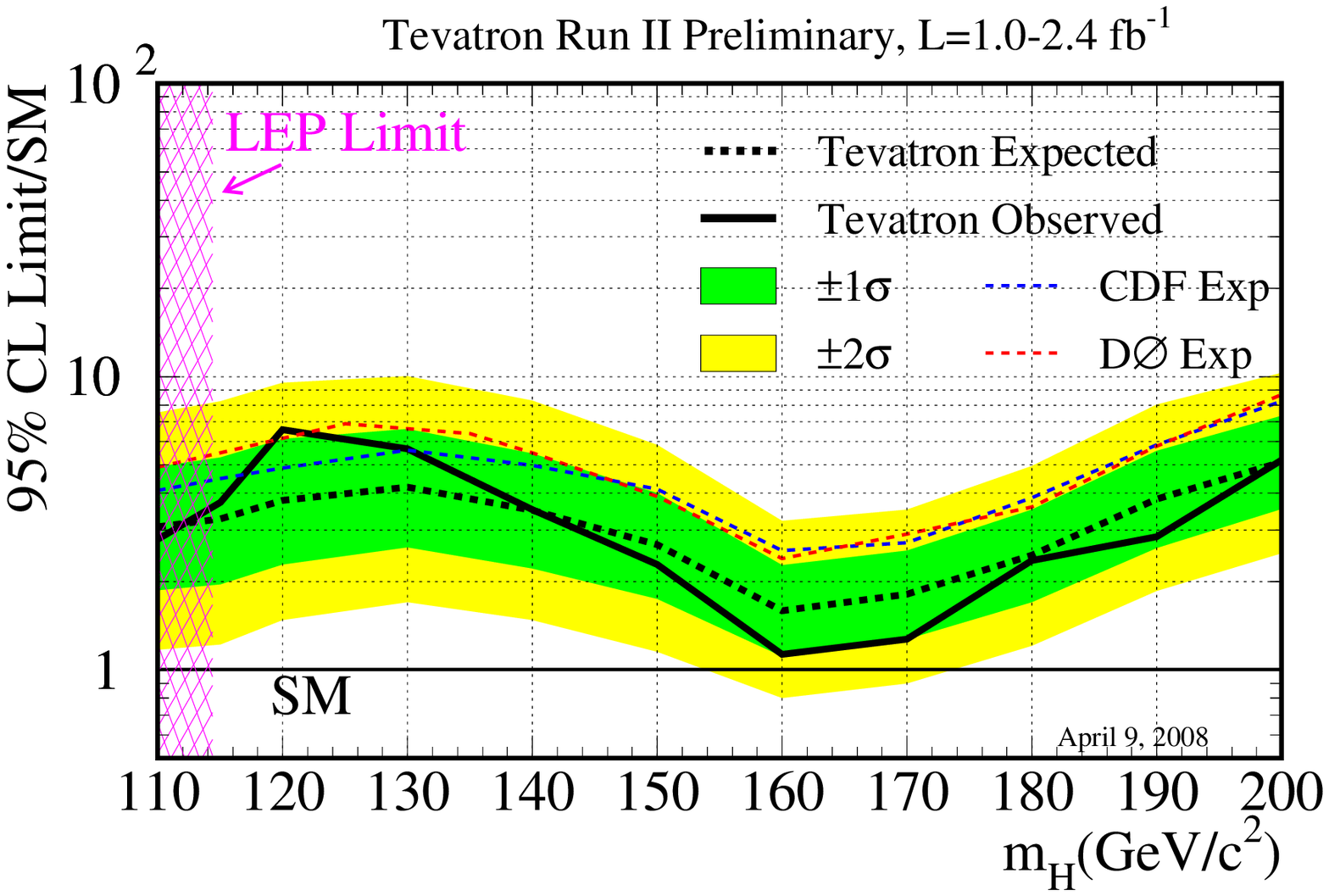}
\caption{\label{fig:CDF_D0_Higgs_comb2.4fb.ps} The new Tevatron
combination~\cite{CDF_D0_Higgs_comb08} presented at the winter
2008 conferences showing the upper bound on the {\tt SM} Higgs
boson cross section as a function of the Higgs boson mass. The
contributing production processes include associated production
(\WH,~\ZH,~\www) or gluon fusion (\hww, \Hgamgam) or vector
boson fusion, and \Htautau\ produced in several modes. The
limits at 95\% confidence level (C.L.) are shown as a multiple
of the {\tt SM} cross section. The solid curve shows the
observed upper bound, while the dashed curves show the expected
upper bounds assuming no signal is present. Analyses are
conducted with integrated luminosities ranging from 1.0~\invfb\
to 2.4~\invfb\ recorded by each experiment. The bands indicate
the 68\% and 95\% probability regions where the limits can
fluctuate in the absence of signal. The expected upper limits
obtained from the CDF and D\O\ experiments are also shown. The
region excluded by the LEP experiments is also
shown~\cite{ref:LEP_Higgs}.}
\end{center}
\end{figure*}

\subsection{$H+X\raa \tau^{+}\tau^{-} + 2$ jets}
The CDF Collaboration recently conducted a search with about 2
\invfb\ of data using the $\tau$ decay mode of the Higgs
boson~\cite{CDF_Htt_2fb}. Several processes are considered:
Higgs production in association with a vector boson ($W/Z$), in
which the vector boson decays into 2 jets ($Br$ are 67/70\% for
$W/Z$), vector boson fusion production in which the 2 jets
coming from the proton and antiproton tend to have a large
rapidity value and, finally, gluon fusion production. The
analysis requires at least 2 jets, one hadronic tau with
$\pt>15~\GeV$ and one leptonic tau identified as an isolated
central electron (or muon) with $\pt>10~\GeV$. In this
analysis, the final variable for setting limits is a
combination of several neural network discriminants.

The expected (observed) sensitivity is roughly about 50\%
(30-40\%) compared to that of searches for associated Higgs
production. The expected cross section limit is 24 times the
{\tt SM} cross section for $m_H=120~\GeV$. Combined with all
other analyses, this result improves the global low mass Higgs
search sensitivity.

\begin{figure*}
\begin{center}
\includegraphics[width=.68\textwidth,height=0.49\textwidth,angle=0]{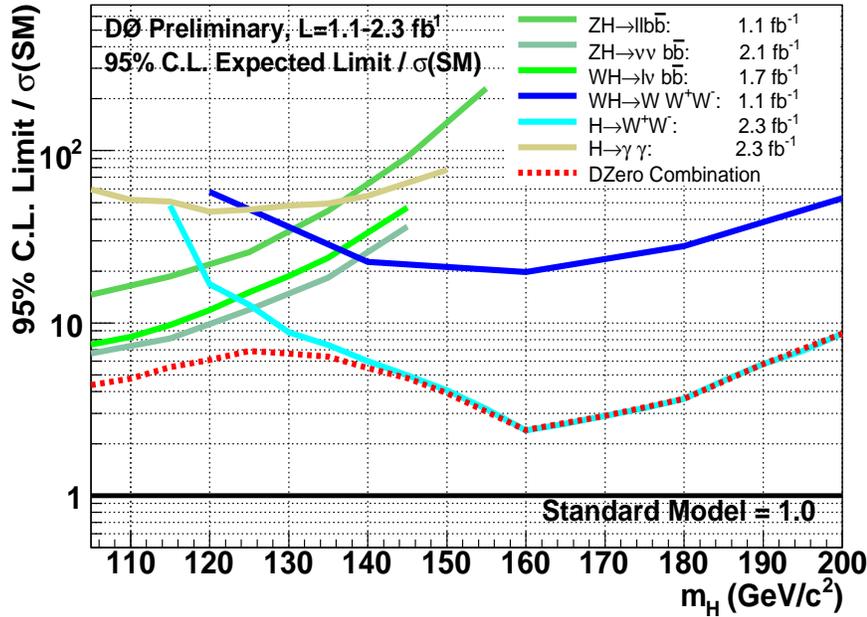}
\caption{\label{fig:D0_All_H_expectedLimits.eps} Expected 95\%
C.L. cross section ratios for the individual $WH/ZH/H, H\raa
b\bar{b}/\gamma\gamma/W^{+}W^{-}$ analyses in the $100~\GeV <
m_H < 200~\GeV$ mass range for D\O\ alone using 1.1-2.3 \invfb\
of data~\cite{D0_Higgs_WIN08}.}
\end{center}
\end{figure*}

\subsection{Combined upper limits}
All individual channels of both Tevatron experiments are
combined to maximize the sensitivity to the Higgs boson. The
combination of D\O\ results based on 0.44 \invfb\ of data has
been published~\cite{D0_Higgs_Pub0.4fb}.

All new results included in this review have been used for the
combination presented at the winter 2008 conferences. These
results have an improved sensitivity compared to the previous
Tevatron combination presented in December
2007~\cite{CDF_D0_Higgs_comb07}. The new Tevatron combination
(21 April 2008, see Ref.~\cite{CDF_D0_Higgs_comb08}) includes
all results from searches for a {\tt SM} Higgs boson produced
in association with vector bosons ($p\bar{p}$ $\raa$ $WH$
$\raa$ $\ell \nu b\bar{b}$, $p\bar{p}$ $\raa$ $ZH$ $\raa$
$\nu\bar{\nu} b\bar{b}$, $p\bar{p}$ $\raa$ $ZH$ $\raa$
$\ell\ell b\bar{b}$ and $p\bar{p}$ $\raa WH$ $\raa$ $WWW^{*}$
$\raa$ $\ell^{\pm} \ell^{\pm}$) or through gluon-gluon fusion (\ppHWW\
and \ppHgamgam) or vector boson fusion, and $H\raa \tau \tau$
produced in several modes, in data corresponding to integrated
luminosities ranging from 1.0-2.4~\invfb\ at CDF
\cite{CDF_Higgs_LP07} and 1.1-2.3~\invfb\ at
D\O~\cite{D0_Higgs_WIN08}. This is the first time that searches
for Higgs bosons decaying to two photons or two tau leptons are
included in the combination.

\subsubsection{Method used for the combination}
To simplify their combination, the searches are separated into
twenty nine mutually exclusive final states (thirteen for CDF,
sixteen for D\O). In addition, several types of combinations
are performed using the Bayesian~\cite{CDF_Higgs_LP07} and
Modified Frequentist~\cite{ref:cls} approaches to check that
the final result does not depend on the details of the
statistical method used for the combination. The Modified
Frequentist approach is sometimes called the LEP $CL_s$ method,
which is based on the log-likelihood ratio ($LLR$) test
statistic
$$LLR(n)=-2\ln(Q),$$
$$ Q = L(s+b)/L(b)
=\frac{e^{-(s+b)}(s+b)^n}{n!}/\frac{e^{-b} (b)^n}{n!},$$
where $s$ and $b$ are the expected numbers of signal and
background events while $n$ is the number of data events. For
all channels, the $LLR$ values per bin are added in order to
use the shape information of the final discriminating variable
and to combine the different channels. In addition, the $CL_s$
method has been extended to involve a fit of the nuisance
parameters for each event, which maximizes the sensitivity in
the context of large background, small signals and large
uncertainties~\cite{emu-wadeLim}. Both $CL_s$ and Bayesian
methods use Poisson likelihoods and agree within $\sim10\%$.
They rely on distributions of the final discriminants like NN
output, matrix-element likelihoods, or dijet mass and not only
on the event counting. The uncertainty on the expected number
of events as well as the shape of the discriminant variables
are included in the systematic uncertainties.

\subsubsection{Systematic uncertainties}
The correlations of systematic uncertainties between channels,
rates and shapes, signals and backgrounds, within and between
experiments are considered. Many sources are measured using
data. Sometimes they are large but correspond to backgrounds
with small contribution. Furthermore, the uncertainties are
constrained by fits to the nuisance parameters and do not
necessarily affect the result significantly.

The following dominant sources of systematic uncertainties are
taken into account in deriving the final results. Each
experiment has a luminosity uncertainty of $\approx$6\%, of
which $\approx$4\% is correlated. Depending on the channel, the
range of uncertainties from {\tt SM} cross sections used to
normalize the simulation vary between 4\% and 20\%. The
uncertainty needed to describe heavy flavor (HF) fractions in
$W/Z$+jets and the effect of the {\tt NLO} HF cross section
normalisation are evaluated to be 20\%-50\%, depending on how
the background is estimated, from data or simulation. The scale
factor used to adjust the $b$-tagging efficiencies in the
simulation results in 4\%-25\% uncertainties, depending on the
number of tagged jets required. The uncertainty from the jet
energy scale ranges between 1\% and 20\%, depending on the
background process. The uncertainties resulting from the
modelling of {\tt QCD} multijet background are dominated by
statistics in the samples from which the estimates are derived
and can reach 30\%. The uncertainties from lepton
identification and reconstruction efficiencies ranges from
negligible to 13\%. Finally, the uncertainty from the trigger
efficiency is less than 5\%.

\subsubsection{Combined results}
The cross section limits on {\tt SM} Higgs boson production
$\sigma\times Br(H\rightarrow X)$ obtained by combining CDF and
D\O\ results with up to 2.4~\invfb\ of
data~\cite{CDF_D0_Higgs_comb08} are displayed in
Fig.~\ref{fig:CDF_D0_Higgs_comb2.4fb.ps}. The result is
normalized to the {\tt SM} cross section, where a value of one
would indicate a Higgs mass excluded at 95\% C.L. The observed
(expected) upper limits are a factor of 3.7~(3.3) higher than
the expected {\tt SM} Higgs boson cross section at
$m_H=115~\GeV$ and 1.1~(1.6) at $m_H=160~\GeV$.

The combined CDF and D\O\ result represents a 40\% improvement
in expected sensitivity over each single experiment. The
observed (expected) limits on the {\tt SM} ratios are 5.0~(4.5)
for CDF and 6.4~(5.5) for D\O\ at $m_H=115~\GeV$, and 1.6~(2.6)
for CDF and 2.2~(2.4) for D\O\ at $m_H=160~\GeV$. The
sensitivity of each individual analysis from D\O\ alone is
given in Fig.~\ref{fig:D0_All_H_expectedLimits.eps}. As shown
in this figure, the associated production and gluon fusion
processes play an important complementary role to improve the
sensitivity for the intermediate mass region around
$m_H=135~\GeV$, which is the most difficult mass to probe at
Tevatron.

\subsection{SM Higgs boson prospects}
It is expected that the sensitivity needed to reach a 95\% C.L.
limit exclusion, at a mass $m_H \approx 160~\GeV$, will be
reached by the end of 2008 by the combined Tevatron
experiments.

Since the first CDF and D\O\ combination in 2006, a lot of
progress has been made, resulting in better sensitivity in all
channels ({\it i.e.}, neural network $b$-tagger, improved
selections, matrix-element techniques). Many of these
improvements led to an equivalent gain of more than twice the
luminosity, which means that the sensitivity has progressed
faster than one would expect from the square root of the
luminosity gained.

Recent projections in sensitivity have been made based on
achievable improvements of the current analyses. These include
progress on the existing improved lepton identification
efficiency, heavy-flavor taggers and $b$-tagging enhancement
from the D\O\ layer L0, upgraded trigger acceptance, increased
usage of advanced analysis techniques, jet resolution
optimization, reduced systematics, and inclusion of additional
channels.

With the Tevatron running well, up to $\approx$~6 {\tt SM}
Higgs bosons events per day are produced per experiment, and
the CDF and D\O\ Collaborations constantly improve their
ability to find them. Combining CDF and D\O, about 4~\invfb\
could be sufficient to exclude the {\tt SM} Higgs boson for
$m_H = 115~\GeV$ and $m_H=160~\GeV$ at 95\%~C.L. Assuming
7~\invfb\ of data analyzed by the end of the Tevatron running,
all {\tt SM} Higgs boson masses -except for the real mass
value- could be excluded at 95\%~C.L. up to 180~\GeV.

\begin{figure}\begin{center}
\includegraphics[width=.5\textwidth,height=0.43\textwidth,angle=0]{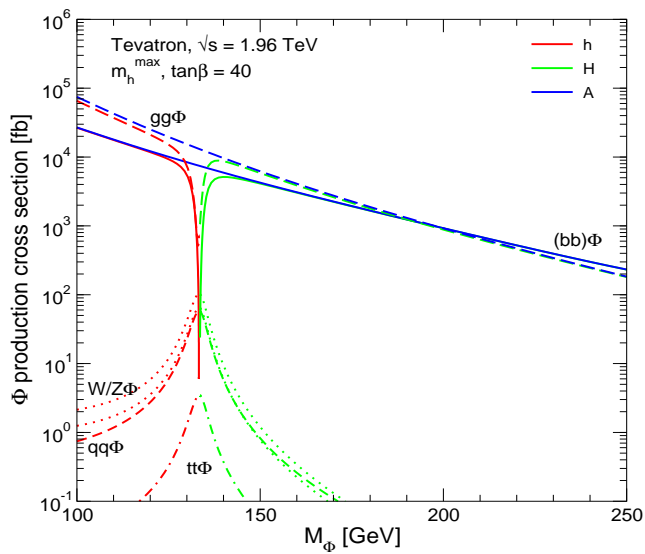}
\caption{\label{fig:mh_max_MSSM_higgs} {\tt MSSM} Higgs boson production cross section
as function of $m_\phi$ ($\phi =h,H,A$) for various production modes at $\tan \beta =40$
in the $m_h^{max}$ scenario~\cite{ref:TeV4LHC}.}
\end{center}\end{figure}

\section{Higgs bosons in the MSSM}
In the minimal supersymmetric extension of the standard model,
two Higgs doublets are necessary to cancel triangular anomalies
and to provide masses to all particles. After electroweak
symmetry breaking, the {\tt MSSM} predicts 5 Higgs bosons.
Three are neutral bosons: $h$, $H$ (scalar) and $A$
(pseudo-scalar), and two are charged bosons: $H^+$ and $H^-$.

An important prediction of the {\tt MSSM} is the theoretical
upper limit $m_h \lesssim 135~\GeV$ on the mass of the lightest
Higgs boson once the radiative loop corrections have been taken
into account~\cite{ref:MSSM_loop,ref:feyn_Higgs}. All other
relations between the Higgs masses and coupling are also
significantly modified by the radiative corrections, which are
dominated by the top- and stop-loop
contributions~\cite{ref:MSSM_loop,ref:theo_tbH}. For large
masses of the pseudo-scalar boson $A$, the light scalar Higgs
becomes {\tt SM}-like. The main difference between the {\tt
MSSM} Higgs bosons and the {\tt SM} Higgs boson is the
enhancement of the production cross section  by a factor
proportional to $\tan^2 \beta$, where $\tan \beta=v_2/v_1$ is
the ratio of the vacuum expectation values associated with the
two neutral components of the scalar Higgs fields. In contrast
to the {\tt SM} Higgs boson, the widths of the {\tt MSSM} Higgs
bosons do not exceed several tens of \GeV\ in most of the
scenarios.

At tree level, only the mass $m_A$ and $\tan \beta$ are
necessary to parameterize the Higgs sector in the {\tt MSSM}.
For $\tan \beta >1$, decays of $h$ and $A$ to $b\bar{b}$ and
$\tau^+\tau^-$ pairs are dominant with branching fraction of
about 90\% and 8\%, respectively. Although the branching
fraction into $\tau$'s is much smaller than the branching
fraction into $b$'s, the $\tau$ mode results in a much cleaner
signature than the $b$ mode, as the latter suffers from a huge
heavy-flavor multijet background which is poorly modeled by
simulation.

Although most of the experimental searches at Tevatron assume
{\it CP} conservation ({\it CPC}) in the {\tt MSSM} sector,
{\it CP}-violating ({\it CPV}) effects can lead to sizable
differences for the production and decay properties of the
Higgs bosons compared to the {\it CPC}
scenario~\cite{ref:theo_susy_MFV}. An observation of a new {\it
CPV} mechanism may yield insight into the observed abundance of
matter over anti-matter in the universe.

\begin{figure}\begin{center}
\includegraphics[width=.5\textwidth,height=0.40\textwidth,angle=0]{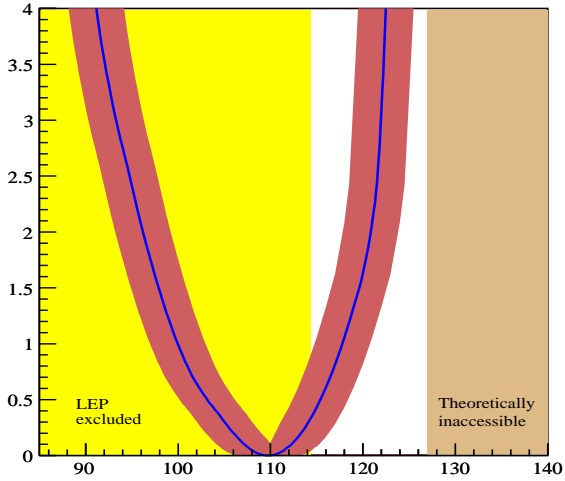}
\caption{\label{fig:MSSM_mh} Scan of the lightest Higgs boson mass versus $\Delta \chi^2$
derived from precision electroweak measurements in the context of a constrained {\tt MSSM}
model. The band around the $\Delta \chi^2$ curve represents the total theoretical uncertainty from
unknown higher-order corrections and the dark shaded area on the right is theoretically inaccessible.
The top mass $m_t=170.9\pm1.8$ was used for this analysis~\cite{ref:MSSM_scan}.}
\end{center}\end{figure}

\subsection{Search strategy}

At the Tevatron, {\it CP} invariance is assumed for the
searches. Both experiments have presented results on searches
for neutral Higgs bosons in the two most promising final
states:
\begin{eqnarray}
\label{eq:hbb} \ppbbphi, \\
\label{eq:htautau} \pphtautau.
\end{eqnarray}

The first process (\ref{eq:hbb}) corresponds to a neutral Higgs
boson decaying into $b\bar{b}$ and produced in association with
bottom quarks. The fourth $b$ is not required in the search,
since a large fraction of the cross section produces a $b$-jet
that does not pass the jet \ET\ threshold. The second topology
investigated at the Tevatron is the gluon fusion process
(\ref{eq:htautau}) where only the $\tau^+\tau^-$ mode is
promising due to the overwhelming $b\bar{b}$ background.

The search strategy for charged {\tt MSSM} Higgs bosons depends
on their mass. For masses $m_{H^\pm}<m_t-m_b$, the charged
Higgs can be produced in the decay of the top quark $t\raa
bH^{+}$, which, in addition to the {\tt SM} $t\raa bW^+$ decay,
leads to the relevant production mode at the Tevatron:
$$ p\bar{p} \raa t\bar{t} \raa bH^+\bar{b}W^-.$$
The charged Higgs may decay to a variety of channels, with $H^+
\raa \tau^+ \nu_\tau$ dominating for large values of $\tan
\beta$. For values of the charged Higgs mass larger than the
top mass, the dominant mode is charged Higgs production in
association with a top and a bottom quark.

\begin{figure}\begin{center}
\includegraphics[width=.5\textwidth,height=0.20\textwidth,angle=0]{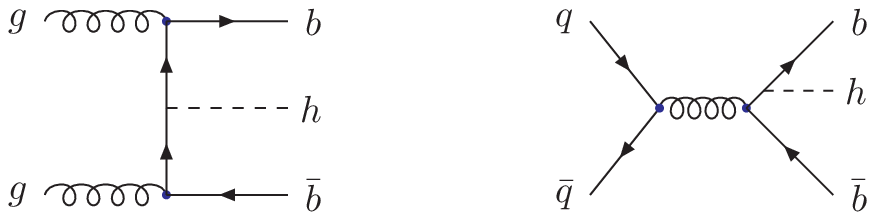}
\caption{\label{fig:bbh} Feynman diagrams contributing to the {\tt LO} $gg\rightarrow b\bar{b}h$ and
  $q\bar{q}\rightarrow b\bar{b}h$ {\tt MSSM} Higgs boson production in association with bottom
quarks. }
\end{center}\end{figure}

\subsection{Benchmark scenarios}

The choice of mechanism for mediating {\tt SUSY} breaking and
of the soft {\tt SUSY} breaking terms governs the main
phenomenological features of {\tt SUSY} models. However, more
than one hundred free parameters remain in the {\tt MSSM},
rendering a complete scan virtually impossible. Several
benchmark scenarios with simplifying assumptions have been
therefore developed to interpret the experimental results.

The preliminary limits from CDF and D\O\ are available in the
($\tan \beta$,$m_A$) plane and are usually summarized for two
{\tt SUSY} scenarios~\cite{ref:MSSM_hbb}. The $m_h^{max}$
scenario is designed to maximize the allowed values of $m_h$
and therefore yields conservative exclusion limits. The
no-mixing scenario differs by the value (set to zero) of the
parameter $X_t$ which controls the mixing in the stop sector,
and hence leads to better limits.

Moreover, if one demands that the values of the bottom and
$\tau$ Yukawa couplings remain in the perturbative regime up to
energies of the order of the unification scale, the region
$\tan \beta \gg 50$ in the {\tt MSSM} is theoretically
disfavoured~\cite{ref:TeV4LHC}. In recent
studies~\cite{ref:mu_neg_disfavoured}, negative values of the
Higgsino mass parameter ($\mu$) are also disfavoured.

The {\tt MSSM} Higgs production cross section is shown in
Fig.~\ref{fig:mh_max_MSSM_higgs} for $\tan \beta=40$ in the
$m_h^{max}$ scenario. At large $\tan \beta$, the pseudoscalar
$A$ boson becomes degenerate with either the light ($h$) or
heavy ($H$) scalar bosons and the bottom-Higgs coupling is
enhanced. A production cross section in the 10~\pb\ range for
the $(bb)\phi$ process is expected and could be observed at
Tevatron.

Direct searches at LEP have placed lower mass limits on both
the lightest scalar and pseudoscalar Higgs bosons at
$m_{h,A}>93$~\GeV\ at 95\% C.L.~\cite{ref:LEP_higgs_WG_2006}.

Similarly to the statistical analysis of precision electroweak
measurements for the {\tt SM} Higgs boson~\cite{ref:LEPEWWG}, a
fit is performed in the context of a constrained {\tt MSSM}
model~\cite{ref:MSSM_scan}. The result is given in
Fig.~\ref{fig:MSSM_mh} as a one parameter scan in the lightest
Higgs boson mass. The predicted value $m_h = 110^{+8}_{-10}
~(exp.) \pm 3 ~(theo.) ~\GeV$ agrees with the direct
experimental lower limit from LEP of
114.4~\GeV~\cite{ref:LEP_Higgs} and the upper theoretical
bound.

\begin{figure}\begin{center}
\includegraphics[width=.5\textwidth,height=0.40\textwidth,angle=0]{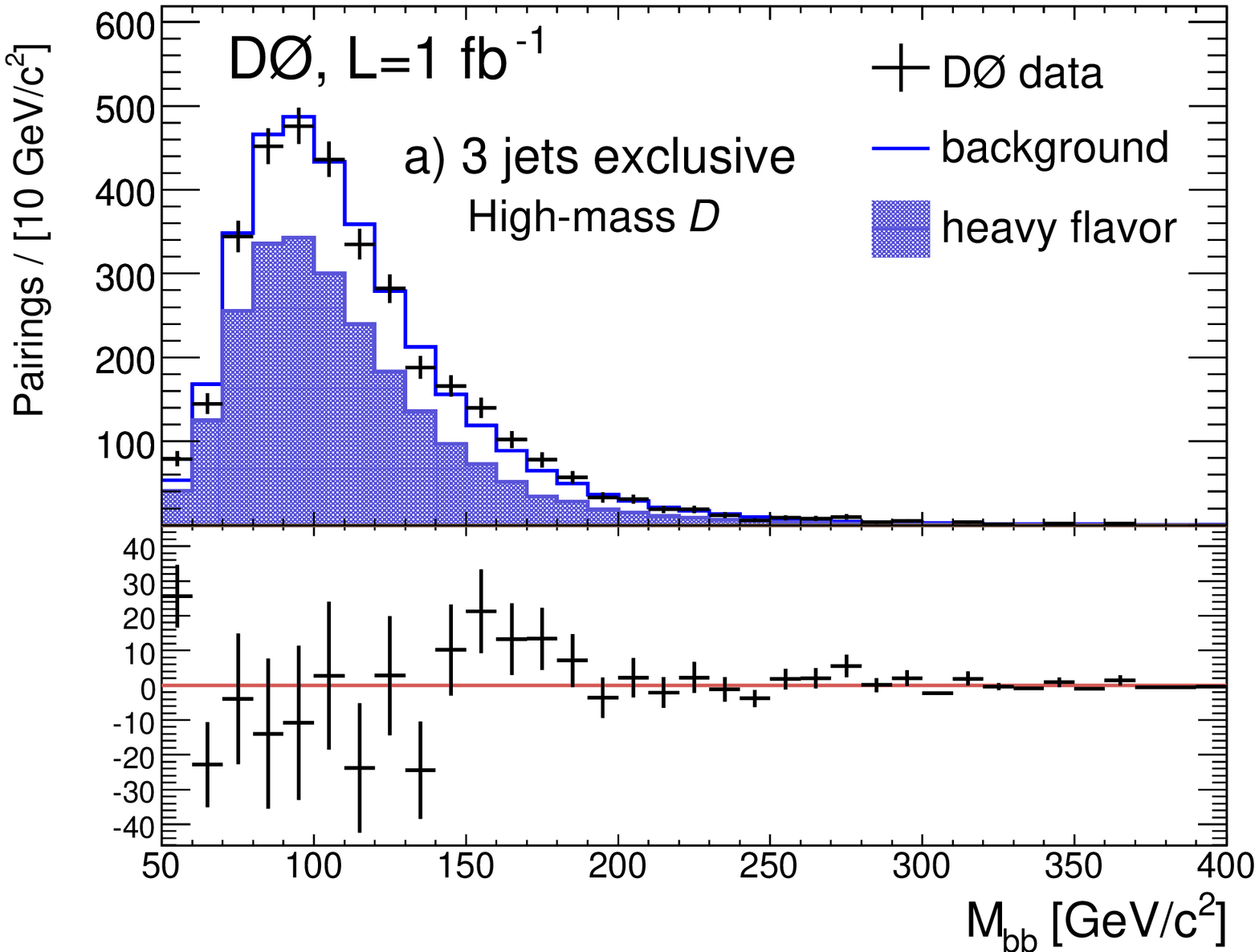}
\caption{\label{fig:D0_hbb_invmass_1fb} Invariant mass for the high-mass likelihood region for the exclusive
three-jet sample used to search for \pphbb\ in 1~\invfb\ of D\O\ data~\cite{D0_hbb_pub_1fb}.
The data are compared to the sum of total background processes (solid line) after all selections. The shaded
region represents the heavy flavor component ($b\bar{b}b, b\bar{b}c, c\bar{c}b$). The ratio between the data
and the total background expectation is also shown.}
\end{center}\end{figure}

\subsection{MSSM neutral Higgs bosons}

\subsubsection{\pphbb}
At tree level (see Fig.~\ref{fig:bbh}), the cross
section~\cite{ref:TeV4LHC,ref2:hbb,ref1:hbb} for production of
{\tt MSSM} neutral Higgs bosons in association with bottom
quarks is almost entirely dominated by the process $gg \raa
b\bar{b}h$, with only a small contribution from $qq \raa
b\bar{b}h$.

A search for \pphbb\ has been previously published by D\O\
based on 260~\invpb~\cite{D0_hbb_pub_260pb}. The results have
been updated with 880~\invpb\ by the D\O\
Collaboration~\cite{D0_hbb_conf_880pb} and 980~\invpb\ by the
CDF Collaboration~\cite{CDF_hbb_conf_980pb}. For the winter
2008 conferences, CDF has presented a result with 1.9~\invfb\
of data~\cite{CDF_hbb_conf_1.9fb}. The D\O\ Collaboration has
just released for publication an improved analysis based on
1~\invfb\ of data~\cite{D0_hbb_pub_1fb}.

In this analysis, CDF and D\O\ search for an event signature of
at least three b-jets with \pt\ greater than 15~\GeV. The
events are triggered by using silicon tracking and jet
requirements. The dijet mass spectrum of the two leading jets
is used to separate the Higgs signal from background events.

A combination of data and simulation is used to model the
background shape. The background in the three-tag sample is
essentially all {\tt QCD} heavy flavor multijet production. The
sample consists of a mix of events: at least two real $b$-jets
with the additional tagged jet being any of a mistagged light
jet ($\approx$ 30\% are $b\bar{b}j$ events where $j$ denotes a
light parton: $u, d, s$ quark or gluon), a $c$-tag ($\approx$
20\% are $b\bar{b}c+bc\bar{c}$ events), or another $b$-jet
($\approx$ 50\% are $b\bar{b}b$ events). In the three-jet
sample, the double $b$-tagged events are found to be
predominantly made of two real $b$-jets. This data sample of
two $b$-tagged jets is exploited to predict the expected triple
$b$-tagged background shape. Both CDF and D\O\ use only the
shape, and not the normalization, of the final discriminating
variable.

\begin{figure}\begin{center}
\includegraphics[width=.24\textwidth,height=0.37\textwidth,angle=0]{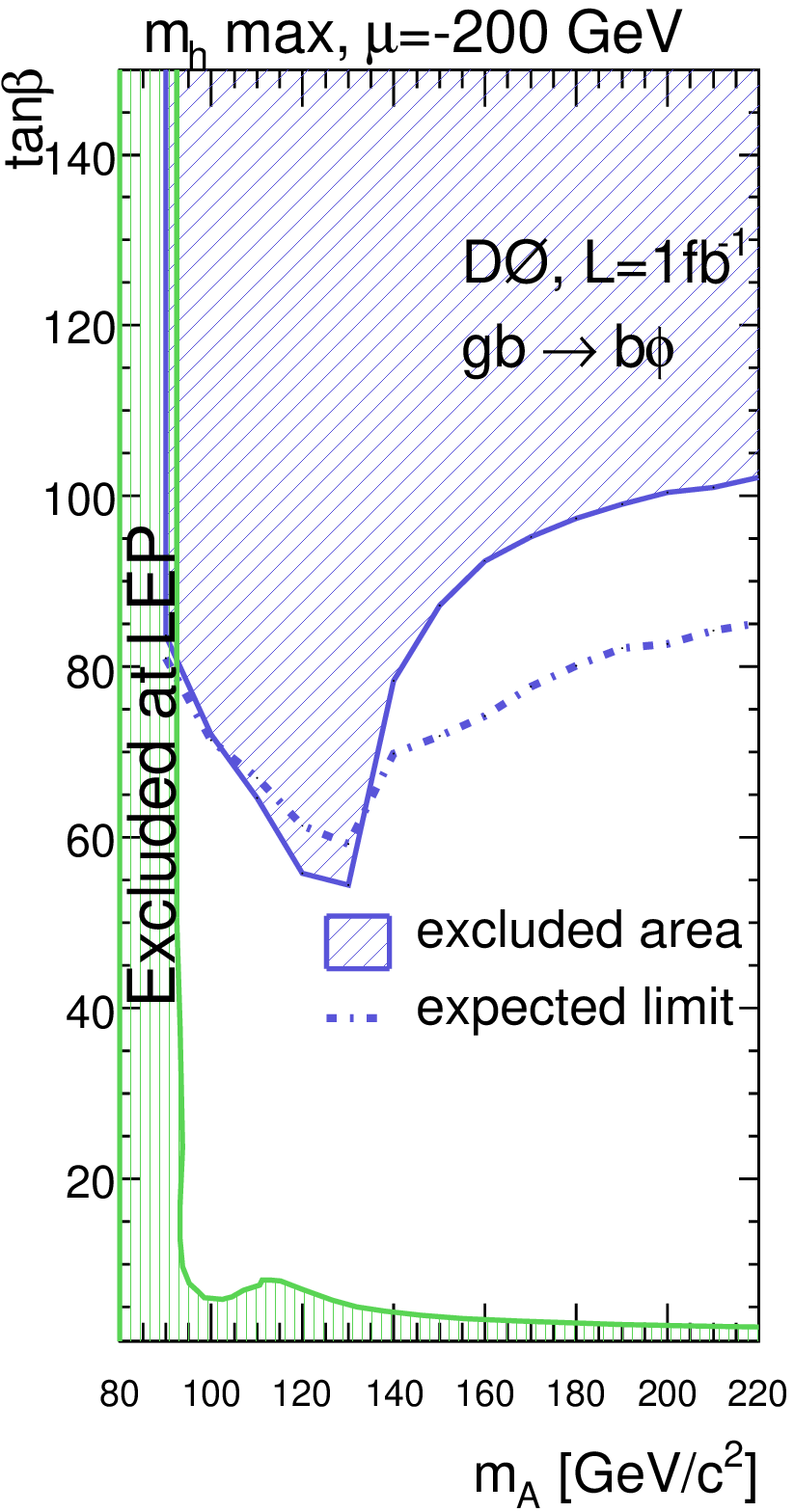}\hfill
\includegraphics[width=.24\textwidth,height=0.37\textwidth,angle=0]{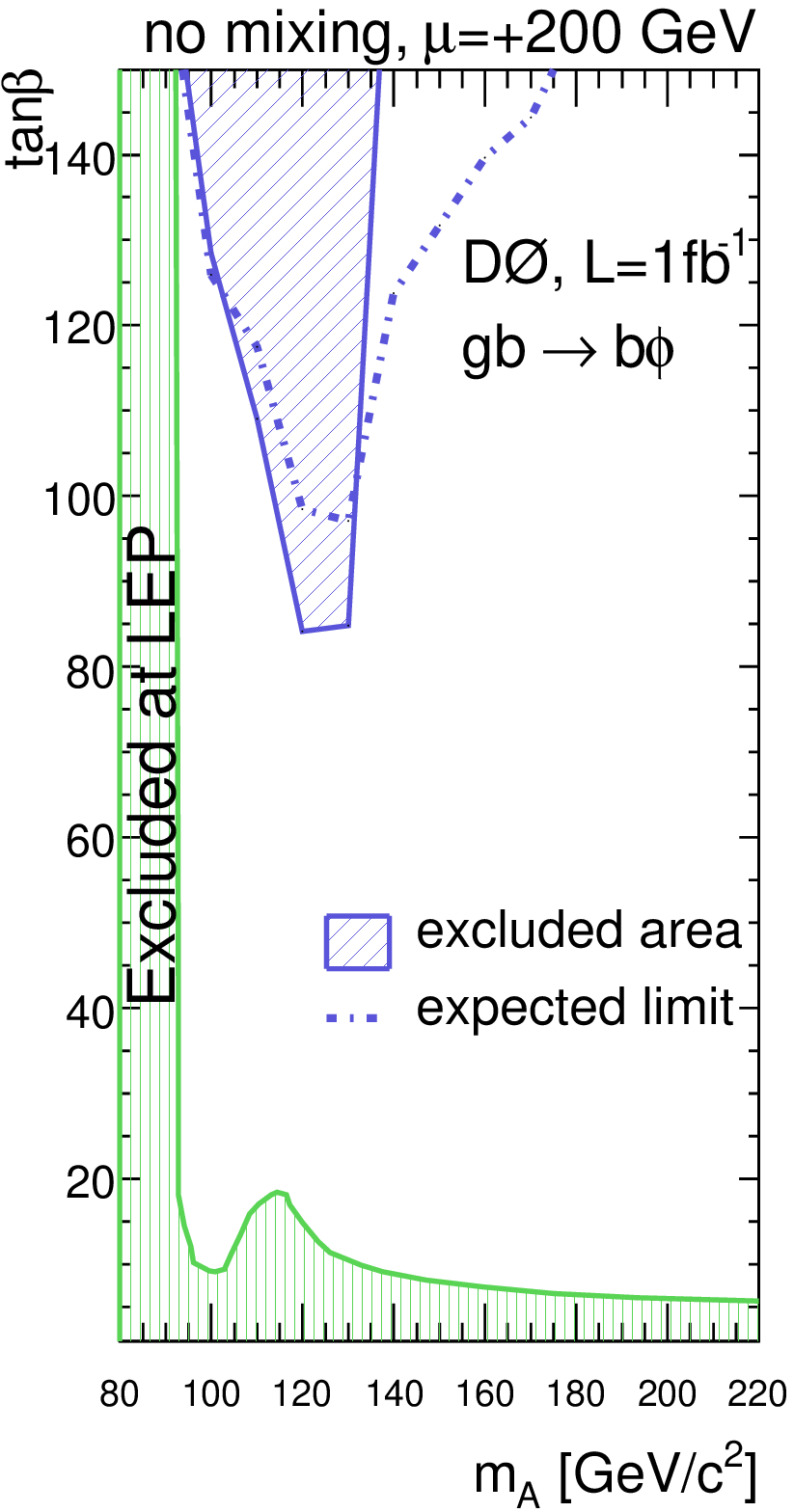}
\caption{\label{fig:D0_hbb_tan_beta_mA}
D\O\ observed and expected 95\% C.L. limits on $\tan \beta$ as a function of $m_A$
using 1~\invfb\ of data~\cite{D0_hbb_pub_1fb}, assuming $\tan^2 \beta$ cross section
enhancement. The effect of the Higgs width is included. The results are given for
two scenarios: $m_h^{max}$ with $\mu=-200$~\GeV, and no-mixing with $\mu=+200$~\GeV. The width of $\phi$
is larger than 70\% above $\tan \beta = 100$ in the $m_h^{max}$ scenario with
$\mu = -200$~\GeV. The exclusions from LEP are also displayed~\cite{ref:LEP_higgs_WG_2006}.}
\end{center}\end{figure}

All Higgs signal events are simulated using the leading order
{\sc pythia}~\cite{ref:pythia} event generator. The cross
section is corrected using next-to-leading order calculations
from {\sc MCFM}~\cite{ref:mcfm} for the Higgs+$b$ process. In
addition, D\O\ corrects the signal acceptance to {\tt NLO}.
Weights obtained with {\sc MCFM} are applied to the signal
samples as function of \pt\ and $\eta$ of the leading $b$-jet
which is not from the decay of the Higgs boson.

A likelihood discriminant ($D$) based on six kinematical
variables built from the two leading jet-pair combinations is
used by the D\O\ search to separate the signal from the
background. The invariant mass distribution for the exclusive
three-tag sample is shown in Fig.~\ref{fig:D0_hbb_invmass_1fb}
for the high-mass optimized likelihood cuts. To further
increase the sensitivity, the analysis is also optimized in the
four-jet and five-jet exclusive samples. The CDF search uses a
binned maximum-likelihood fit of two-dimensional templates in
the mass of the two leading jets versus a variable sensitive to
the flavor of the jet (based on the mass of the tracks forming
the displaced vertex in the jets). The event selection
efficiency varies as a function of the mass of the Higgs boson
and is typically below 1\%.

\begin{figure}\begin{center}
\includegraphics[width=.5\textwidth,height=0.40\textwidth,angle=0]{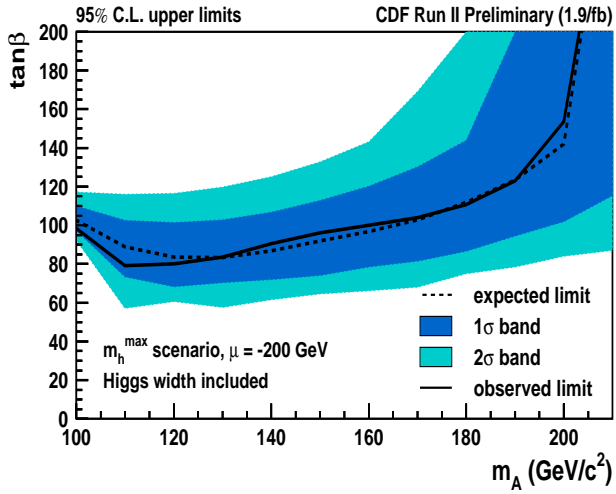}
\caption{\label{fig:CDF_hbb}
The {\tt MSSM} exclusion limit at 95\% C.L. obtained by the CDF experiment with 1.9~\invfb\ of data
on searches for neutral Higgs
bosons produced in association with bottom quarks and decaying into $b\bar{b}$,
projected onto the ($\tan \beta$,$m_A$) plane assuming
standard model cross section times branching fraction (90\%) with $\tan^2 \beta$
enhancement~\cite{CDF_hbb_conf_1.9fb}. The error bands indicate the
$\pm 1 \sigma$ and $\pm 2 \sigma$ range of the expected limit.}
\end{center}\end{figure}

The results have been interpreted in the context of {\tt MSSM}
models since no significant excess has been observed. Limits
are placed on $\tan \beta$ versus the pseudoscalar mass $m_A$.
The D\O\ and CDF exclusion contours, shown in
Fig.~\ref{fig:D0_hbb_tan_beta_mA} and Fig.~\ref{fig:CDF_hbb},
respectively, are based on the following approximate
formula~\cite{ref2:MSSM_hbb}:
\[
\begin{array}{c} \sigma(b \bar b A) \times {\rm BR}(\phi \to
b \bar b)  = \\
\sigma(b \bar b A)_{\rm SM} \; \frac{\tan^2\beta}{\left(1 +
\Delta_b \right)^2} \times \frac{ 9}{ \left(1 + \Delta_b
\right)^2 + 9},
\end{array}
\]
where $\sigma(b\bar{b}A)_{\rm SM}$ denotes the value of the
corresponding {\tt SM} Higgs boson production cross section for
a Higgs boson mass equal to $m_A$. The dependence of the
exclusion bounds in the ($\tan \beta$,$m_A$) plane on the
parameters entering through the most relevant supersymmetric
radiative corrections has been investigated. The loop effects,
incorporated into the $\Delta_b $ parameter in the formula
above, are discussed in Ref.~\cite{ref:TeV4LHC}. The bottom
line is that their inclusion can enhance the cross section by
$\approx\tan^2 \beta$ depending upon the {\tt MSSM} scenario
and significantly modify the bounds obtained. Negative values
of $\mu$ will result in stronger limits on $\tan \beta$ since
the $\Delta_b $ parameter is proportional to the product of
$\tan \beta$ and $\mu$. In addition, CDF and D\O\ take into
account the effect of the Higgs boson width which is calculated
with {\sc feynhiggs}~\cite{ref:feyn_Higgs} and included in the
simulation as a function of the mass and $\tan \beta$ by
convoluting a relativistic Breit-Wigner function with the {\tt
NLO} cross section. Currently the interpretation of these
results within the {\tt MSSM} framework is carried out
using the program {\sc feynhiggs}. The cross section of
{\sc feynhiggs} are
based on a rescaling of the {\tt SM} cross section by the
corresponding {\tt MSSM} factors of the Yukawa couplings. In
future version of these analyses, comparisons with exact {\tt
NLO} calculations of the {\tt MSSM} cross section for $gg \raa$
Higgs should be considered using, for instance, the {\sc higlu}
program \cite{ref:IGLU}.

In the $m_h^{max}$ scenario with $\mu$ negative, the enhanced
production through loop effects allows exclusion of $\tan
\beta$ values greater than 60-100 over the mass range
90-210~\GeV\ for $m_A$.

\begin{figure}\begin{center}
\begin{minipage}{\linewidth}
  \begin{center}
\includegraphics[width=1.\textwidth,height=1.\textwidth,angle=0]{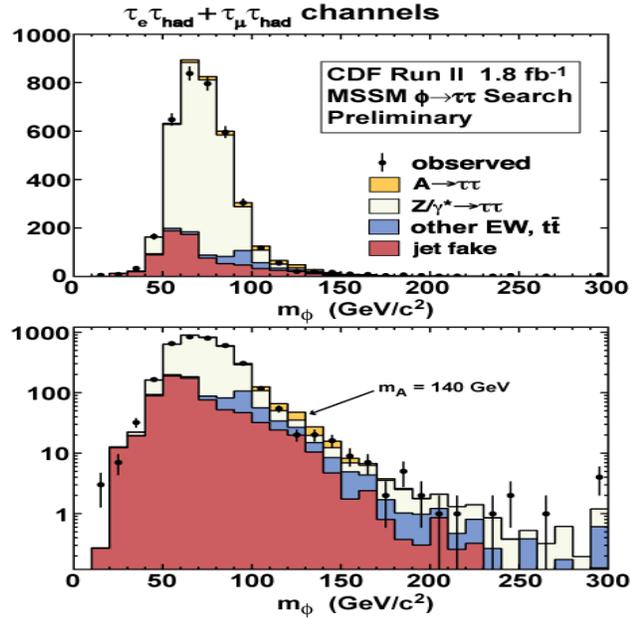}
  \end{center}
\caption{\label{fig:CDF_htautau}
Partially reconstructed ditau mass ($M_{vis}=\sqrt{\pt_\ell + \pt_\tau + \met}$) from the
CDF search for neutral {\tt MSSM} Higgs boson production in the $\tau^+\tau^-$ final state
using 1.8~\invfb\ of data~\cite{CDF_htautau_conf_1.8fb}.
Data (points with error bars) and expected backgrounds (filled histogram) are compared.
The expected contribution from a signal at $m_A=140~\GeV$ is shown.}
\end{minipage}
\end{center}\end{figure}

The observed limits are within 2 standard deviations of the
expectations over the mass region from 90 to 210~\GeV, with the
largest excess occurring around 160~\GeV\ and 180~\GeV\ in the
CDF and \DO\ searches, respectively.

\subsubsection{\pphtautau}
The channels with $\tau^+ \tau^-$ final states have smaller
signal branching fractions, but the searches do not suffer from
the large multijet backgrounds that affect $\phi \raa b
\bar{b}$. In addition, compensations between large corrections
in the Higgs production and decay reduce the impact of
radiative corrections~\cite{ref:TeV4LHC}.

The published Run II CDF \cite{CDF_htautau_pub_310pb} (D\O\
\cite{D0_htautau_pub_348pb}) results use 310 (348) \invpb\ of
data. The CDF Collaboration has recently released a new
preliminary analysis with 1.8 \invfb\
\cite{CDF_htautau_conf_1.8fb}. The D\O\ Collaboration has just
submitted for publication the result corresponding to 1 \invfb\
of data \cite{D0_htautau_pub_1fb}.

Both CDF and D\O\ searches for inclusive production of neutral
{\tt MSSM} Higgs bosons are performed in three final states:
$\tau_e \tau_h$, $\tau_\mu \tau_h$, and $\tau_e \tau_\mu$,
where $\tau_e$, $\tau_\mu$, and $\tau_h$ are notations which
stand for $\tau \raa$ $e\nu_e \nu_\tau$, $\tau \raa$
$\mu\nu_\mu \nu_\tau$, and $\tau \raa$ $\hbox{hadrons}~
\nu_\tau$, respectively. The decay products in $\tau_h$ appear
as narrow jets with low track and $\pi^0$ multiplicity. The
dominant and irreducible background in the final sample of
selected events is from $Z/\gamma^*$ production with subsequent
decays to $\tau$ pairs. Other sources of backgrounds are
$W$+jets, di-bosons, and fake jets. The D\O\ search uses neural
networks to improve tau lepton identification. These neural
networks make use of input variables that exploit the tau
signature such as longitudinal and transverse shower shapes and
isolation in the calorimeter and the tracker.

\begin{figure}\begin{center}
\begin{minipage}{\linewidth}
  \begin{center}
\includegraphics[width=1.\textwidth,height=.9\textwidth,angle=0]{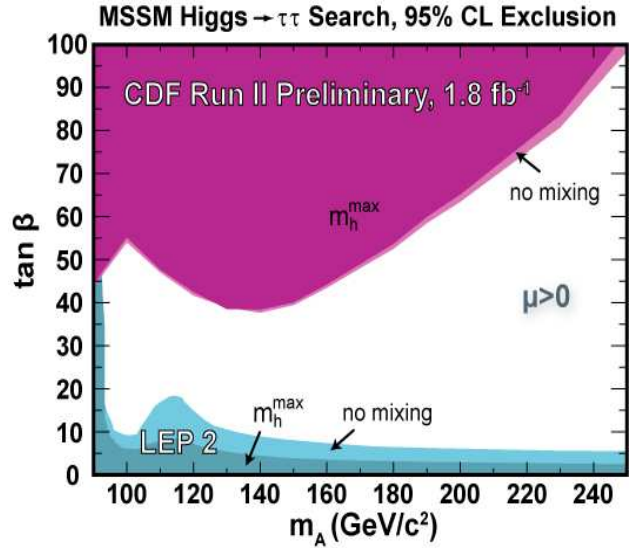}
  \end{center}
\caption{\label{fig:CDF_htautau_tan_beta_m_A}
{\tt MSSM} neutral Higgs boson preliminary results in the $\pphtautau$ channels
using 1.8~\invfb\ of CDF data \cite{CDF_htautau_conf_1.8fb}. The excluded regions
in the ($\tan \beta$,$m_A$) plane are shown for $\mu >0$. The Tevatron excluded domains in the
$m_h^{max}$ and no-mixing scenario are similar. Also shown are the regions
excluded by LEP for these two scenarios~\cite{LEPHIGGS_MSSM}.}
\end{minipage}
\end{center}\end{figure}

\begin{figure}\begin{center}
\begin{minipage}{\linewidth}
  \begin{center}
\includegraphics[width=1.\textwidth,height=.9\textwidth,angle=0]{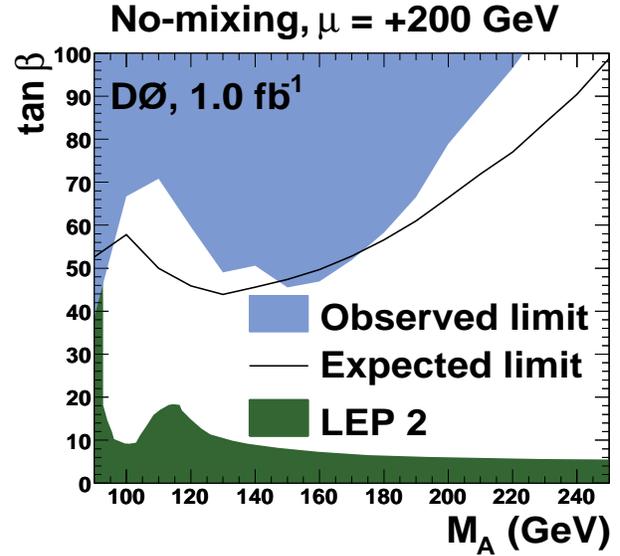}
  \end{center}
\caption{\label{fig:D0_htautau_no-mixing_1fb}
D\O\ {\tt MSSM} exclusion limits at 95\% C.L. obtained with 1~\invfb\ of data
on searches for neutral {\tt MSSM} Higgs bosons decaying to tau pairs~\cite{D0_htautau_pub_1fb}.
The ($\tan \beta$,$m_A$) plane is shown for the no-mixing scenario. The excluded region at LEP is
also represented~\cite{ref:LEP_higgs_WG_2006}.}
\end{minipage}
\end{center}\end{figure}

\begin{figure*}\begin{center}
\begin{minipage}{\linewidth}
  \begin{center}
\includegraphics[width=1.\textwidth,height=.35\textwidth,angle=0]{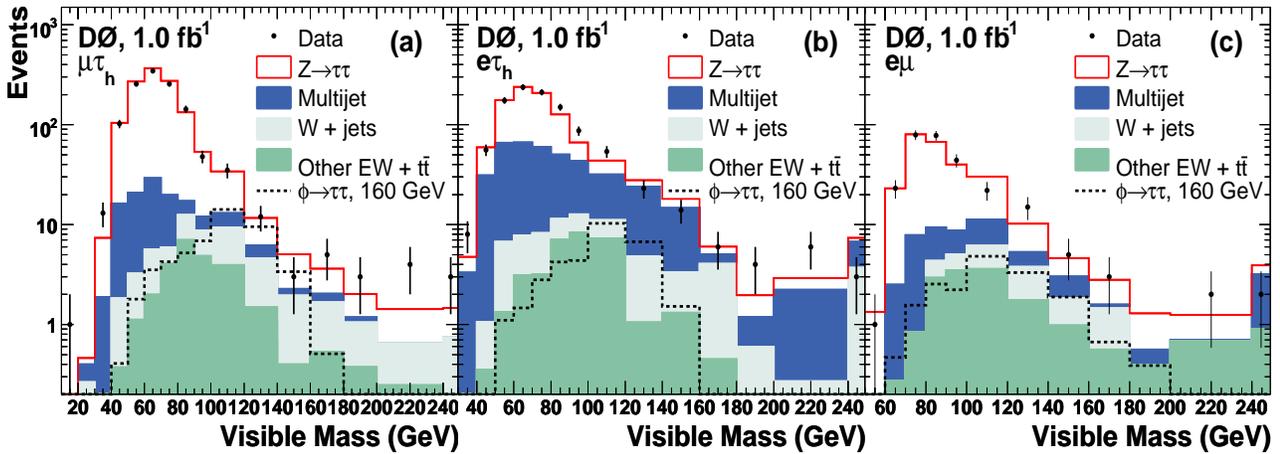}
  \end{center}
\caption{\label{fig:D0_htautau_mvis_1fb} D\O\ visible mass
($m_{vis}$) distributions used to search for neutral {\tt MSSM}
Higgs bosons decaying to tau pairs in the (a) $\mu \tau_h$, (b)
$e\tau_h$, and (c) $e\mu$ channels with 1~\invfb\ of
data~\cite{D0_htautau_pub_1fb}. The signal $\phi \raa \tau^+
\tau^-$ corresponds to $m_\phi=160$~\GeV\ and is normalized to
a cross section of 3 \pb. The highest bin includes the
overflow.}
\end{minipage}
\end{center}\end{figure*}

The CDF search probes for a possible Higgs signal by using a
binned likelihood ratio of the partially reconstructed mass of
the ditau system ($m_{vis}$), defined as the invariant mass of
the visible tau decay products and the \met.
Figure~\ref{fig:CDF_htautau} shows $m_{vis}$ in the
$\tau^+\tau^-$ final state based on 1.8~\invfb\ of data
collected by the CDF experiment. Since the data are consistent
with a background only observation, limits are derived on the
cross section for Higgs boson production times the branching
fraction into tau leptons. The background contributions are
allowed to float within limits set by Gaussian constraints
corresponding to the systematic uncertainties. The
corresponding excluded regions in the ($\tan \beta$, $m_A$)
plane are shown in Fig.~\ref{fig:CDF_htautau_tan_beta_m_A} for
the case $\mu >0$. The dependence of the $\tan \beta$ bounds on
the sign of $\mu$ can be as large as $\Delta \tan \beta \sim
30$ for the $m_h^{max}$ scenario, while in the case of the
no-mixing scenario its effect is smaller, of the order of
$\Delta \tan \beta \sim 10$~\cite{ref:TeV4LHC}. The cross
sections are taken from~\cite{ref2:MSSM_hbb} and are obtained
from {\tt SM} calculations and scaling factors $\sigma_{\tt
MSSM}/\sigma_{\tt SM}$ accounting for the modified Higgs
couplings.

The D\O\ search based on 1~\invfb\ of data exploits the full
$m_{vis}$ spectrum for the likelihood ratio limit calculation.
Limits on $\tan \beta$ as a function of $m_A$ are derived for
the $m_h^{max}$ and no-mixing scenarios, where only positive
values of $\mu$ are considered. The result corresponding to the
no-mixing scenario is represented in
Fig.~\ref{fig:D0_htautau_no-mixing_1fb}. A sensitivity of $\tan
\beta \approx 50$ for $m_A$ below 180~\GeV\ is obtained. The
difference between the observed and expected limits are within
two standard deviations or slightly above for $m_A>250$~\GeV.
It is mainly caused by a data excess in the $\mu \tau_h$
channel for $m_{vis}>160$~\GeV. The distributions of $m_{vis}$
in the three final states are shown in
Fig.~\ref{fig:D0_htautau_mvis_1fb}.

The combination of the LEP and Tevatron searches for neutral
Higgs bosons is expected to probe vast regions of the ($\tan
\beta$,$m_A$) plane by the end of the Run~II. In the no-mixing
scenario shown in Fig.~\ref{fig:CDF_htautau_tan_beta_m_A} and
Fig.~\ref{fig:D0_htautau_no-mixing_1fb}, the lower limits on
$\tan \beta$ obtained at LEP~\cite{LEPHIGGS_MSSM} will slightly
increase because of the assumed top mass $m_t = 174.3~\GeV$,
which is higher than the currently measured value $m_t = 172.6
\pm 1.4 ~\GeV$~\cite{CDF_DO_mtop}. The Tevatron results are not
sensitive to the precise value of the top mass. The upper
limits will also extend with the growing data samples, together
with the improvement of the CDF and D\O\ searches. The
projected sensitivity on the excluded domain would potentially
allow exclusion of $\tan \beta > 20$ for values of $m_A$ up to
few hundred \GeV.

\subsection{Charged Higgs bosons}

\subsubsection{$t \raa b H^+$}

At the Tevatron, direct production of single charged Higgs
bosons is expected to have negligible rate, and the direct
production of $H^+H^-$ via the weak interaction is expected to
have a relatively small cross section, on the order of
0.1~\pb~\cite{ref:theo_tbH}. However, more significant
production could be obtained in the decay of the top quark $t
\raa b H^+$, which would compete with the {\tt SM} process $t
\raa b W^+$. The only recent search for $t \raa b H^+$ at
Run~II is from the CDF Collaboration, which has published a
result based on 193~\invpb\ of
data~\cite{CDF_charged_Higgs_193pb}. The CDF search excludes
the top quark branching fraction to a charged Higgs boson and
$b$-quark $BR(t\raa H^+ b)
> 0.4$ at 95\% C.L. in the region $80~\GeV < m_{H^\pm} <
160~\GeV$, assuming $BR(H^+ \raa \tau^+ \nu_\tau ) = 1$.
Another search, interpreted in the context of the {\tt MSSM}
model, was for anomalous production of high transverse momentum
tau leptons in the decay products of pair-produced top quarks
using 335~\invpb\ of data taken with the CDF detector in Run~II
~\cite{CDF_charged_Higgs_335pb}. An upper limit on $BR(t\raa
H^+ b) > 0.34$ at 95\% C.L. is set for a charged Higgs mass of
120~\GeV.

\begin{figure}\begin{center}
\includegraphics[width=.5\textwidth,height=0.40\textwidth,angle=0]{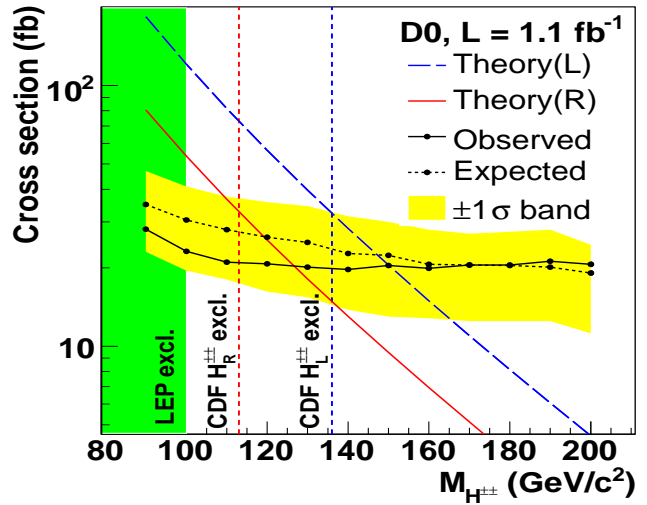}
\caption{\label{fig:DO_charged_higgs} Cross section limit as a function of
the doubly charged Higgs mass $M_{H^{\pm\pm}}$ at the 95\% C.L. in the
$\mu^+\mu^+\mu^-\mu^-$ final state using 1~\invfb\ of data collected by the
D\O\ detector at Run II~\cite{DO_HH_mmmm_1.1fb}.
The mass regions excluded by CDF \cite{CDF_Doubly_Charged_240pb}
and LEP \cite{LEP_double_Charged_Higgs} are also shown. The
$\pm 1 \sigma$ uncertainty on the expected limit is represented by the band.}
\end{center}\end{figure}

\subsubsection{Doubly charged Higgs bosons}
Doubly charged Higgs bosons $H^{\pm\pm}$ are predicted in many
scenarios, such as left-right symmetric models
\cite{ref:Left-Right}, Higgs triplet models and little Higgs
models~\cite{ref:doubly-charged,ref:little-higgs}.

Limits on doubly charged Higgs bosons have been published in
the $ee$, $e\mu$, and $\mu\mu$ channels by the
CDF~\cite{CDF_Doubly_Charged_240pb}
(D\O~\cite{DO_Doubly_Charged_113pb}) experiment based on
240~\invpb\ (113~\invpb) of data collected at Tevatron Run~II.
The search for $p\bar{p} \raa H^{++} H^{--}$ $\raa \ell^+
\tau^+ \ell^- \tau^-$ final states has also been performed by
CDF \cite{CDF_Doubly_Charged_350pb} with 350~\invpb. The D\O\
Collaboration has recently submitted for publication a search
for $H^{\pm\pm}$ in the $\mu^+\mu^+\mu^-\mu^-$ final state
using 1~\invfb\ of data \cite{DO_HH_mmmm_1.1fb}, and sets lower
bounds for right- ($H^{\pm\pm}_R$) and left-handed
($H^{\pm\pm}_L$) bosons at 126~\GeV\ and 150~\GeV,
respectively, at 95\%~C.L. This result is shown in
Fig.~\ref{fig:DO_charged_higgs}. In addition, CDF
has published~\cite{CDF_Doubly_Charged_292pb_long_lived} with
292~\invpb\ of data the case where the doubly charged Higgs
boson lifetime is long ($>3$ m), such that it decays
outside the detector. The lower mass bound on long-lived doubly
charged $H^{\pm\pm}_L$ and $H^{\pm\pm}_R$ bosons are 133~\GeV\
and 109~\GeV, respectively. When the two states are degenerate
in mass, the limit is increased to $m_{H^{\pm\pm}}< 146$~\GeV\
at 95\%~C.L.

\begin{figure}\begin{center}
\includegraphics[width=.5\textwidth,height=0.40\textwidth,angle=0]{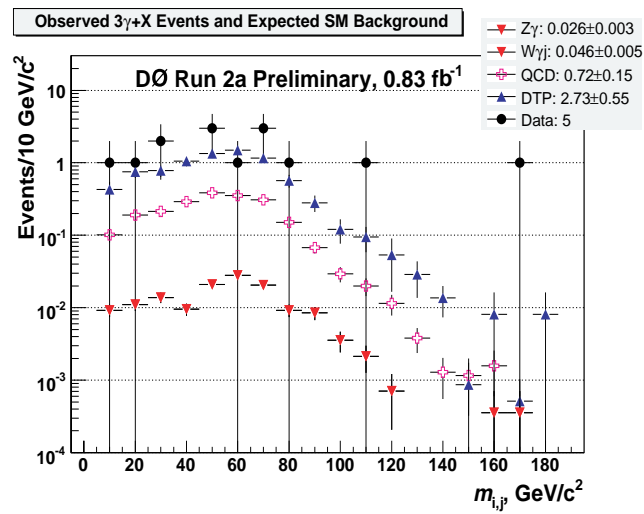}
\caption{\label{fig:DO_fermiophobic_gamgamgam_0.83}
Distribution of two-body invariant mass for $\gamma \gamma \gamma +X$ events observed
in 0.83~\invfb\ of D\O\ data along with the expected {\tt SM} background~\cite{D0_Higgs_fermiophobic_0.83fb}.
The $2.73\pm0.55$ events are estimated from direct
triphon production (DTP). This analysis is used to search for fermiophobic Higgs
via the process $p\bar{p} \raa h_f H^{\pm} \raa h_f h_f \raa \gamma \gamma \gamma (\gamma) + X$.}
\end{center}\end{figure}

\section{Extended Higgs models\label{sec:fermiophobic}}
In a more general framework, one may expect deviations from the
{\tt SM} predictions to result in significant changes to the
Higgs boson discovery signatures. One such example is the
so-called ``fermiophobic'' Higgs
boson~\cite{ref:Fermiophobic1,ref:Fermiophobic2,ref:Fermiophobic3},
which has suppressed couplings to all fermions.

Experimental searches for fermiophobic Higgs at LEP and
Tevatron have yielded negative results so far. In fermiophobic
models, the decay $H^{\pm} \raa h_f W^*$ can have a larger
branching fraction than the conventional decays $H^{\pm} \raa
tb,\tau\nu$. This would lead to double $h_f$ production.
Searches have been conducted via the process $p\bar{p} \raa h_f
H^{\pm} \raa h_f h_f W^{\pm} \raa \gamma \gamma \gamma (\gamma) + X$ by
the D\O\ experiment using
0.83~\invfb\ of data \cite{D0_Higgs_fermiophobic_0.83fb}.
Figure~\ref{fig:DO_fermiophobic_gamgamgam_0.83} shows the
distribution of the diphoton invariant mass in data and from
the expected backgrounds, where each event contributes three
histogram entries since they are three possible photon-photon
combinations. This analysis select 5 events in the data. The
$2.73\pm0.55$ background events from direct triphon production
(DTP), {\it i.e.}, direct diphoton production (DDP) along with
the {\tt FSR/ISR} photon, are estimated by scaling the
corrected number of diphoton events observed in data with the
rate at which one would expect to observe a third photon in DDP
processes from {\sc pythia}. The background from events in
which jets or electrons were misidentified as photons is
estimated in data and represents $0.8\pm0.15$ event. In absence
of excess, a limit is set at $m_{h_f} > 80~\GeV$ for
$m_{H^{\pm}}<100~\GeV$ and $\tan \beta =30$ at 95\% C.L.

\begin{figure}\begin{center}
\includegraphics[width=.5\textwidth,height=0.40\textwidth,angle=0]{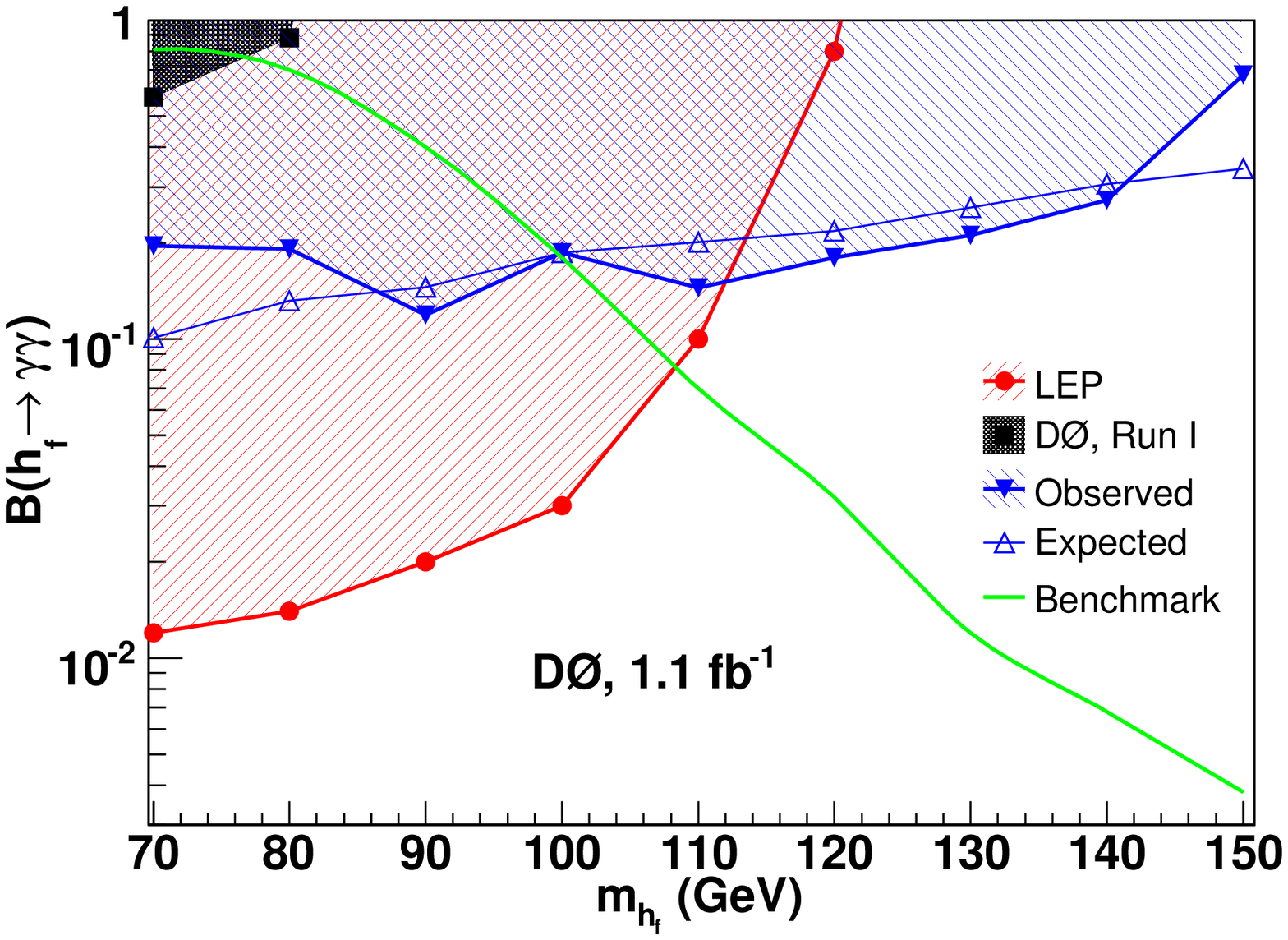}
\caption{\label{fig:D0_hgg_1.1fb} Excluded branching fractions
$B(h_f\raa \gamma \gamma)$ as function of the fermiophobic Higgs boson mass $m_{h_f}$
for the D\O\ search with 1.1~\invfb\ of data \cite{D0_Hgamgam}. The signal is the sum of $p\bar{p} \raa VV \raa h_f$
and $p\bar{p} \raa h_f V$ processes.
The theoretical curve uses the benchmark scenario assuming that the coupling $h_fVV$ ($V=W,Z$) has the
same strength as in the {\tt SM} and that all fermion branching
fractions are exactly zero.}
\end{center}\end{figure}

Another D\O\ search for fermiophobic Higgs bosons has been
recently submitted for publication using 1.1 \invfb\ of
data~\cite{D0_Hgamgam}. This analysis searches for inclusive
production of diphoton final states via Higgsstrahlung
$p\bar{p} \raa h_f V \raa \gamma \gamma + X$ and vector boson
fusion $p\bar{p} \raa VV$ $\raa \gamma \gamma + X$ processes
($V=W,Z$). The benchmark model used to set mass limits assumes
that the coupling $h_fVV$  has the same strength as in the {\tt
SM} and that all fermion branching fractions are exactly zero.
The study shown in Fig.~\ref{fig:D0_hgg_1.1fb} excludes
fermiophobic Higgs bosons of mass up to 100~\GeV\ at the 95\%
C.L. and represents the most stringent limits to date at a
hadron collider.

Technicolor models~\cite{ref:Technicolor} provide an
alternative dynamical explanation of electroweak symmetry
breaking through a new strong gauge interaction acting on new
fermions, called ``techni-fermions''. The D\O\ Collaboration
published a search corresponding to 390 \invpb\ of
data~\cite{D0_technicolor_390pb} in the final state containing
one electron and two jets coming from the decay of vector
techni-mesons ($p\bar{p} \raa \rho_T/\omega_T$) to a $W$ boson
and a techni-pion $\pi_T$, followed by the decays $W\raa e \nu$
and $\pi_T \raa b\bar{b}, b\bar{c},$ or $c\bar{c}$. As no
significant excess in the data was observed, limits have been
set. For instance, a mass of $m_{\rho_T}\approx 210~\GeV$ is
excluded for the corresponding $m_{\pi_T}\approx 120~\GeV$ at
95\% C.L. Similarly, CDF presented a search for technicolor
particles decaying into $b\bar{b}$, $b\bar{c}$ or $bu$ and
produced in association with $W$ bosons using 1.9~\invfb\ of
data~\cite{CDF_technicolor_1.9fb}. Events matching the
$W+2$~jets signature are selected by requiring the electron or
muon to be isolated with \ET\ or \pt $>20$~\GeV,
$\met>20$~\GeV, and at least one $b$-tagged jet. The number of
tagged events and the invariant mass distributions of $W$ + 2
jets and dijet events are consistent with the {\tt SM}
expectations. For $m_{\rho_T}\approx 250$~\GeV, the excluded
mass range is $135~\GeV < m_{\pi_T} < 145~\GeV$ at 95\% C.L.

%% file: BSM.tex
\section{Beyond the standard model}
Beyond the elucidation of the mechanism of electroweak symmetry
breaking, there are many compelling and well-motivated models
that can be tested at the Tevatron. But what are the CDF and
D\O\ Collaborations looking for at Tevatron? By far, the most
widely studied theory beyond the {\tt SM} involves new
particles predicted by low energy supersymmetry. Searches are
therefore often divided into {\tt SUSY} and non-{\tt SUSY}
categories. This succinct summary covers the following topics:
\begin{itemize}
 \item[$\bullet$] Extension beyond the Poincar\'{e} group, {\it
     i.e.}, supersymmetry, such as searches for electroweak
     gauginos with leptonic decay, and squark and gluino
     production resulting in multijet+\met\ topologies.
 \item[$\bullet$] Existence of a new symmetry leading to
     massive particles with a lifetime comparable to the
     typical transit time through the detector.
 \item[$\bullet$] Particle substructure or compositeness,
     such that history repeats itself, leading to {\it
     e.g.}, leptoquark particles or excited fermion states.
 \item[$\bullet$] An enlarged gauge group resulting in
     exotic $Z'$ or $W'$ bosons.
 \item[$\bullet$] An increase in the number of spacial
     dimensions, {\it i.e.}, extra-dimension models with
     real Kaluza-Klein gravitons produced in association
     with a jet, or virtual Kaluza-Klein gravitons
     exchanged in the production of fermion or vector-boson
     pairs.
 \item[$\bullet$] A search for an excess in data without a
     specific model in mind. These are grouped into
     so-called signature-based searches.
\end{itemize}

\section{Charginos and neutralinos}
Charginos and neutralinos are respectively the charged and
neutral partners of gauge and Higgs bosons. The primary search
modes are pair production of charginos ($\chi^+_1 \chi^-_1$) or
associated chargino-neutralino ($\chi^\pm_1 \chi^0_2$)
production, where $\chi^\pm_1$ is the lightest chargino and
$\chi^0_2$ is the second lightest neutralino. A search for {\tt
SUSY} can therefore be performed via the associated production
of charginos and neutralinos, where the $\chi^\pm_1$ and
$\chi^0_2$ are assumed to decay either via exchange of vector
bosons or via sleptons into {\tt SM} fermions while the
lightest neutralino remains undetected.

\begin{figure}\begin{center}
\includegraphics[width=1.0\linewidth,height=0.42\textwidth,angle=0]{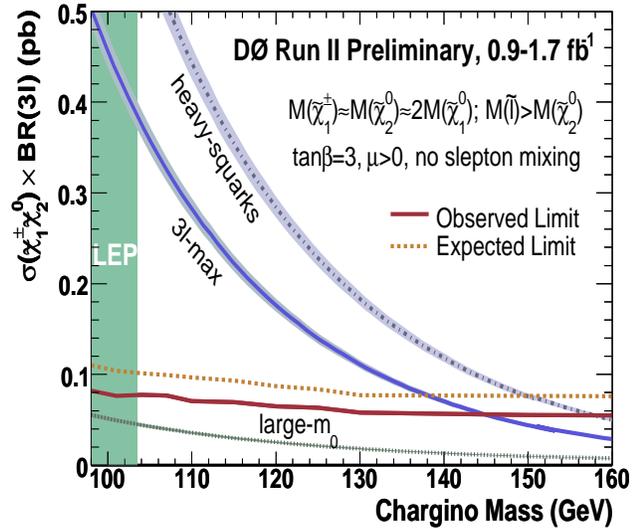}
\caption{\label{fig:D0_trilep} D\O\ 95\% C.L. limits on the total cross section
for associated chargino and neutralino production with leptonic final states as a
function of the chargino ($\chi_{1}^{\pm}$) mass, in comparison with the
expectation for several {\tt SUSY} scenarios using 0.9-1.7 \invfb\ of data \cite{DO_tril_0.9-1.7fb}. The line
corresponds to observed minimal {\tt SUGRA} limit. {\tt PDF}
and renormalization/factorization scale uncertainties are shown
as shaded bands. The lower mass limit at 103.5~\GeV\ is from LEP searches~\cite{LEPSUSY_Chargino}.}
\end{center}\end{figure}

The CDF and D\O\ Collaborations have searched in the trilepton
final state. Published results are based on 320~\invpb\ for
D\O~\cite{DO_pub_trilepton_320pb} and 1.1~\invfb\ for
CDF~\cite{CDF_pub_tril_1fb,CDF_pub_dil_1fb,CDF_pub_trilepton_1.1fb}.
The $eel$ channel was updated to $1.7~\invfb$ by D\O\ and, in
combination with results in four other trilepton search
channels based on approximately 1~\invfb\ of data, new limits
on the associated production of charginos and neutralinos have
been set by the D\O\ Collaboration \cite{DO_tril_0.9-1.7fb}.
The CDF results were updated with 2~\invfb\ of
data~\cite{CDF_tri_2fb}.

The trilepton final state has long been suggested to be one of
the most promising channel for discovery of {\tt SUSY} at a
hadron collider. However, these searches suffer from a cross
section below 0.5~\pb\ with leptons that are difficult to
reconstruct due to their low transverse momenta, rendering the
analyses challenging. Furthermore, many channels need to be
combined to achieve sensitivity. The selection consists of two
well identified and isolated electrons ($e$) or muons ($\mu$)
with a \pt\ cut above $\approx10$ \GeV. An additional isolated
track provides sensitivity to the third lepton ($l$) and, by
not requiring explicit lepton identification, efficiency is
maximized. The presence of neutrinos and neutralinos in the
final state results in some missing transverse energy. Finally,
since very few {\tt SM} processes are capable of generating a
pair of isolated like-charge leptons, the same analysis is
performed with this looser criterion.

The results are interpreted in minimal {\tt SUGRA} inspired
scenarios ({\tt mSUGRA}) where gravity mediates {\tt SUSY}
breaking from the grand unification theory ({\tt GUT}) scale to
the electroweak scale. With {\it R}-parity conservation (see
section \ref{sec:R-parity}), {\tt mSUGRA} can be completely
characterized by five parameters: a common scalar mass ($m_0$),
a common gaugino mass ($m_{1/2}$), a common trilinear coupling
value ($A_0$), the ratio of the vacuum expectation values of
the two Higgs doublets ($\tan \beta$), and the sign of the
Higgsino mass parameter ($\mu$). Direct searches at LEP set a
lower limit on the mass of the chargino ${\chi^\pm_1}$ at
103.5~\GeV\ for sneutrino masses larger than
200~\GeV~\cite{LEPSUSY_Chargino}.

\begin{figure}
\begin{center}
\includegraphics[width=1.0\linewidth,height=0.4\textwidth,angle=0]{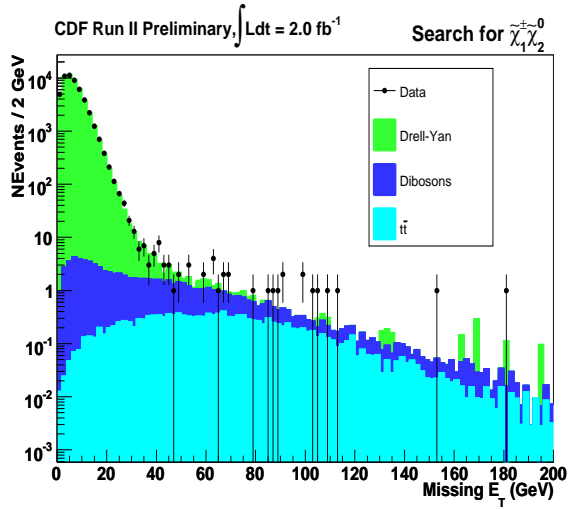}
\caption{\label{fig:CDF_trileptons_met} \met\ distribution for di-lepton events with
invariant mass $76~\GeV < m_{\ell \ell} < 106$~\GeV. This control region is used to test {\tt SM} predictions
of the CDF search for charginos and neutralinos using 2~\invfb\ of data~\cite{CDF_tri_2fb}.}
\end{center}\end{figure}

As a guideline, D\O\ results are interpreted in this model with
chargino $\chi_{1}^{\pm}$ and neutralino ($\chi_{2}^{0},
\chi_{1}^{0}$) masses following the relation $m_{\chi_{1}^{\pm}}
\simeq m_{\chi_{2}^{0}} \simeq 2m_{\chi_{1}^{0}}$. The leptonic
branching fraction of chargino and neutralino depends on the
relative contribution from the slepton- and W/Z-exchange
graphs, which varies as a function of the slepton masses. Three
{\tt mSUGRA} inspired scenarios were used for the
interpretation as shown in Fig.~\ref{fig:D0_trilep}. Two of
them are with enhanced leptonic branching fractions (``heavy
squarks'' and ``3$l$-max'' scenarios). For the 3$l$-max
scenario, the slepton mass is just above the neutralino mass
($m_{\chi_{2}^{0}}$), leading to maximum branching fraction
into leptons. The heavy squark scenario is characterized by
maximal production cross section. Finally, the large universal
scalar mass parameter ($m_0$) scenario is not yet sensitive
because the $W/Z$ exchange dominates. The new D\O\
result~\cite{DO_tril_0.9-1.7fb} in the $eel$ channel using
$1.7~\invfb$ of data observes no events after final selection,
with $1.0 \pm 0.3$ events expected from the {\tt SM} background
and between 0.5-0.2 events for the signal. In the dataset
corresponding to 1~\invfb, no candidates have been found in the
$e\mu\ell$ channel with an expected background of
$0.9^{+0.4}_{-0.1}$ events, while two candidates are found in
the $\mu\mu\ell$ channel consistent with the background
expectation of $0.3^{+0.7}_{-0.1}$ events. In the $ee\ell$
channel, no candidates have been found, with an expected
background of $0.8\pm0.7$ events. The observation of one event
in the data is consistent with the $1.1\pm0.4$ events expected
from the background in the $\mu^{\pm}\mu^{\pm}$ channel. Since
no evidence for {\tt SUSY} is reported, all results are
combined to extract limits on the total cross section, taking
into account systematic and statistical uncertainties including
their correlations. The D\O\ combination excludes chargino
masses below 145~\GeV\ at 95\% C.L. for the 3$l$-max scenario.

\begin{figure}
\begin{center}
\includegraphics[width=1.0\linewidth,height=0.4\textwidth,angle=0]{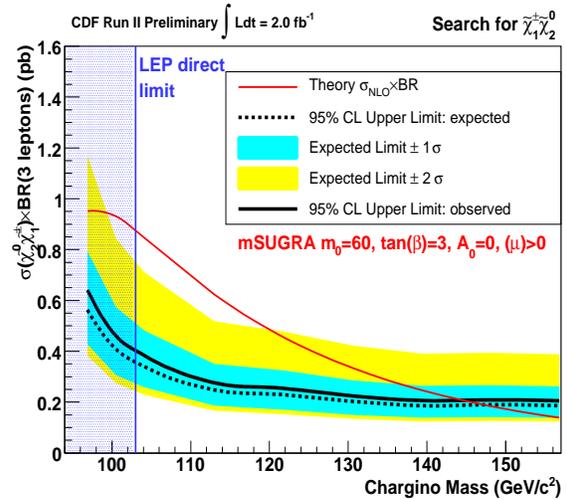}
\caption{\label{fig:CDF_trileptons} CDF 95\% C.L. limits on the total cross section for associated chargino
and neutralino production with leptonic final states using 2 \invfb\ of data \cite{CDF_tri_2fb}. The expected limit
corresponds to the dashed line, with $\pm 1 \sigma$ and $\pm 2 \sigma$ uncertainty
bands shown. The next-to-leading order ({\tt NLO}) production cross section
corresponds to an {\tt mSUGRA} model with the universal scalar mass parameter fixed to $m_0=60~\GeV$.}
\end{center}\end{figure}

Similar analyses have been performed by CDF but interpreted
with slightly different scenarios and with a total integrated
luminosity corresponding to $2~\invfb$ of
data~\cite{CDF_tri_2fb}. After selecting dilepton events, a
control region with $76~\GeV < m_{\ell \ell} < 106$~\GeV\ is
used to establish a good understanding of the data and to test
the {\tt SM} predictions. The \met\ distribution for this
control region is displayed in
Fig.~\ref{fig:CDF_trileptons_met}. The low and high invariant
mass regions are also explored. The search is splitted into
five exclusive channels and optimized for a benchmark signal
point in the minimal {\tt SUGRA} scenario, corresponding to
$m_0 = 60$~\GeV, $m_{1/2} = 190$~\GeV, $\tan \beta = 3$, $A_0 =
0$, and $\mu>0$. In this case, the masses
$m_{\chi_{1}^{\pm}}=119.6$ \GeV, $m_{\chi_{1}^0}=122$~\GeV,
$m_{\chi_{2}^0}=67$~\GeV\ have been computed with {\sc
isajet}~\cite{ref:isajet} and the corresponding $\chi_{1}^{\pm}
\chi_{2}^0$ production cross section 0.327~\pb\ with {\sc
prospino-2}~\cite{ref:prospino2}. A total of $0.9 \pm 0.1$
background events in the trilepton channels are expected for
$4.5\pm0.4$ events from the signal, and $5.5 \pm 1.1$
background events for the dilepton+track channels for
$6.9\pm0.6$ signal events. CDF observes 1 event in the
trilepton channel and 6 events in the dilepton+track channels.
No excess is therefore observed and the resulting cross section
limit shown in Fig.~\ref{fig:CDF_trileptons} is given as a
function of the chargino mass for the benchmark {\tt mSUGRA}
scenario defined above but varying $m_{1/2}$. This scenario
enhances the branching fraction of chargino and neutralino into
leptons, and excludes chargino masses below 140~\GeV\ for a
sensitivity (expected limit) of 142~\GeV\ at 95\% C.L. Other
models are being investigated for upcoming analyses with
increased luminosity.

The results between the two experiments cannot be directly
compared since the fixed low $m_0$ value leads to a two-body
decay for the CDF analysis, while for the D\O\ analysis a
sliding window of $m_0$ is used to keep the slepton mass
slightly above the ${\chi_{2}^{0}}$ mass which corresponds to a
three-body decay.

\section{Squarks and gluinos}

\subsection{Generic $\tilde{q}$ and $\tilde{g}$ searches}

In $p\bar{p}$ collisions, squarks ($\tilde{q}$) and gluinos
($\tilde{g}$), the superpartners of quarks and gluons, are
expected first to be abundantly produced if they are
sufficiently light, and second to largely exceed the mass reach
achieved at LEP. However, these searches have large background
at the Tevatron. The final states are studied within the
framework of {\tt mSUGRA} assuming {\it R}-parity conservation.
All {\tt SUSY} particles, except the lightest neutralino, are
unstable and will therefore decay into their {\tt SM}
counterparts right after being produced, leading to a cascade
decay with a final state consisting of several jets from the
squarks and the gluinos, plus missing transverse energy coming
from the neutralinos. Note that to interpret the results of
this search, the ten {\tt SUSY} partners of the five light
quarks flavors were considered to be degenerate in mass by D\O.
In the following, the squark mass is therefore defined as the
average mass of all squarks other than the superpartners of the
top. The CDF analysis assumes that only the first and second
generation masses are degenerated.

\begin{figure}\begin{center}
\includegraphics[width=1.0\linewidth,height=0.46\textwidth,angle=0]{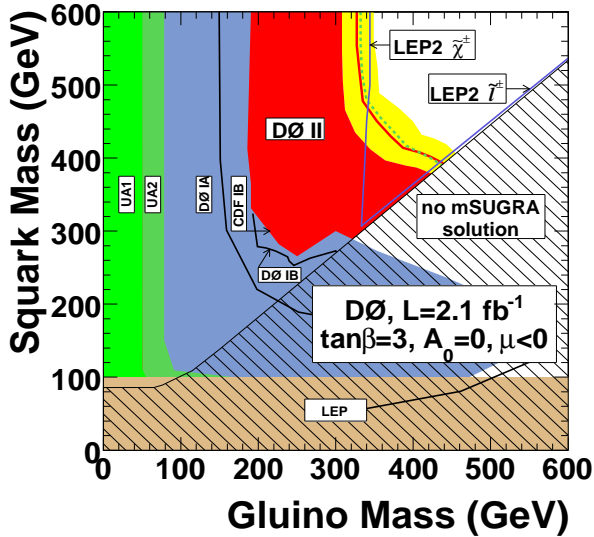}
\caption{\label{fig:D0-SqGl}  D\O\ Run~II exclusion plane for squark and gluino masses at 95\% C.L.
using 2.1~\invfb\ of Run~II data~\cite{DO_pub_sqgl_2.1fb}, in the mSUGRA framework. The region
excluded by previous results and this analysis is shown as the ``D\O\ II'' shaded area. The
thick (dotted) line is the limit of the observed (expected) excluded
region for the nominal theoretical cross section. The band around these limits
shows the effect of the {\tt PDF} choice and of the variation of renormalisation/factorisation scale
by a factor of two.}
\end{center}\end{figure}

The most constraining direct limits on squark and gluino masses
are published by the D\O\ Collaboration
\cite{DO_pub_sqgl_2.1fb}, based on an analysis of 2.1~\invfb\
of data. A preliminary result with similar sensitivity was
shown by CDF~\cite{CDF_conf_sqgl_2fb} during the winter 2008
conferences with 2~\invfb\ of data.

Three different scenarios have been probed by both CDF and D\O\
experiments. The first one corresponds to pair production of
squarks, each decaying into a quark and a neutralino
($\tilde{q} \raa q \chi_1^0$), leading to a two jets+\met\
final state. This decay channel is dominant if the gluino is
heavier than the squark ($m_{\tilde{q}} \ll m_{\tilde{g}}$).
The second scenario applies when the squark is heavier than the
gluino, leading to a final state with 4 jets and \met\ from
$\tilde{g} \raa \tilde{q}^{*} \bar{q} \raa q \bar{q} \chi_1^0$.
The third one addresses similar squark and gluino masses, with
a final state of three or more jets arising from
$\tilde{q}\tilde{g}$ associated production.

Table~\ref{Tab:sqgl:cutflow1} from D\O\ illustrates the
selection criteria used for these searches. The data show good
agreement with {\tt SM} expectations after requiring dedicated
multijet+\met\ triggers and tight cuts on \met\ and the scalar
\pt\ sum ($\HT$). No signal is seen and cross section upper
limits at 95\% C.L. have been obtained for the sets of minimal
{\tt SUGRA} parameters considered ($\tan \beta = 5~(3)$, $A_0
=0~(-2m_0)$, $\mu <0$ for CDF (D\O)). The two Collaborations
show the results translated into the excluded regions in the
($m_{\tilde{g}},m_{\tilde{q}}$) and ($m_0,m_{1/2}$) planes.

\begin{figure}\begin{center}
\includegraphics[width=1.1\linewidth,height=0.43\textwidth,angle=0]{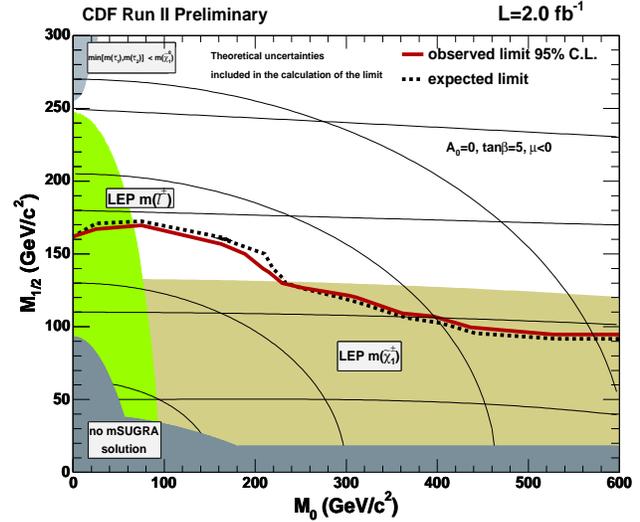}
\caption{\label{fig:CDF-SqGl} Regions excluded by the squark and gluino analyses at the 95\% C.L.
in the ($m_0,m_{1/2}$) plane, in the framework of {\tt mSUGRA} assuming {\it R}-parity conservation, using 2~\invfb\
of CDF Run II data \cite{CDF_conf_sqgl_2fb}. The regions excluded by LEP chargino and slepton searches are shown.
The nearly horizontal black lines are the iso-mass curves for gluinos corresponding to masses of
150, 300, 450 and 600~\GeV. The other lines are iso-mass curves for squarks, corresponding to
masses of 150, 300, 450 and 600~\GeV.}
\end{center}\end{figure}

\begin{table*}
\begin{minipage}{0.6\linewidth}
\begin{tabular}{cccc}
Preselection Cut & \multicolumn{3}{c}{All Analyses} \\
\hline
$\met$				        & \multicolumn{3}{c}{$\geq 40$}			\\
$|\mathrm{Vertex}\ z\ {\rm pos.}|$	& \multicolumn{3}{c}{$<60$ cm}			\\
Acoplanarity			        & \multicolumn{3}{c}{$< 165^\circ$}		\\
\hline
Selection Cut			& $\tilde{q}\tilde{q} \raa q \chi_1^0 q \chi_1^0$	& $\tilde{q}\tilde{g} \raa q \chi_1^0 q
\bar{q} \chi_1^0$	& $\tilde{g}\tilde{g} \raa q \bar{q}
\chi_1^0 q \bar{q} \chi_1^0$ 		\\
\hline
Trigger				& dijet		& multijet	& multijet		\\
${\rm jet}$ $p_T$     & $\geq 35$ 	& $\geq 35$ 	& $\geq 35$		\\
\hline
Electron veto			& yes		& yes		& yes			\\
Muon veto			& yes		& yes		& yes			\\
\hline
$\Delta\phi (\met,\mathrm{jet_1})$ 		& $\geq 90^\circ$ & $\geq 90^\circ$ & $\geq 90^\circ$ 	\\
$\Delta\phi (\met,\mathrm{jet_2})$ 		& $\geq 50^\circ$ & $\geq 50^\circ$ & $\geq 50^\circ$ 	\\
$\Delta\phi_{\rm{min}}(\met,\mathrm{any\,jet})$ & $\geq 40^\circ$	& $-$ & $-$				\\
\hline
$\HT$				& $\geq$ 325  & $\geq$ 375  & $\geq$ 400 		\\
$\met$				& $\geq$ 225  & $\geq$ 175  & $\geq$ 100 		\\
\hline
$N_{\mathrm{backgrd.}}$ & $11.1 \pm 1.2 ^{+2.9}_{-2.3}$ & $10.7 \pm 0.9 ^{+3.1}_{-2.1}$ & $17.7 \pm 1.1 ^{+5.5}_{-3.3}$    \\
$N_{\mathrm{sig.}}$     & $ 10.4 \pm 0.6 ^{+1.8}_{-1.8}$ & $ 12.0 \pm 0.7 ^{+2.5}_{-2.3}$ &  $ 17.0 \pm 1.2 ^{+3.3}_{-2.9}$  \\
$N_{\mathrm{obs.}}$     &  11 & 9 & 20\\
\hline
\end{tabular}
\end{minipage} \hfill
\begin{minipage}{0.33\linewidth}
\begin{flushleft}
\caption{\label{Tab:sqgl:cutflow1} Selection criteria for the
three squark and gluino analyses published by the D\O\
Collaboration~\cite{DO_pub_sqgl_2.1fb} with 2.1~\invfb\ of data
(all energies and momenta in \GeV). $|\mathrm{Vertex}\ z\ {\rm
pos.}|$ is the longitudinal position of the interaction
collision with respect to the detector center. The acoplanarity
is defined as the azimuthal angle between the two leading jets.
First and second (third and fourth) jets are also required to
be central $|\eta|<0.8$ ($|\eta|<2.5$) with $\pt\geq 35$~\GeV\
($\pt \geq 20$~\GeV\ for the fourth jet). The missing
transverse energy and scalar \pt\ sum are denoted \met\ and
\HT, respectively. The numbers of events observed and expected
from {\tt SM} backgrounds and from signal are given for each
analysis (the first uncertainty is statistical and the second
is systematic).}
\end{flushleft}
\end{minipage}
\end{table*}

The observed and expected mass limits derived for D\O\ using
$2.1~\invfb$ are given in Fig.~\ref{fig:D0-SqGl} as functions
of the squark and gluino masses, improving on previous
published limits. Lower limits at 95\% C.L. of 379~\GeV\ and
308~\GeV\ on the squark and gluino masses, respectively, are
derived in the most conservative hypothesis by D\O. The
corresponding expected limits are 377~\GeV\ and 312~\GeV. For
the particular case $m_{\tilde{q}} \simeq m_{\tilde{g}}$,
squark and gluino masses below 390~\GeV\ are excluded. The
observed limit becomes 408~\GeV\ for the {\tt NLO} nominal
signal cross section computed with the {\tt CTEQ6.1M} {\tt
PDF}~\cite{ref:cteq} and for the renormalization and
factorization scale $\mu_{r,f} = Q$, where $Q$ is taken to be
equal to $m_{\tilde{g}}$ for $\tilde{g}\tilde{g}$ production,
$m_{\tilde{q}}$ for $\tilde{q}\tilde{q}$, and
$(m_{\tilde{g}}+m_{\tilde{q}})/2$ for $\tilde{q}\tilde{g}$
production. The factor of two on the renormalization and
factorization scale reduces or increases the nominal signal
cross sections by 15-20\%. The {\tt PDF} and $\mu_{r,f}$
effects were added in quadrature to compute minimum and maximum
signal cross sections. If one considers the less conservative
scenario (maximum signal cross section), the observed lower
mass limit for $m_{\tilde{q}} \simeq m_{\tilde{g}}$ is
427~\GeV\ at 95\% C.L.

The CDF search with 2~\invfb\ excludes masses up to 392~\GeV\
in the region where gluino and squark masses are similar,
gluino masses up to 280~\GeV\ for every squark mass, and gluino
masses up to 423~\GeV\ for squark masses below 378~\GeV.
Figure~\ref{fig:CDF-SqGl} shows the results of this analysis
translated into the excluded regions in the {\tt mSUGRA}
($m_0,m_{1/2}$) plane. This search improves on the limit from
indirect LEP searches for $m_0$ values between 75 and 250~\GeV\
and for $m_{1/2}$ values between 130 and 170~\GeV. However, the
LEP Higgs search limits remain more constraining in a purely
{\tt mSUGRA} scenario~\cite{LEPSUSY_LSP}.

The D\O\ and CDF \ limits use slightly different model
parameters and methods to compute the excluded masses. Thus,
they are not directly comparable. However, it was verified that
similar results hold for a large class of parameter sets.

A complementary search for squarks has been performed by D\O\
in the topology of multijet events accompanied by large missing
transverse energy and at least one tau lepton decaying
hadronically using 1~\invfb\ of
data~\cite{DO_conf_sqgl_tau_1fb}. Lower limits on the squark
mass up to 366~\GeV\ are derived in the framework of {\tt
mSUGRA} with parameters enhancing final states with taus. This
analysis has the advantage of providing additional sensitivity
for squark searches, mainly at large values of $\tan \beta$.

\subsection{Stop and sbottom searches}

\begin{figure}
\begin{center}
\includegraphics[width=0.98\linewidth,height=0.48\textwidth,angle=0]{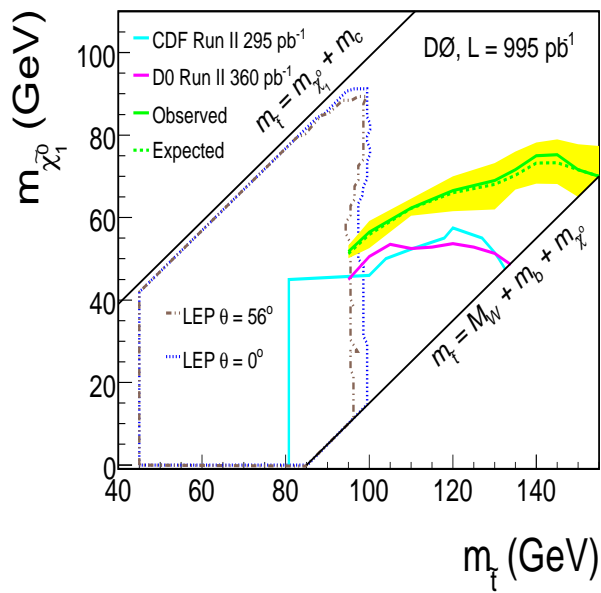}
\caption{\label{fig:D0_stop} D\O\ 95\% C.L. exclusion contours in the stop and neutralino mass plane,
assuming a stop branching ratio of 100\% into a charm quark and a neutralino. This search is based on 1~\invfb\
of data~\cite{DO_pub_stop_995pb}.}
\end{center}\end{figure}

For the third generation, mass unification is broken in many
{\tt SUSY} models due to potentially large mixing effects. This
can result in a sbottom or stop with much lower mass than the
other squarks and gluinos. In addition, the lightest stop quark
could well be the lightest of all quarks because of the impact
of the large top Yukawa coupling on the renormalization group
equations.

Dedicated searches are conducted in a general {\tt MSSM}
framework assuming the decays $\tilde{t}\to c\tilde{\chi}^0_1$
and $\tilde{b}\to b\tilde{\chi}^0_1$ are the only ones
kinematically allowed. Despite a much smaller cross section for
$\tilde{t}$ and $\tilde{b}$ production compared to previous
generic squark searches, heavy-flavor tagging can be used to
reduce the important {\tt SM} backgrounds. However, in the mass
range of interest, the jets are much softer compared to the
generic squark search, and therefore the {\tt QCD} multijet
background is much larger, reducing the stop and sbottom masses
which can be excluded.

\begin{figure}\begin{center}
\includegraphics[width=1.1\linewidth,height=0.53\textwidth,angle=0]{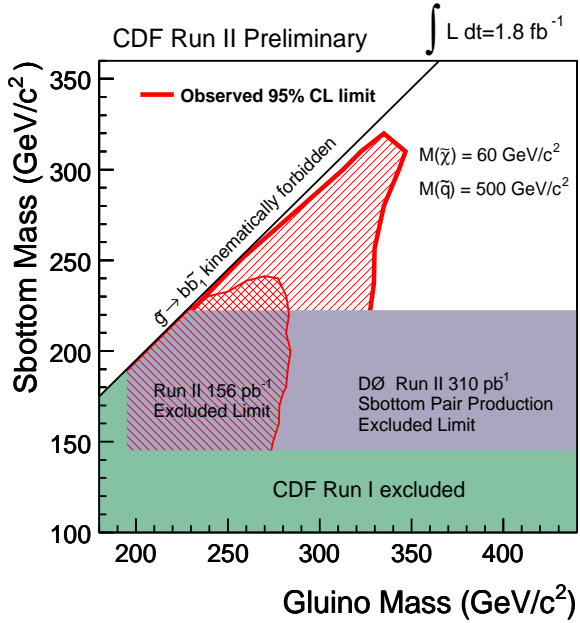}
\caption{\label{fig:CDF_sbottom_gluino_4b_1.8fb} CDF 95\% C.L. exclusion contours in the sbottom
and neutralino mass plane, assuming gluino pair production where the gluino decays to $b\tilde{b}$
with subsequent sbottom decay to a $b$-quark and the lightest neutralino.
This search is based on 1.8~\invfb\ of data~\cite{CDF_sbot_gluino_1.8fb}.}
\end{center}\end{figure}

D\O\ has recently submitted for publication an update using
1~\invfb\ of data~\cite{DO_pub_stop_995pb} compared to previous
D\O~\cite{DO_pub_stop_360pb} and
CDF~\cite{CDF_pub_stop_sbo_295pb} published results with about
300~\invpb\ of the case where the stop decays with a branching
ratio of 100\% into a charm quark and a neutralino. Good
agreement between the data and the {\tt SM} prediction is
obtained. The derived limits at 95\% C.L. on the stop mass are
shown in Fig.~\ref{fig:D0_stop}. With the theoretical
uncertainty on the $\tilde{t}$ pair production cross section
taken into account, the largest limit on $m_{\tilde{t}}$ is
150~\GeV, for $m_{\chi^0_1} = 65~\GeV$.

At large values of $\tan \beta$, the mixing can be enhanced in
the sbottom sector. The analysis of the decay channel
$\tilde{b}\to b\tilde{\chi}^0_1$ is similar to the one applied
for the stop except that higher masses can be excluded because
heavy-flavor tagging is more efficient for $b$-jets than for
$c$-jets. Supersymmetric bottom quark masses up to 193~\GeV\
for a neutralino mass of 40~\GeV\ are excluded by CDF with
295~\invpb\ of data~\cite{CDF_pub_stop_sbo_295pb}. For the D\O\
analysis~\cite{DO_pub_sbot_310pb} using 310~\invpb, the maximum
\MSB\ excluded is 222~\GeV, which is the most restrictive limit
on the sbottom mass to date from direct $\tilde{b}$ pair
production.

The CDF Collaboration also considered the scenario where the
sbottom could be produced through the decay of gluinos into
bottom and sbottom quarks, yielding a signature consisting of
four $b$-jets and two neutralinos from the sbottom decay
$\tilde{b}\to b\tilde{\chi}^0_1$. Requiring inclusive double
$b$-tagging, CDF observes 4 events where $2.6\pm0.7$ are
expected in 156~\invpb\ of Run~II
data~\cite{CDF_pub_sbot_gluino_156pb}. Exclusion lower limits
have been published on the masses of the gluino and sbottom up
to 280 and 240~\GeV, respectively. This result has been
recently updated with 1.8~\invfb\ of CDF
data~\cite{CDF_sbot_gluino_1.8fb}.  At least one $b$-tagged jet
was required and two different signal regions were optimized.
In the small (large) $\Delta m =
m_{\tilde{g}}-m_{\tilde{\chi}^0_1}$ region, 19 (25) events are
observed for $22.0\pm3.6$ ($22.7\pm4.6$) events expected from
{\tt SM} processes. The ($m_{\tilde{b}}$,$m_{\tilde{g}}$)
exclusion contour plot at 95\% C.L. is shown in
Fig.~\ref{fig:CDF_sbottom_gluino_4b_1.8fb}. A lower gluino mass
limit $m_{\tilde{g}} > 340$~\GeV\ is set for $m_{\tilde{b}} =
300$~\GeV, $m_{\tilde{\chi}^0_1} =60$~\GeV, and $m_{\tilde{q}}
= 500$~\GeV.

\begin{figure}\begin{center}
\includegraphics[width=1.0\linewidth,height=0.47\textwidth,angle=0]{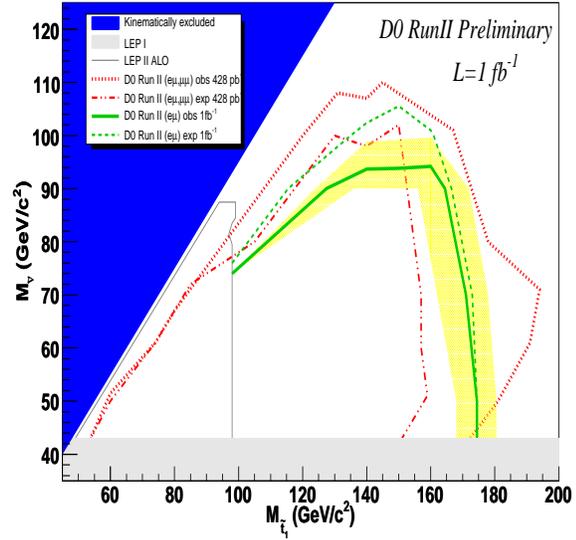}
\caption{\label{fig:D0_stop_blsnu} D\O\ 95\% C.L. exclusion contours in the stop and
sneutrino mass plane,
assuming $\tilde{t} \raa b \ell \tilde{\nu}$ and
$\tilde{\nu} \raa \nu \chi^0_1$ decay modes using 1~\invfb\ of data \cite{DO_conf_stop_blsnu_1.1fb}.}
\end{center}\end{figure}

The D\O\ Collaboration has searched for a light stop in the
lepton+jets channel using two scenarios. The first one uses the
stop decay modes $\tilde{t}_1 \raa t \chi_{1}^{0}$ $\raa$ $b
W^+ \chi_{1}^{0}$. In this analysis, kinematic differences
between stop pair production and the dominant $t\bar{t}$
process are used to separate the two possible contributions.
The preliminary results~\cite{D0_conf_stop_admixture_1fb} with
1~\invfb\ set upper cross section limits at 95\% C.L. on
$\tilde{t}_1\tilde{t}_1$ production that are a factor of about
7-12 higher than expected for the {\tt MSSM} model for stop
masses ranging between 145-175~\GeV. The second scenario
considers the pair production of the stop decaying into a
$b$-quark and the supersymmetric partner of the neutrino,{\it
i.e.}, the sneutrino ($\tilde{\nu}$). This decay $\tilde{t}
\raa b \ell \tilde{\nu}$ is then followed by the $\tilde{\nu}
\raa \nu \chi^0_1$ involving only invisible particles. The
ensuing final state consists of two leptons, two $b$-jets and
missing transverse energy. The result combines $e\mu$+jets and
$\mu\mu$+jets final states and limits are set in the plane
($m_{\tilde{t_1}},m_{\tilde{\nu}}$) as shown in
Fig.~\ref{fig:D0_stop_blsnu}. The published
result~\cite{DO_pub_stop_blsnu_400pb} with 428~\invpb\ of data
has been recently updated~\cite{DO_conf_stop_blsnu_1.1fb} with
1.1~\invfb\ and there is now good agreement between the
expected and observed limits.

\begin{figure}\begin{center}
\includegraphics[width=0.98\linewidth,height=0.40\textwidth,angle=0]{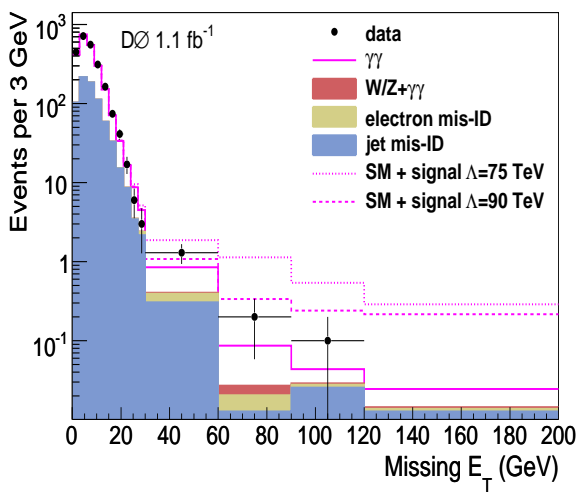}
\caption{\label{fig:D0_GMSB_MET} D\O\ \met\ distribution in 1.1 \invfb\ of $\gamma\gamma$ data,
along with expected background processes \cite{DO_GMSB_pub_1.1fb}.
The expected \met\ distribution for {\tt GMSB} {\tt SUSY} signal
with $\Lambda = 75~\TeV$ and 90~\TeV\ are presented as dotted and dashed lines, respectively.}
\end{center}\end{figure}

\section{Gauge mediated SUSY breaking}
In gauge mediated {\tt SUSY} breaking models ({\tt GMSB}), the
gravitino, with a mass less than few \keV, is the {\tt LSP}.
The phenomenology of these models is therefore determined by
the nature and the lifetime of the next-to-lightest
supersymmetric particle ({\tt NLSP}), which can be either a
neutralino or the lightest stau, depending on the choice of
model parameters~\cite{theo:SUSY_ref1}.

\subsection{Long-lived final state}

It is possible for a stau {\tt NLSP} in these models to be
long-lived~\cite{ref:stau_GMSB_long_live}. Stau pair production
has been searched for by D\O\ using 390~\invpb\ of
data~\cite{DO_conf_champs_stau_AMSB_396pb}. These long-lived
particles loose energy principally by ionization and can
traverse the entire detector, registering in the muon
detectors. The search is not yet sensitive enough to set a stau
lower limit. In anomaly mediated supersymmetry breaking ({\tt
AMSB}), or in models that do not have gaugino mass unification,
the signature is the same for long-lived charginos that escape
the detector~\cite{ref:long_lived_susy_signature}. The larger
production cross section allows a preliminary lower limit of
140~\GeV\ on higgsino-like charginos and 174~\GeV\ on
gaugino-like charginos, both at the 95\% C.L.

\subsection{Diphoton final state}

Final states with two photons and \met\ can be produced in {\tt
GMSB} models. In such a scenario, the lightest neutralino
($\chi_1^0$) decays into a photon and a weakly interacting
stable gravitino ($\tilde{G}$). Most of the searches assume the
prompt decay $\chi_1^0 \to \gamma \tilde{G}$. The CDF
Collaboration, however, also searched in 570~\invpb\ of data
for non-prompt decays and a $\chi_1^0$ with a lifetime that is
on the order of nanoseconds or
more~\cite{CDF_long_lived_photon_570pb}. Two candidate events,
consistent with the background estimate of 1.3$\pm$0.7 events,
are selected based on the arrival time of the photon at the
calorimeter. This result allows for setting both
quasi-model-independent cross section limits and for an
exclusion region of {\tt GMSB} models in the $\chi_1^0$
lifetime versus mass plane ($\tau_{\chi_1^0}$,$m_{\chi_1^0}$),
with a mass reach of 101~\GeV\ for $\tau_{\chi_1^0}$ = 5~ns.

As for the prompt decay $\chi_1^0 \to \gamma \tilde{G}$, CDF
\cite{CDF_GMSB_202pb} and D\O\ \cite{DO_GMSB_263pb} combined
their published results based on 200-260~\invpb\ of data. The
combined limit~\cite{Tevatron_GMSB_260pb} excludes a chargino
mass of less than 209~\GeV, for {\tt GMSB} parameters following
the ``Snowmass benchmark scenario''~\cite{ref:snowmass}:
messenger mass $m_M = 2\Lambda$ where $\Lambda$ is the
effective {\tt SUSY}-breaking scale, $\tan \beta = 15$,
$\mu>0$, and the number of messenger fields $N_M = 1$. The D\O\
Collaboration recently published an update using 1.1~\invfb\ of
data~\cite{DO_GMSB_pub_1.1fb}. The \met\ distribution for the
$\gamma\gamma$ sample is given in Fig.~\ref{fig:D0_GMSB_MET}
with the expected signal contribution for two different values
of the effective energy scale~$\Lambda$. After determination of
all backgrounds from data, D\O\ observes no excess of such
events and thus sets 95\% C.L. limits: the masses of the
lightest chargino and neutralino are found to be larger than
229~\GeV\ and 125~\GeV, respectively. These results represent
the most stringent limits to date on this particular {\tt GMSB
SUSY} model.

\begin{figure}\begin{center}
\includegraphics[width=0.98\linewidth,height=0.40\textwidth,angle=0]{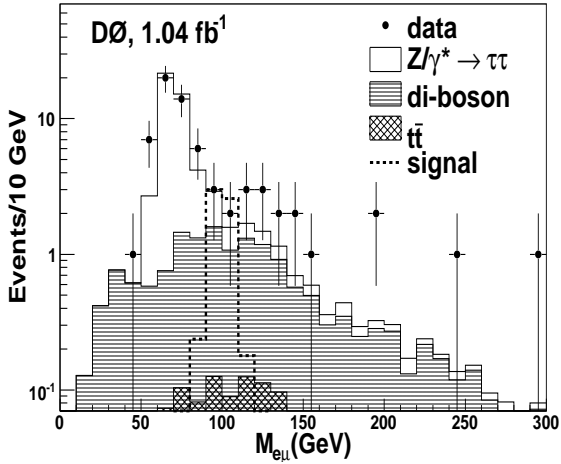}
\caption{\label{fig:D0_sneutrino_tau} Invariant mass of the electron-muon system in
1~\invfb\ of data collected by the D\O\ experiment~\cite{D0_sneutrino_emu_1fb}. The search is performed
in the context of {\it R}-parity-violating production and decay.
The dashed line indicates the signal hypothesis for a third-generation sneutrino
($\tilde{\nu}_\tau$) with a mass of 100~\GeV\ and $\sigma\times Br$ of 0.057~\pb.}
\end{center}\end{figure}

\section{{\it R}-parity violation}\label{sec:R-parity}
The {\tt MSSM} superpotential is minimal in the sense that it
is sufficient to produce a phenomenologically viable model.
However, the most general gauge-invariant and renormalizable
superpotential would include additional terms like
$$W_{\hbox{RPV}} = \lambda_{ijk} L_i L_j \bar{E}_k +
\lambda_{ijk}^{'} L_i Q_j \bar{D}_k + \lambda_{ijk}^{''}
\bar{U}_i \bar{D}_j \bar{D}_k,  $$ where $i = 1,2,3$ are the
family indices. The lepton and quark doublet superfields (weak
isospin singlet superfields) are denoted $L$ and $Q$ ($E,U$ and
$D$), respectively. These terms violate both baryon number (B)
and lepton number (L), which is in contradiction with
experimental observations. The most obvious experimental
constraint comes from the non-observation of proton decay,
which would violate both B and L by 1 unit. Therefore, these
new couplings in the trilinear terms, if present in nature,
must be extremely small. In the {\tt MSSM}, a new symmetry is
thus introduced to eliminate the possibility of B and L
violating terms in the superpotential. This new symmetry,
called {\it R}-parity~\cite{theo:SUSY_ref1}, is a
multiplicatively conserved quantum number defined as
$P=(-1)^{3(B-L)}$, which takes a value of +1 for {\tt SM}
particles and -1 for {\tt SUSY} particles.

The CDF and D\O\ Collaborations have considered a number of
scenarios under the hypothesis that {\it R}-parity violation
({\tt RPV}) can occur. The experimental consequences are
characterized by less missing transverse energy and more
leptons and jets in the final states, due to the decay of the
{\tt LSP} into {\tt SM} particles. In addition, sparticles may
be resonantly produced by {\tt RPV} couplings as single
sparticles, by virtue of which the {\tt LSP} cannot be a
candidate for dark matter.

Searches for gaugino pair production via $\lambda_{121}$ and
$\lambda_{122}$ with a signature of at least four charged
leptons and two neutrinos are published by
CDF~\cite{CDF_RPV_pub_346pb} (D\O~\cite{DO_RPV_pub_360pb})
using 346~\invpb\ (360~\invpb) of data. The case in which a
$\tau$ appears in the final state via $\lambda_{133}$ coupling
has been included in the D\O\ analysis. Using the {\tt mSUGRA}
model with $m_0=1~\TeV$, $\tan \beta=5$, and $\mu>0$, D\O\
obtains 95\% C.L. lower limits on the $\chi_1^0$ ($\chi_1^{\pm}$)
masses of 119, 118, 86 \GeV\ (231, 229, 166 \GeV) for the
corresponding $\lambda_{121}$, $\lambda_{122}$, and
$\lambda_{133}$ couplings, respectively. For the CDF analysis,
the $\chi_1^0$ mass limits range from 98 to 110~\GeV, while the
chargino mass limits range from 185 to 203~\GeV\ at 95\% C.L.,
depending on the choice of model parameters.

A search for pair production of scalar top quarks decaying via
the {\it R}-parity-violating $\lambda_{333}^{'}$ coupling to a
$\tau$ lepton and a $b$ quark has been presented by the CDF
Collaboration based on 322~\invpb\ of
data~\cite{CDF_stop_RPV_LQ3}. A lower mass limit
$m_{\tilde{t}_1}>151~\GeV$ has been set in the final state of
either an electron or a muon from the first $\tau$ decay, a
hadronic decay for the second $\tau$, and two or more jets.

Resonant slepton production has also been probed at the
Tevatron. A single slepton could be produced in hadron
collisions by $LQD$ interactions followed by decays into {\tt
SM} di-lepton final states via $LLE$ interactions, leading to a
high-mass di-lepton resonance. The CDF
\cite{D0_sneutrino_emu_344fb} and D\O\
\cite{D0_sneutrino_emu_1fb} Collaborations have reported a
search for resonant production of sneutrinos decaying into an
electron and a muon using 344~\invpb\ and 1~\invfb\ of data,
respectively. The invariant mass of the electron-muon system in
the D\O\ search is shown in Fig.~\ref{fig:D0_sneutrino_tau}.
For a sneutrino with mass of 100~\GeV, $\lambda_{311}^{'}>
1.6\times10^{-3}$ is excluded by D\O\ at 95\% C.L. when the
$\lambda_{312}$ coupling constant is fixed at 0.01. CDF
excludes $\lambda_{311}^{'}$ values above 0.01 for a
$\tilde{\nu}_\tau$ mass of 300~\GeV\ and $\lambda_{132}>0.02$.
In addition, CDF published sneutrino mass limits  depending on
the $\lambda^{'}$ and $\lambda$ couplings considered in the
other $ee$, $\mu\mu$, $\tau\tau$ final states based on
$\approx200$~\invpb~\cite{CDF_sneutrino_ee_mm_200pb,CDF_sneutrino_tautau_195pb}.

Results from D\O\ have been published~\cite{D0_slepton_380pb}
based on 380~\invpb\ on the production and decay of resonant
smuons and muon-sneutrinos in the channels
$\tilde{\mu}\to\tilde{\chi}^0_1\,\mu$,
$\tilde{\mu}\to\tilde{\chi}^0_{2,3,4}\,\mu$, and
$\tilde{\nu}_\mu\to\tilde{\chi}^\pm_{1,2}\,\mu$. A lower limit
on the slepton mass $m_{\tilde{l}} \le 363$~\GeV\ is set for
$\lambda_{211}^{'}\ge0.10$, $\tan\beta=5$, $A_0=0$ and $\mu<0$.

\begin{figure}\begin{center}
\includegraphics[width=.5\textwidth,height=0.40\textwidth,angle=0]{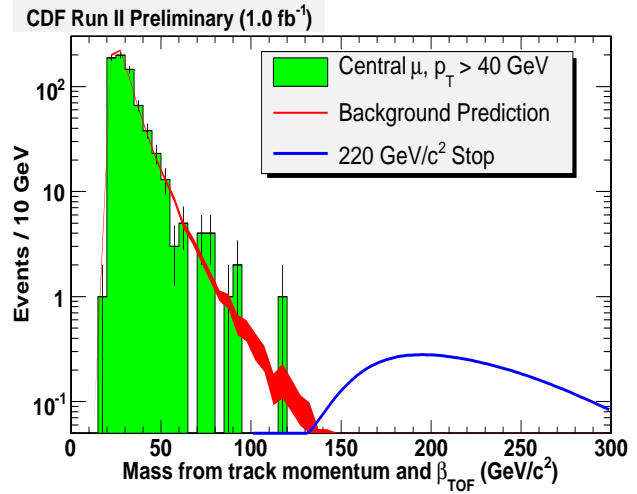}
\caption{\label{fig:D0_CHAMP_stop} Mass distribution measured in 1 \invfb\ of data by the
time-flight flight and transverse momentum of tracks in events collected
by the CDF experiment using the high transverse momentum muon trigger.
The expected contribution from stable stop pair production
is shown for a stop mass of 220~\GeV~\cite{CDF_CHAMP_1fb}.}
\end{center}\end{figure}

\section{Long-lived particles}
Although cosmological considerations put strict limits on new
particles that are absolutely stable, these restrictions do not
apply to particles that live long enough to decay outside the
detector~\cite{ref:cosmo_long_lived}. Several models, outlined
below, predict charged or neutral long-lived particles decaying
inside or outside the detector.

\subsection{Neutral long-lived particles}
The existence of neutral long-lived particles decaying into two
leptons that arise from a highly displaced vertex is expected
in ``hidden valley'' theories~\cite{ref:hidden_valey} or {\tt
SUSY} models with {\it R}-parity
violation~\cite{theo:SUSY_ref1}. Motivated by the excess of
di-muon events observed by the Fermilab neutrino experiment
NuTeV~\cite{ref:NuTeV}, D\O\ has
published~\cite{D0_Long_Lived_RPV_mumu_380pb} a search based on
380~\invpb\ of data assuming a benchmark model where the
$\chi^0_1$ has traveled at least 5~cm and decays via {\tt RPV}
to $\mu^+\mu^-\nu$. The background has been estimated to be
about one event and, since no candidates were observed, a 95\%
C.L. upper cross section limit of 0.14~\pb\ is set on
pair-production of neutral long-lived particles with a mass of
10~\GeV\ and a lifetime of $4\times10^{-11}$.

\subsection{Charged long-lived particles}
If a long-lived particle has a large mass and is charged
(CHAMP) \cite{ref:CHAMP}, it will appear in the detector as a
slowly moving, highly ionizing particle with large transverse
momentum that can be observed in the muon detectors. CDF has
performed a model independent search by measuring the
time-of-flight of particles from muon triggers in 1~\invfb\ of
data~\cite{CDF_CHAMP_1fb}. As shown in
Fig.~\ref{fig:D0_CHAMP_stop}, the result is consistent with
muon background expectation. Within the context of stable stop
pair production, CDF infers an upper mass limit of 250~\GeV\ at
95\%~C.L.

The introduction of a fourth generation quark $b^{'}$ provides
another possibility for such long-lived
particles~\cite{ref:bprim}. The CDF Collaboration
reported~\cite{CDF_bprim_163pb} a result, based on 193 \invpb\
of data, for such particles using $Z$ boson decays to muons and
reconstructing di-muon vertices in the tracker. D\O\ also
performed a $b^{'}$ search, using the capability of its
detector to reconstruct the direction of electromagnetic
showers, and thus enhancing its sensitivity to long-lived
particles. Although limits have been set within the framework
of the $b^{'}$ model, loose requirements are imposed to limit
potentially model-dependent selection. No evidence of such
excess is found in 1~\invfb\ of data by
D\O~\cite{DO_bprim_1fb}. Limits are set on the production cross
section and lifetime of such long-lived particles that decay
into a $Z$ boson or any final state with a pair of electrons or
photons with mass above 75~\GeV\ at 95\% C.L.

In a variant of {\tt SUSY} known as split
supersymmetry~\cite{ref:split-susy}, gluino decays into squarks
and the neutralino {\tt LSP} are suppressed, leading to a long
lived gluino. At the Tevatron, such colorless bound states
(R-hadrons) of a gluino and other quarks or gluons could be
pair produced through strong interactions. As studied in
Ref.~\cite{ref:stopped-gluino}, some charged R-hadrons have the
potential to become ``stopped gluinos'', by losing all of their
momentum through ionization and come to rest in the
calorimeter. No excess is observed above the primary source of
background coming from cosmic muons in 410~\invpb\ of D\O\
data~\cite{DO_pub_split_susy_410pb}. Their main decay mode is
expected to be $\tilde{g} \raa g \chi_1^0$ with a lifetime
assumed to be long enough such that the decay of the gluino
occurs during a bunch-crossing adequately later than the one
which has produced it (about $30~\mu$s). Limits are therefore
placed on the gluino cross section times the stopping
probability as a function of the gluino and $\chi_1^0$ masses,
for gluino lifetimes from $30~\mu$s to 100~hours. This analysis
excludes $m_{\tilde{g}}<270~\GeV$ for a $\chi_1^0$ mass of
50~\GeV, assuming a 100\% branching fraction for $\tilde{g}
\raa g \chi_1^0$, a gluino lifetime less than 3 hours, and a
neutral to charged R-hadron conversion cross section of 3~\mb.

\section{Leptoquarks}
\begin{figure}\begin{center}
    \includegraphics[width=.23\textwidth,height=0.20\textwidth,angle=0]{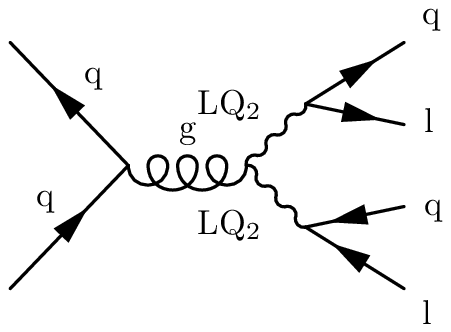}
    \includegraphics[width=.23\textwidth,height=0.20\textwidth,angle=0]{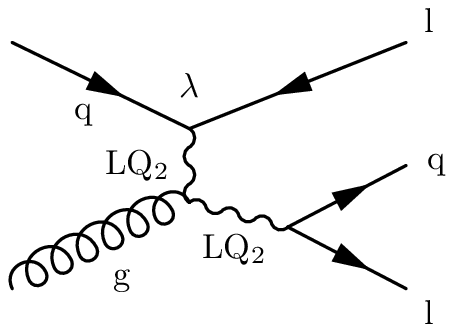}
\caption{\label{fig:LQ} Examples of leading-order Feynman graphs for pair-production (left)
    and single production (right) of leptoquarks.}
\end{center}\end{figure}

Leptoquarks ($LQ$) are colored bosons that were postulated to
explain the parallels between the families of quarks and
leptons~\cite{ref:LQ}. They are predicted in many extensions of
the standard model, such as SU(5) grand
unification~\cite{ref:su5}, superstring~\cite{ref:superstring},
and compositeness models~\cite{ref:compositness}.
Figure~\ref{fig:LQ} shows mechanisms for leptoquark production
and decay in $p\bar{p}$ collisions, where leptoquarks can be
pair produced via the strong interaction. Single leptoquark
production can also occur in association with a lepton.

\subsection{LQ pair production}
At the Tevatron, $LQ$ states would be predominately pair
produced with larger cross sections predicted for vector (spin
1) than for scalar (spin 0) leptoquarks. They are expected to
decay into a quark and a charged lepton with a branching
fraction $\beta$, or into a quark and a neutrino with a
branching fraction ($1-\beta$). Experimental limits on lepton
number violation, on flavor-changing neutral currents, and on
proton decay motivate the assumption that there would be three
different generations of leptoquarks, where each leptoquark
generation couples to only one generation of quarks and
leptons.

Results have been published by CDF for first generation scalar
leptoquarks ($LQ_1$) using about 200 \invpb\ of data. No
evidence is observed of such particles in the topologies
arising from $LQ_1 \overline{LQ}_1 \raa$ $eqeq$, $LQ_1
\overline{LQ}_1 \raa$ $eq\nu q$, and $LQ_1 \overline{LQ}_1
\raa$ $q\nu q\nu$ \cite{CDF_LQ1_203pb,CDF_LQ1_2_qvqv_191pb}.
Lower mass limits are derived: 236, 205 and 145~\GeV\ for
$\beta = 1$, $\beta = 0.5$ and $\beta = 0.1$, respectively. The
$LQ_1$ mass limits which are published in the $eqeq$ and $eq\nu
q$ final states by D\O\ with 252~\invpb\ of data are 256 and
234~\GeV, for $\beta=1$ and 0.5,
respectively~\cite{D0_LQ1_252pb}. In $LQ_1\overline{LQ}_1\raa
q\nu q\nu$, D\O\ has published \cite{D0_LQ1_nunuqq_310} a
result with 310~\invpb\ and a lower mass limit of 136~\GeV\ is
set at the 95\% C.L. Recently, CDF has released a search based
on 2~\invfb\ of dijet+\met\ data~\cite{CDF_LQ_nunuqq_2fb}. Two
separate analyses are performed. The first one requires two
jets with $\pt>30$~\GeV, no third jet with $\pt>15$~\GeV,
$\met>80$~\GeV, and $\HT>125$~\GeV. The second one search in
the high kinematic region defined by $\HT>225$~\GeV\ and $\met>
100$~\GeV. In both regions, CDF compares the expected {\tt SM}
backgrounds with data and no excess is observed. A scalar
leptoquark model is used to place a limit of
$m_{\overline{LQ}_1}> 177$~\GeV, for $\beta = 0$ at 95\% C.L.

Second generation scalar leptoquarks ($LQ_2$) have also been
searched for at the Tevatron. The CDF Collaboration has
published a result in the dimuons+jets and muon+missing
energy+jets topologies using 198 \invpb\ of data
\cite{CDF_LQ2_198pb}. Combining the results with those from the
\met+jets channel topology~\cite{CDF_LQ1_2_qvqv_191pb}, CDF
excludes $LQ_2$ with masses below 226~\GeV\ for $\beta=1$,
208~\GeV\ for $\beta = 0.5$, and 143~\GeV\ for $\beta = 0.1$ at
95\% C.L. The D\O\ Collaboration has published limits in the
channel $LQ_2\overline{LQ}_2\raa$ $\mu q \mu q$ using an
integrated luminosity of 294~\invpb~\cite{D0_LQ2_290pb}. In
combination with previous D\O\ measurements, lower mass limits
of $m_{LQ_2}> 251~\GeV$ for $\beta =1$ and $m_{LQ_2}> 204~\GeV$
for $\beta=0.5$ are set. The first D\O\ search performed in
Run~II in the channel $LQ_2\overline{LQ}_2\raa \mu q \nu q$,
which has maximal sensitivity for $\beta=0.5$, is based on
1~\invfb\ of data~\cite{D0_LQ2_muq_nuq}. From this analysis
alone, a lower mass limit for scalar second generation
leptoquarks of $m_{LQ_2}>214~\GeV$ at $\beta=0.5$ is set at
95\% C.L. Using 2~\invfb\ of dijet+\met\ data, CDF excludes
$LQ_2$ masses below 177~\GeV\ at 95\%
C.L~\cite{CDF_LQ_nunuqq_2fb}.

A search for third generation scalar $LQ_3$ pair production has
been performed in the $\tau b \tau b$ channel using 1~\invfb\
of data collected at D\O~\cite{D0_LQ3_tau_b_tau_b_1fb}. To
increase the search sensitivity, advantage is taken of the
presence of heavy-flavor jets in the signal. No evidence of
signal has been observed, and limits are set on the production
cross section as a function of the leptoquark mass. Assuming
$\beta$, the branching fraction of the leptoquark into $\tau
b$, equal to 1, the limit on the mass is 180~\GeV\ at 95\% C.L.
With a smaller dataset of 0.4~\invfb, assuming a decay into
$b\nu$, the limit is 229~\GeV~\cite{D0_LQ3_bnubnu_400pb}. If
leptoquark decays into a $\tau$ lepton and a top quark are
taken into account, and if equal couplings are assumed, a mass
limit of $m_{LQ_3}>221~\GeV$ is set by D\O\ at 95\%
C.L.~\cite{D0_LQ3_bnubnu_400pb}. The CDF Collaboration has
performed a similar analysis with 322~\invpb\ of data but in
the context of vector leptoquarks ($VLQ_3$). Assuming
Yang-Mills (minimal) couplings, CDF obtaines the most stringent
upper limit on the $VLQ_3$ pair production cross section of
344~fb (493~fb) and lower limit on the $VLQ_3$ mass of
317~\GeV\ (251~\GeV) at 95\%~C.L in $\tau b$
decay~\cite{CDF_VLQ3_btaubtau_322pb}. Finally, a mass limit of
$m_{\overline{LQ}_3}> 167$~\GeV\ is set for third generation of
scalar leptoquark using 2~\invfb\ of CDF dijet+\met\
data~\cite{CDF_LQ_nunuqq_2fb}. In this case, the efficiency for
third generation events to pass a dijet plus missing \ET\
selection criteria is smaller due to lepton rejection
requirements, and therefore the mass limits set are lower than
those for the first and second generation.

\begin{figure}\begin{center}
    \includegraphics[width=.55\textwidth,height=0.47\textwidth,angle=0]{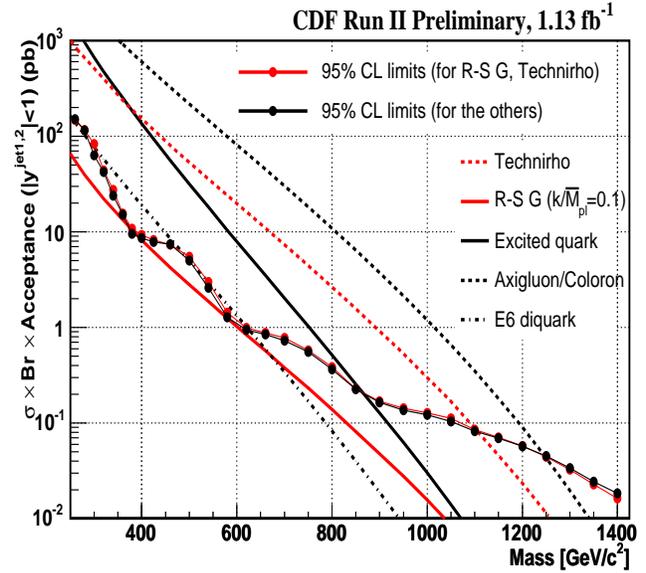}
\caption{\label{fig:CDF_jj_1.13fb} CDF 95\% C.L. limits based on 1.1~\invfb\ of
data~\cite{D0_jj_reso_1.13fb} on the Randall-Sundrum graviton ($G\raa q\bar{q},gg$~\cite{ref:ED}),
color-octet techni-$rho$ production ($\rho_T \raa q\bar{q},gg$~\cite{ref:techni_rho}), excited quark ($q^{*} \raa
qg$~\cite{ref:excited_quark}),
axigluon and flavor-universal coloron ($A\raa q\bar{q}$~\cite{ref:axigluon}), and $E_6$ di-quark ($D(D^c)\raa (qq)\bar{q}\bar{q}$~\cite{ref:superstring}), compared with
the theoretical predictions for production of these particles.}
\end{center}\end{figure}

\subsection{Single LQ production}
The production of single leptoquarks leads to final states
consisting of two leptons and one jet. The D\O\ Collaboration
has published a search in the $\mu\mu j$ final state using
300~\invpb\ of data~\cite{D0_single_LQ_300pb}. Compared to the
search for leptoquarks which considered only pair-production,
the mass limits are improved to $m_{LQ}>274~\GeV$ for $\beta
=1$ and $\lambda = 1$, where $\lambda$ is the
leptoquark-lepton-quark coupling. For $\beta =0.5$, a lower
limit on the mass of a second generation scalar leptoquark
$m_{LQ}$ $>226~\GeV$ is set at 95\% C.L.

\section{Compositeness}
In the {\tt SM}, the quarks and leptons are treated as
fundamental particles. However, one proposed explanation for
the three generations is a compositeness model of the known
leptons and quarks~\cite{ref:composit1}.

\subsection{Excited lepton}
Compositeness models comprise a large spectrum of excited
states. The CDF \ and \ D\O\ Collaborations have searched for
excited electron ($e^*$) in the process $p\bar{p} \raa e^* e$,
with the $e^*$ subsequently decaying to an electron plus
photon. The agreement observed by D\O\ in 1 \invfb\
\cite{D0_estar_1fb} and CDF in 202~\invpb\
\cite{CDF_estar_200pb} of data with the {\tt SM} backgrounds
are interpreted in the context of a model that describes
production by four-fermion contact interactions (CI) and
excited electron decay via electroweak processes. Choosing the
scale for CI to be $\Lambda = 1~\TeV$, $e^*$ masses below
756~\GeV\ are excluded at 95\% C.L. by the D\O\ analysis. To
make a comparison with LEP results, CDF also reinterprets its
search in the gauge-mediated model and excludes $126~\GeV <
m_{e^*}$ $< 430$ \GeV\ at 95\% C.L. for the phenomenological
coupling $f/\Lambda \approx 10^{-2}~\GeV$
\cite{CDF_estar_200pb}.

Similarly, searches for excited muons ($\mu^*$) subsequently
decaying to a muon plus photon have been carried out in a data
sample corresponding to an integrated luminosity of 371 \invpb\
for CDF \cite{CDF_mustar_371pb} and 380 \invpb\ for D\O\
\cite{D0_mustar_380pb}. CDF excludes in the contact interaction
model $107~\GeV < m_{\mu^*} < 853$ \GeV\ for $\Lambda =
m_{\mu^*}$ and in the gauge-mediated model $100~\GeV <
m_{\mu^*}< 410~\GeV$ for $f/\Lambda \approx 10^{-2}~\GeV$ at
95\% C.L. Choosing the scale for contact interactions to be
$\Lambda = 1~\TeV$, masses below 618~\GeV\ are excluded by the
D\O\ search.

\subsection{Excited quark}
The D\O\ Collaboration has published a search within the
framework of a quark substructure model
\cite{ref:excited_quark}. In 370 \invpb\ of data, no indication
for resonances in the $Z$+jet channel has been observed, where
the $Z$ boson is detected via its $Z \raa e^+e^-$ decay mode.
This analysis leads to a mass limit $m_{q^*} > 510~\GeV$ at
95\% C.L., assuming the decay mode $q^* \raa q + Z$ for the
excited quark~\cite{D0_qstar_370pb}. The CDF preliminary result
based on 1.1~\invfb\ of data excludes the mass region $260~\GeV
< m_{q^{*}} < 870~\GeV$ at 95\%~C.L. assuming the decay mode
$q^{*} \raa qg$~\cite{D0_jj_reso_1.13fb}. The obtained limits
are shown in Fig.~\ref{fig:CDF_jj_1.13fb} for various models
corresponding to the production of new particles that decay
into dijet.

In the same vein, CDF searched for new particles that lead to a
$Z$ boson plus jets but, this time, in the context of a fourth
generation model~\cite{ref:bprim}. In a data sample of
1~\invfb, the $Z$ boson decays to $ee$ and $\mu\mu$ are used to
set a lower limit on $b^{'}$ quark masses below 268~\GeV\ at
95\% C.L., assuming the decay mode $b^{'} \raa b +
Z$~\cite{CDF_bprim_resonance_1fb}.

In 1.9~\invfb\ of lepton+jet data, CDF also investigated the
existence of a massive gluon and set limits on the coupling
strength of this particle as function of its
mass~\cite{CDF_conf_tt_reso_1.9fb}. A search for the heavy top
($t^{'}$) quark pair production decaying to $Wq$ final states
in 2.3~\invfb\ of CDF lepton+jets data excludes a
fourth-generation $t^{'}$ quark with a mass below 284~\GeV\ at
95\% C.L~\cite{CDF_tprim_2.3fb}.

\begin{figure}\begin{center}
\includegraphics[width=1.0\linewidth,height=0.47\textwidth,angle=0]{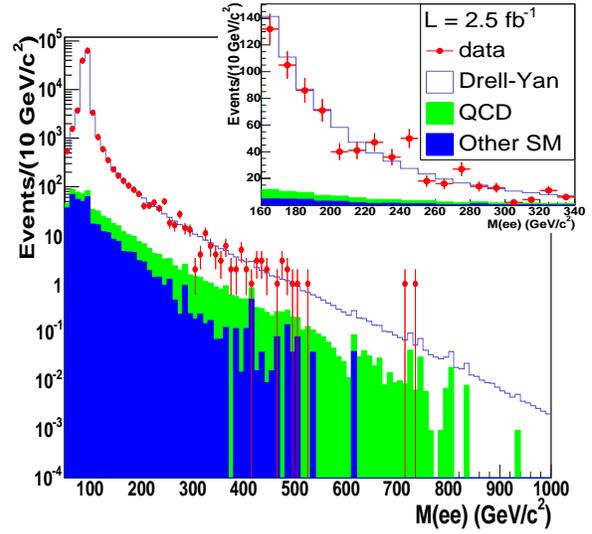}
\caption{\label{fig:CDF-Zprim} The dielectron mass measured by CDF and the expected background
in 2.5~\invfb\ of data~\cite{CDF_conf_reso_ee_2.5fb}. A slight excess in the data is observed
in the region $m_{e^+e^-} \approx 240~\GeV$.}
\end{center}\end{figure}

\section{Extra gauge bosons}
Multiple extensions of the {\tt SM} predict extra gauge bosons.
For instance, $Z^{'}$ are predicted in $E_6$ {\tt GUT}s models
\cite{ref:ExtaBoson}, and $W^{'}$ bosons appear in models such
as left-right-symmetric theories \cite{ref:Left-Right-Sym}. The
new gauge group can comprise a new mixing angle and new
couplings depending on the models considered.

\subsection{$Z^{'}$ bosons}
\begin{figure}\begin{center}
\includegraphics[width=1.0\linewidth,height=0.4\textwidth,angle=0]{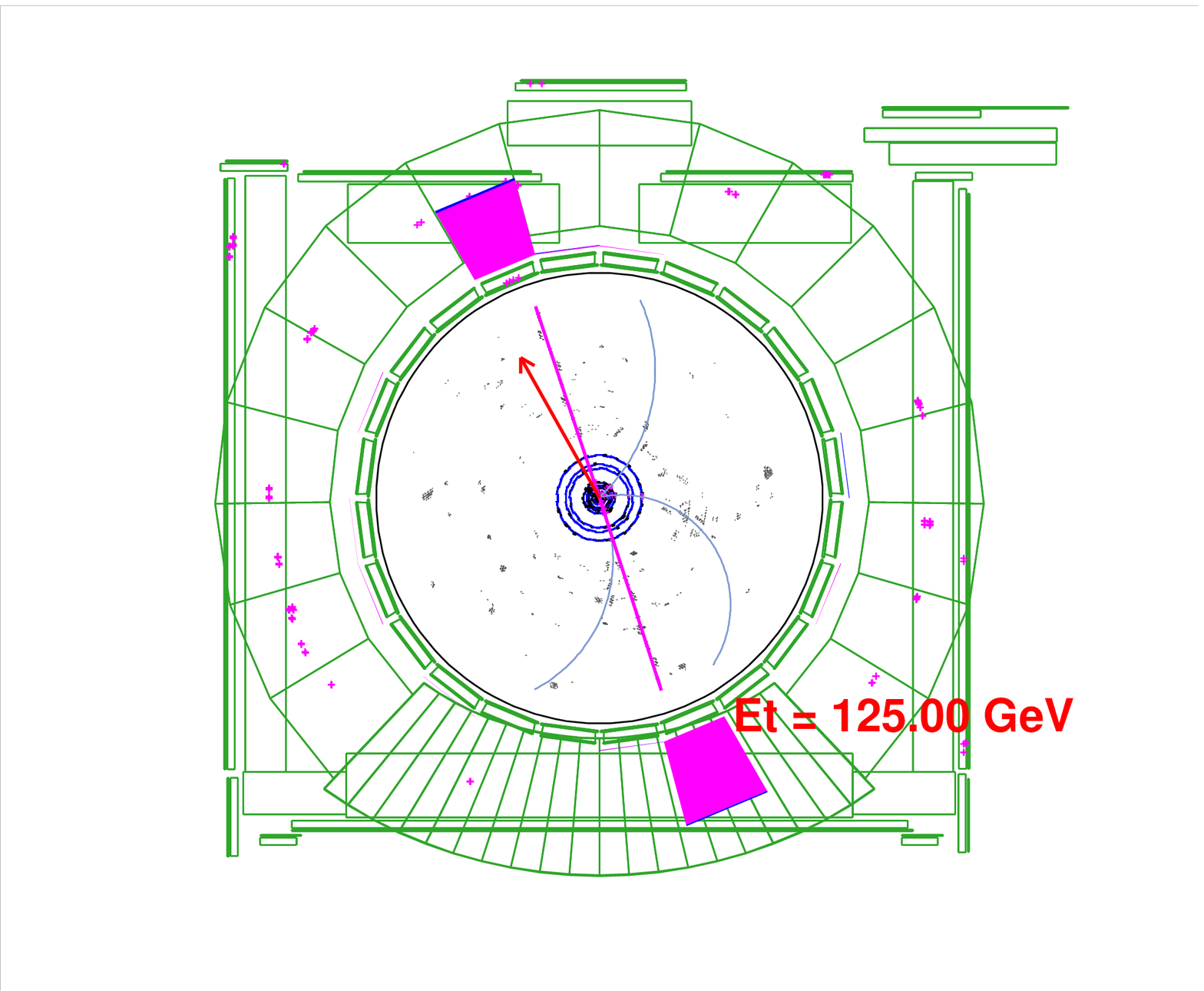}
\caption{\label{fig:CDF-diem_display} Event display of a typical dielectron event,
as measured by CDF in Run II~\cite{CDF_conf_reso_ee_2.5fb}.}
\end{center}\end{figure}

The CDF Collaboration has recently released a new preliminary
result in the search for dielectron resonances using
2.5~\invfb\ of data~\cite{CDF_conf_reso_ee_2.5fb}. The previous
CDF searches have been published with 0.2-1.3 \invfb\ of data
\cite{CDF_sneutrino_ee_mm_200pb,CDF_pub_reso_ee_Zprim_0.45fb,CDF_pub_reso_ee_Z_RS_1.3fb}.
Resonance with dilepton in the final states have always been
leading channels for early discovery searches due to low
backgrounds. In addition, lepton energy and momentum can be
measured precisely by combining calorimeter and tracking
information. The searches are performed by reconstructing the
dielectron mass, as shown in Fig.~\ref{fig:CDF-Zprim}. The $Z$
mass peak and the Drell-Yan tail at high mass is well
reproduced by the {\tt SM} background prediction. However, in
the region $m_{e^+e^-} \approx 240~\GeV$, an excess of data
over background of $3.8\sigma$ is observed, with a 0.6\%
probability that it is caused by the background fluctuation,
given that the search probes the mass range 150~\GeV-1~\TeV. A
typical di-electron event is displayed in
Fig.~\ref{fig:CDF-diem_display}.

By performing a scan for high-mass resonances, CDF sets limits
depending on the model considered. For instance, a lower mass
limit of 966~\GeV\ can be set assuming {\tt SM}-like couplings
of the $Z^{'}$, with a somewhat lower mass limit for $E_6$
$Z^{'}$ bosons with masses below 737/933 \GeV\
(lightest/heaviest) \GeV\ excluded at 95\%~C.L. The D\O\
Collaboration has presented a preliminary
results~\cite{D0_reso_ee_Zprim_0.2fb} with 200~\invpb\ of data
and excludes lower masses for $Z^{'}$ boson with {\tt SM}-like
couplings to fermions of 780~\GeV\ at 95\%~C.L.

The $e^+e^-$ final state can also easily be reinterpreted in
the context of technicolor models, which predict large amounts
of techniparticle production at the Tevatron
\cite{ref:technicolor_tev}. Degenerated technihadrons
($\rho_T,\omega_T$) decaying into $e^+e^-$ are excluded for
certain model parameters with masses below 367 \GeV, based on
200 \invpb\ of D\O\ data \cite{D0_technihadron_200pb}.

Similar analyses in the dimuon channel have been carried out by
CDF based on 200~\invpb\ of
data~\cite{CDF_sneutrino_ee_mm_200pb} and by D\O\ with
250~\invpb\ of data~\cite{D0_reso_mumu_Zprim_0.25fb}. The
signal process $Z^{'} \raa \tau^+ \tau^-$ has been searched by
CDF with 195 \invpb\ of data and a lower mass limit of
399~\GeV\ ({\tt SM} couplings) has been
set~\cite{CDF_sneutrino_tautau_195pb}.

A general search for resonances decaying to a neutral $e\mu$
final state based on 344~\invpb\ of CDF data has been
interpreted in the context of lepton family number violating
(LFV) $Q^l_{12}$ couplings~\cite{ref:LFV_Z_prim} of the $Z^{'}$
boson and $E_6$-like models of $U(1)^{'}$
symmetry~\cite{ref:E6_U1_prim}.

A search for a narrow-width heavy resonances decaying into top
quark pairs $X\raa t\bar{t}$ based on 913 \invpb\ has been
submitted for publication by D\O\ \cite{D0_pub_tt_reso_1fb}.
This result was recently updated with 2.1~\invfb\ of data for
the winter 2008 conferences \cite{D0_conf_tt_reso_2.1fb}.
Within a topcolor-assisted technicolor model
\cite{ref:topcolor}, the existence of a leptophobic $Z^{'}$
boson with mass $m_{Z^{'}} < 760~\GeV$ and width
$\Gamma_{Z^{'}} = 0.012m_{Z^{'}}$ are excluded at 95\% C.L.
Similarly, CDF performed several searches for a $t\bar{t}$
resonance in the lepton+jets final state using 680
\invpb~\cite{CDF_conf_tt_reso_680pb} and 1 \invfb\
\cite{CDF_conf_tt_reso_1fb} of data. A leptophobic $Z^{'}$
predicted by the topcolor theory is excluded below 725~\GeV.

\subsection{$W^{'}$ bosons}
Both CDF~\cite{CDF_pub_Wprim_205pb} and
D\O~\cite{D0_pub_Wprim_1fb} have published results of searches
for a heavy charged vector boson ($W^{'}$) decaying to an
electron-neutrino pair using 205~\invpb\ and 1~\invpb\ of data,
respectively. In the context of a $W^{'}$ with {\tt SM}
coupling to fermions, and for these data samples, the lower
mass limits are 1~\TeV\ for D\O\ and 788~\GeV\ for CDF, at
95\%~C.L. In a similar analysis that exploits 1.1~\invfb\ of
dijet data, CDF excludes at 95\% C.L. the existence of $W^{'}$
in the mass range $280 < m_{W^{'}} <
840$~\GeV~\cite{D0_jj_reso_1.13fb}.

$W^{'}$ bosons that couple to right-handed fermions may not be
able to decay to leptonic final states if the corresponding
right-handed neutrinos ($\nu_R$) are too massive. In this case,
only hadronic decays are possible. To investigate the
possibility of $W^{'}$ decaying into $t\bar{b}$, D\O\ and CDF
have used a similar approach as for their single top searches.
The D\O\ Collaboration has published a result using 230~\invpb\
of data which excludes masses between 200~\GeV\ and 610~\GeV\
for a $W^{'}$ boson {\tt SM} couplings
\cite{D0_pub_Wprim_tb_230pb}. This search has been updated with
approximately 0.9~\invfb\ of data~\cite{D0_pub_Wprim_tb_0.9fb}.
For a left-handed $W^{'}$ boson with {\tt SM} couplings, D\O\
sets a lower mass limit of 731~\GeV. For right-handed $W^{'}$
bosons, the lower mass limits on this hypothetical new particle
at 95\% C.L. are 739~\GeV\ assuming that the $W^{'}$ boson
decays to both leptons and quarks, and 768~\GeV\ if it decays
only to quarks. Assuming {\tt SM} couplings to fermions for the
$W^{'}$, CDF has used 1.9~\invfb\ of data to set limits of
$m_{W^{'}}<800$~\GeV\ when $m_{W^{'}}>m_{\nu_R}$ and
$m_{W^{'}}<825$~\GeV\ when $m_{W^{'}}<m_{\nu_R}$, at 95\% C.L.,
on $W^{'}$ resonances in the $t\bar{b}$ decay
channel~\cite{CDF_Wprim_tb_1.9fb}.

\section{Large extra dimensions}
Models postulating the existence of large extra spatial
dimensions have been proposed to solve the hierarchy problem
posed by the large difference between the electroweak symmetry
breaking scale at 1~\TeV\ and the Planck scale, at which
gravity is expected to become strong.

In the original compactified large extra dimensions model of
Arkani-Hamed, Dimopoulos and Dvali ({\tt ADD} \cite{ref:ADD}),
the effect of the extra spatial dimensions is visible as the
presence of a series of quantized energy states referred to as
graviton ($G$) states Kaluza-Klein ({\tt KK}) towers. However,
the visible states are too close in mass to be distinguished
individually and the coupling is small. Thus, it is only due to
their very large number that the Kaluza-Klein gravitons could
be observed. The direct production of gravitons, which
immediately disappear into bulk space, gives rise to an excess
of events with a high transverse energy jet (or photon) and
large missing transverse energy.

Another way to look for extra dimensions is to search for a
resonance. The first excited graviton mode predicted by the
Randall and Sundrum ({\tt RS}) model \cite{ref:ED} could be
resonantly produced at the Tevatron. The graviton is then
expected to decay to fermion-anti-fermion or diboson pairs.

\begin{figure}\begin{center}
\includegraphics[width=0.98\linewidth,height=0.45\textwidth,angle=0]{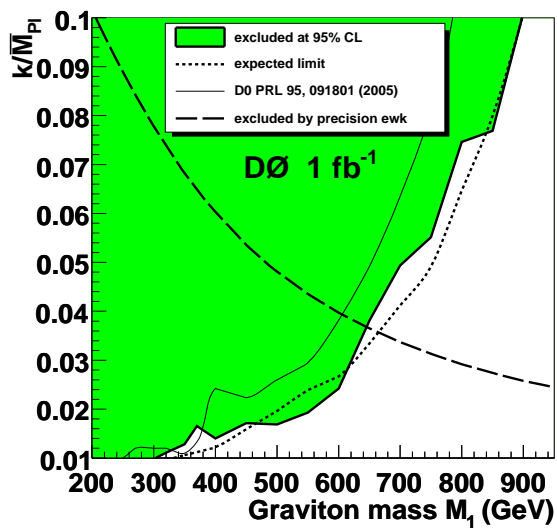}
\caption{\label{fig:D0-LED} D\O\ 95\% C.L. upper limit on $k/M_{\hbox{\rm Pl}}$
versus graviton mass from 1~\invfb\ of data for the $ee+\gamma\gamma$ final
states combined~\cite{D0_pub_ee_gamgam_RS_1fb}.}
\end{center}\end{figure}

\subsection{Graviton resonances}
The CDF and D\O\ Collaborations have searched for resonances in
their data in many different final states. Since the graviton
has spin 2, the branching fraction to the diphoton final state
is expected to be twice that of $e^{+}e^{-}$ final states. The
diphoton background is estimated from misidentified
electromagnetic objects and is extracted from the data. Results
have been published by CDF \cite{CDF_sneutrino_ee_mm_200pb}
(D\O\ \cite{D0_pub_ee_gamgam_RS_260pb}) based on 200 \invpb\
(260 \invpb) of data. The combined $ee+\gamma\gamma$ final
states have been recently published by both experiments based
on 1.2-1.3~\invfb\ of
CDF~\cite{CDF_pub_reso_ee_Z_RS_1.3fb,CDF_pub_gamma_gamma_RS_1.2fb}
(D\O~\cite{D0_pub_ee_gamgam_RS_1fb}) data. Limits obtained are
as a function of the graviton mass and the coupling parameter
($k/M_{Pl}$), as represented in Fig.~\ref{fig:D0-LED} for D\O.
The CDF Collaboration derives a lower limit of 889~\GeV\ on the
graviton mass at the 95\% C.L. for $k/M_{Pl} =$0.1. The D\O\
combined result of both $ee$ and $\gamma\gamma$ channels set
lower masses limits of 300~(900)~\GeV\ at 95\% C.L. for
$k/M_{Pl}$ = 0.01~(0.1)~\cite{D0_pub_ee_gamgam_RS_1fb}.

\begin{figure}\begin{center}
\includegraphics[width=0.98\linewidth,height=0.40\textwidth,angle=0]{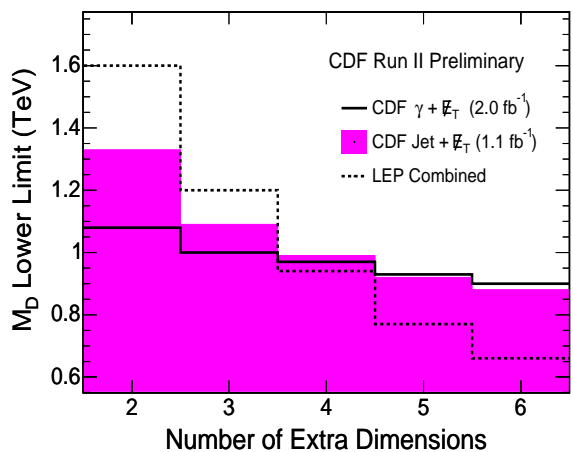}
\caption{\label{fig:CDF-LED} Limits on the fundamental Planck scale $M_D$ for various
numbers of extra dimensions from the CDF $q\bar{q} \raa \gamma+G$ and
$q\bar{q} \raa {\rm jets}+G$~\cite{CDF_ADD_2fb}, along with the LEP limit.}
\end{center}\end{figure}

Recently, CDF has released a new result based on 1.1 \invfb\ of
data in the search for a massive object decaying to a pair of Z
bosons, both of which decay to
$ee$~\cite{CDF_RS_ZZ_eeee_1.1fb}. The cross section times
branching fraction for {\tt RS} gravitons that decay to $Z$
bosons is small, leading to about one $G \raa ZZ \raa eeee$
expected event produced in 2~\invfb\ of data. For this
analysis, CDF relaxed the lepton identification requirements to
optimize the signal sensitivity. The relaxed selection admits
more background, which is then rejected by imposing kinematic
requirements on the invariant masses of the two $Z$ boson
candidates. Finally, a sample in data that is kinematically
similar to the signal has been used to estimate backgrounds in
the signal region. No event are observed with $m_{eeee} >
500$~\GeV\ for an expected background of $0\pm0.02$ in
1.1~\invfb\ of data. The search is not yet sensitive to {\tt
RS} gravitons, so cross section limits of $\sigma \times Br(G
\raa ZZ \raa eeee) \lesssim 4~\pb$ for $500< m_G < 800$~\GeV\
are set on graviton production, assuming {\tt RS} couplings.

\subsection{Jet/$\gamma$+\met}
At the Tevatron, gravitons can be produced recoiling against a
quark or a gluon jet~\cite{ref:monojet}, leading to an excess
of events with a high \pt\ jet and large \met. The resulting
topology is a monojet. Similarly, gravitons can be produced
directly in processes such as $q\bar{q} \raa \gamma+G$.

The D\O\ Collaboration has investigated {\tt KK} graviton
production with a photon and missing transverse energy in
1~\invfb\ of data~\cite{D0_pub_ADD_monophoton_1fb}. At the 95\%
C.L., D\O\ sets limits on the fundamental Planck scale ($M_D$)
from 884~\GeV\ to 778~\GeV\ for 2 to 8 extra dimensions.

The CDF Collaboration published results based on 368~\invpb\ of
monojet data~\cite{CDF_pub_ADD_monojet_368pb}. Recently, the
Collaboration has released a new result based on up to
2~\invfb\ of data which combines the jet/$\gamma+\met$ final
states~\cite{CDF_ADD_2fb}. The optimization for the {\tt ADD}
model yields a photon requirement of $\ET
> 90~\GeV$ with $\met>50~\GeV$ and a jet requirement of $\pt>150~\GeV$ with
$\met>150~\GeV$. The dominant {\tt SM} background to the
monojet search consists of $Z$ or $W$ boson plus jet
production, with the $Z$ decaying to a pair of neutrinos, or
the lepton from $W$ decay escaping detection. From the absence
of an excess in the data, limits on $M_D>$1.42 (0.95) are
derived at 95\%~C.L. for the number of extra dimensions $n_D=$2
(6). The results are shown in Fig.~\ref{fig:CDF-LED}.

The D\O\ Collaboration has searched for monojet in 85 \invpb\
of data \cite{D0_ADD_jets_85pb}. The most recent D\O\ search is
in the $q\bar{q} \raa \gamma+G$ final state with 1~\invfb\ of
data~\cite{D0_ADD_gamma_1fb}. This analysis sets limits at 95\%
C.L. on $M_D$ from 884~\GeV\ to 778~\GeV\ for 2 to 8 extra
dimensions.

\begin{figure}\begin{center}
\includegraphics[width=0.75\linewidth,height=0.47\textwidth,angle=-90]{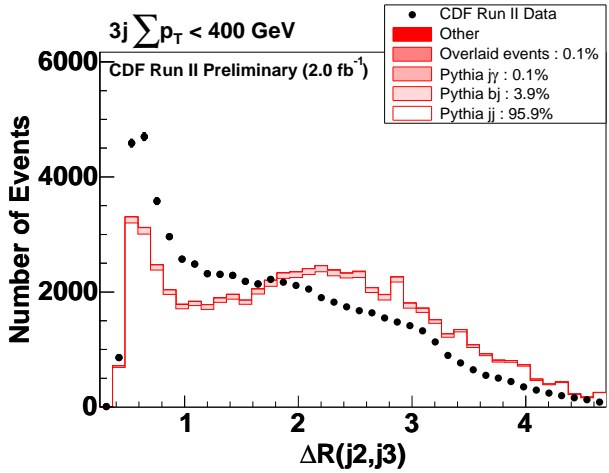}
\caption{\label{fig:CDF-vista-3j} CDF signature-based search using 2 \invfb\ of
data. The distribution illustrates an example of a shape discrepancy found by
{\sc Vista}~\cite{CDF_vista_sleuth_2fb} in the final
state consisting of exactly three jets with
$|\eta| < 2.5$ and $\pt > 17~\GeV$, and with one of the jets satisfying
$|\eta| < 1$ and $\pt > 40~\GeV$. This distribution illustrates
the effect underlying most of the {\sc Vista} shape discrepancies which were attributed
to modeling parton radiation rather than to new physics.}
\end{center}\end{figure}

\section{CDF signature-based searches}
Most of the searches presented so far have been optimized for
signatures within a specific {\tt BSM} model. However, it is
also important to search for discrepancies with {\tt SM}
prediction in a model-independent approach, instead of focusing
only on particular new physics scenarios.

Examples of such searches have been performed by the CDF
Collaboration in the channels $\gamma\gamma+X$, where $X$ could
be an electron, a muon, a photon, a tau, or missing transverse
energy~\cite{CDF_gam_gam_X_1-2fb,CDF_gam_gam_MET_2fb}, or $\ell
\gamma +
\met$~\cite{CDF_pub_l_gam_X_305pb,CDF_pub_l_gam_X_929pb}. Other
preliminary results have been presented for final state such as
$\ell+ \gamma + \met +
b$~\cite{CDF_l_gam_met_b_929pb_new1.9fb}, or $Z$-boson $+X+Y+$
anything, where $X$ and $Y$ can be leptons, photons, missing
energy, or large total transverse
energy~\cite{CDF_z_x_y_929pb}. These searches are based on
0.3-2.0~\invfb\ of data and nothing striking has been observed
yet. In particular, the \met+photon+lepton final state
analysis~\cite{CDF_pub_l_gam_X_929pb} using 1~\invfb\ of Run~II
data has not confirmed the Run I
event~\cite{CDF_Run-I_l_gam_met}.

An even more global analysis of CDF Run~II data has also been
carried out to search for indications of new phenomena in
2~\invfb\ of data~\cite{CDF_vista_sleuth_2fb}. First, a
model-independent approach ({\sc Vista}) focuses on obtaining a
panoramic view of the entire data landscape, and is sensitive
to new large-cross-section physics~\cite{CDF_vista}. It
consists of a standard set of object identification criteria,
which are used to identify isolated and energetic objects
produced in the hard collision. All objects are required to
have $\pt > 17~\GeV$. Events are partitionated into exclusive
final states labelled according to the objects
($e^\pm,\mu^\pm,\tau^\pm,\gamma,j,b,\met$) and compared to {\tt
SM} prediction. The {\tt SM} prediction is obtained with {\sc
pythia} \cite{ref:pythia} for the generation of inclusive $W,
Z, \gamma\gamma$, $\gamma j, jj, WW, WZ,$ and $ZZ$ production,
while {\sc madevent} \cite{MadEvent} provides events modeling
for $W/Z+n$ jets and {\sc herwig} \cite{ref:herwig} is used for
top quark pair production.  Detector response is modelled with
the CDF simulation and a global fit for corrections factors
(such as efficiencies and fake rates) is performed on the data.
In the end, the number of events observed supports the standard
model prediction with few discrepancies (see
Fig.~\ref{fig:CDF-vista-3j}) attributed to modeling the parton
radiation and underlying event in the data. A subset of the
{\sc Vista} comparison is given in Table.~\ref{Tab:CDF-vista}.

 A quasi-model-independent approach
({\sc Sleuth}) emphasizes the high-\pt\ tails and is
particularly sensitive to new electroweak-scale
physics~\cite{CDF_sleuth}. {\sc Sleuth} is a
quasi-model-independent search technique based on the
assumption that new electroweak-scale physics will manifest
itself as an excess of data over the {\tt SM} expectation in a
particular final state at large summed scalar transverse
momentum ($\sum \pt$). An algorithm has also been developed to
search invariant mass distributions for ``bumps" that could
indicate resonant production of new particles. Here again, this
global search for new physics in 2~\invfb\ of \pp\ collisions
reveals no indication of physics beyond the {\tt SM}.

\begin{table*}[p]
\caption{\label{Tab:CDF-vista} A subset of the
model-independent search ({\sc Vista}), which compares CDF
Run~II data with the {\tt SM} prediction in 2~\invfb\ of
data~\cite{CDF_vista_sleuth_2fb}. Events are partitioned into
exclusive final states based on standard CDF particle
identification criteria. Final states are labelled in this
table according to the number and types of objects present, and
are ordered according to decreasing discrepancy between the
total number of events expected and the total number observed
in the data. Only statistical uncertainties on the background
prediction have been included in this Table.}
\includegraphics[width=1.\linewidth,height=1.3\textwidth,angle=0]{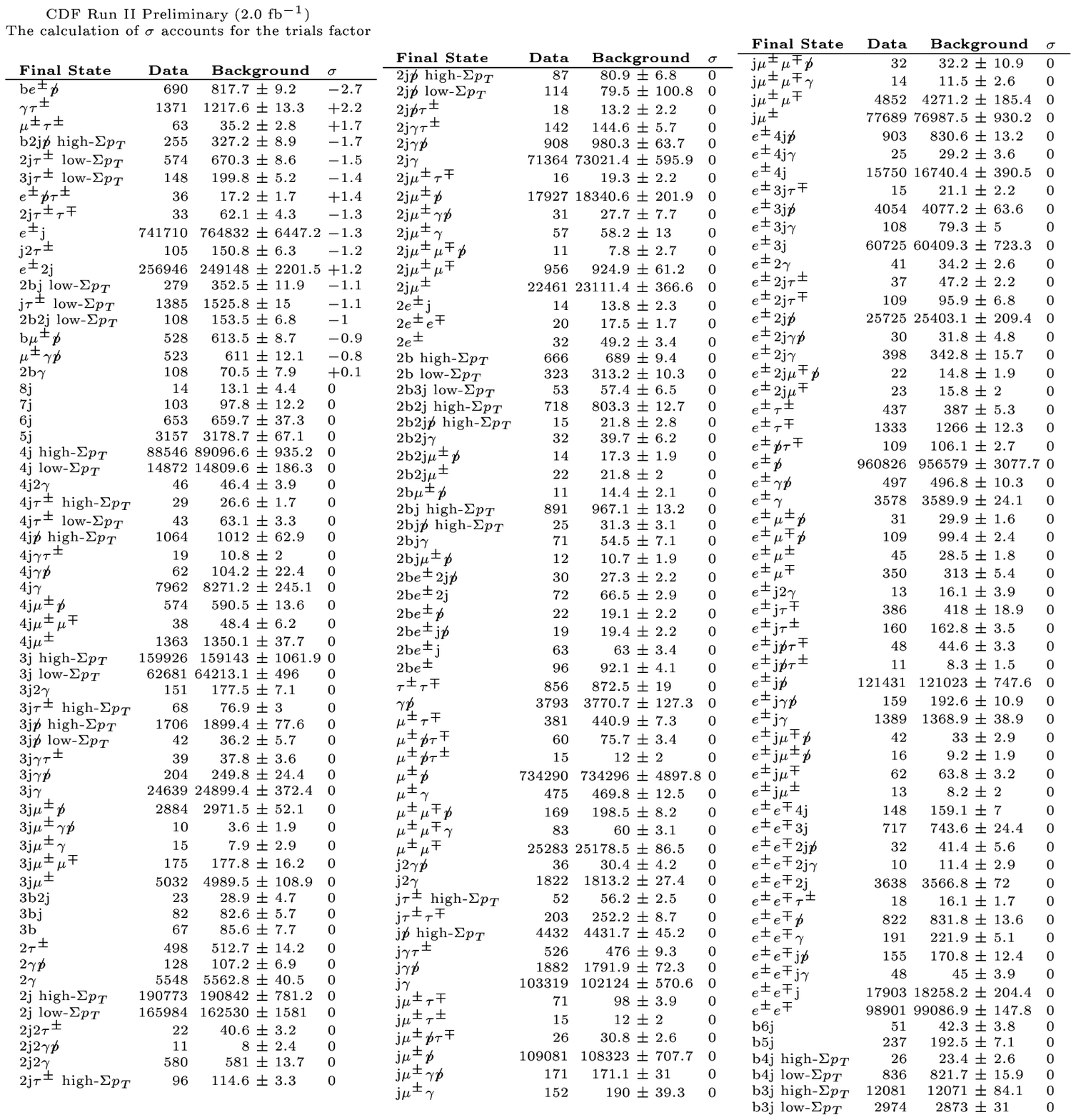}
\end{table*}

These global searches are complementary to targeted searches
with specific signatures. However, it has been demonstrated
that in order to exploit the data sample fully in term of
sensitivity for a specific model with particularly distinct
kinematic features, targeted searches out-perform these global
approaches. For instance, a 115~\GeV\ {\tt SM} Higgs boson
decaying to two $b$-tagged jets in association with a heavy
electroweak gauge boson is better treated using the $b\bar{b}$
invariant mass resonance rather than using the scalar
transverse momentum sum.